%% file: main.tex
\documentclass[11pt,a4paper]{article}

\usepackage{epsfig}
\usepackage{fullpage}
\usepackage{a4wide}

\usepackage{amsmath}
\usepackage{amssymb}

\usepackage{algorithm}
\usepackage{algorithmic}
\usepackage{xcolor}

\usepackage{latexsym}
\usepackage{graphics}
\usepackage{epic}
\usepackage{eepic}
\usepackage{color}

\newtheorem{theorem}{Theorem}[section] 
\newtheorem{corollary}[theorem]{Corollary} 
\newtheorem{claim}[theorem]{Claim} 
\newtheorem{lemma}[theorem]{Lemma} 
\newtheorem{theorem*}{Theorem}

\newtheorem{definition}[theorem]{Definition} 
\newcommand{\qed}{\rule{1em}{0in} \hspace*{\fill}$\square$\vspace{1ex}\par}
\newenvironment{proof}{\noindent {\bf Proof.}}{\hfill \qed \smallskip}
\newenvironment{proofof}[1]{\noindent {\bf Proof of #1.}}{\hfill \qed \smallskip}

\usepackage[indention=0mm,font=sl,labelfont=bf]{caption}

\newcommand{\comment}[1]{} 

\newcommand{\ACo}{\mbox{{\sf AC}$^1$}}
\newcommand{\NCo}{\mbox{{\sf NC}$^1$}}
\newcommand{\Log}{\mbox{{\sf L}}}

\newcommand{\ULcoUL}{\mbox{{\sf UL $\cap$ coUL}}}

\newcommand{\NP}{\mbox{{\sf NP}}}
\newcommand{\DET}{\mbox{{\sf DET}}}
\newcommand{\graph}{\mbox{{\sf graph}}}
\newcommand{\size}[1]{|#1|}

\newcommand{\SPP}{\mbox{{\sf SPP}}}
\newcommand{\T}{\mbox{{\sf T}}}

\begin{titlepage}
\title{Planar Graph Isomorphism is in Log-Space\footnote{
A preliminary version of the paper appeared at arXiv:0809.2319v1.
}
}
\author{Samir Datta$^1$ \and Nutan Limaye$^2$ \and Prajakta Nimbhorkar$^2$ \and \\
Thomas Thierauf$^3$\thanks{Supported by DFG grants Scho 302/7-2 and TO
200/2-2.} ~~~~~~~~ Fabian Wagner$^4$\thanks{Supported by DFG grants Scho 302/7-2 and TO 200/2-2.}\\[3ex]
$^1$ Chennai Mathematical Institute\\
        {\tt sdatta@cmi.ac.in}\\
$^2$ The Institute of Mathematical Sciences\\
        {\tt \{nutan,prajakta\}@imsc.res.in}\\
        $^3$ Fak. Elektronik und Informatik, HTW Aalen\\
        $^4$ Institut f\"ur Theoretische Informatik, \\
        Universit\"at Ulm, 89073 Ulm\\
        {\tt \{thomas.thierauf,fabian.wagner\}@uni-ulm.de} \\
}
\date{\today}
\end{titlepage}

\begin{document}
\maketitle



\begin{abstract}
Graph Isomorphism is the prime example of a computational problem with a wide difference between the best known lower and upper
bounds on its complexity. We bridge this gap for a natural and important special case, planar graph isomorphism, by presenting an
upper bound that matches the known logspace hardness \cite{Lin92}. In fact, we show the formally stronger result that planar graph
canonization is in logspace. This improves the previously known upper
bound of \ACo\ \cite{MR91}.

Our algorithm first constructs the biconnected component tree of a connected planar graph and then refines each biconnected
component into a triconnected component tree. The next step is to logspace reduce the biconnected planar graph isomorphism and
canonization problems to those for 3-connected planar graphs, which are
known to be in logspace by \cite{DLN08}. This
is achieved by using the above decomposition, and by making significant modifications to Lindell's algorithm for tree canonization,
along with changes in the space complexity analysis.

The reduction from the connected case to the biconnected case requires further new ideas, including a non-trivial case analysis and
a group theoretic lemma to bound the number of automorphisms of a
colored $3$-connected planar graph. This lemma is crucial for the
reduction to work in logspace.
\end{abstract}


\input{1-intro.tex}

\input{2-prelim.tex}
\input{3-tridecomp.tex}
\input{4-bicanon.tex}
\input{5-plcanon.tex}


\section{Conclusion}
In this paper, we improve the known upper bound for isomorphism and
canonization of planar graphs from \ACo~to \Log. This implies
\Log-completeness for this problem, thereby settling its complexity. 
An interesting question is to extend it to other important classes of
graphs.

\section{Acknowledgement}
We thank V. Arvind, Bireswar Das, Raghav Kulkarni, Meena Mahajan 
and Jacobo Tor\'{a}n for helpful discussions.


\bibliography{iso}
\bibliographystyle{alpha}

\end{document}

%% file: 1-intro.tex
\section{Introduction}\label{sec:int}
The graph isomorphism problem GI consists of deciding whether there 
is a bijection between the nodes of two graphs, which preserves edges. 
The wide gap between the known lower and upper bounds has kept alive the 
research interest in GI.

The problem is clearly in $\NP$, and, by a 
group theoretic proof, also in $\SPP$~\cite{AK06}. This is the
current frontier of our knowledge as far as upper bounds go.
The inability to give efficient algorithms for the problem would
lead one to believe that the problem is provably hard.
$\NP$-hardness is precluded by a result that states if GI is \NP-hard then the
polynomial time hierarchy collapses to the second level \cite{BHZ87,S88}. 
What is more surprising is that not even {\sf P}-hardness is known for the
problem. The best we know is that GI is hard for $\DET$~\cite{Tor04}, 
the class of problems $\NCo$-reducible to the determinant, defined by 
Cook~\cite{Coo85}. 

While this enormous gap has motivated a study of isomorphism in
\emph{general\/} graphs, it has also induced research in isomorphism
restricted to special cases of graphs, where this gap can be reduced.
Tournaments are an example of directed graphs where the $\DET$ lower bound
is preserved~\cite{Wag07}, 
while there is a quasi-polynomial time upper bound~\cite{BL83}.

Trees are an example of graphs where the lower and upper bounds match
and are $\Log$~\cite{Lin92}.
Note that for trees, the problem's complexity crucially depends on
the input encoding: if the trees are presented as strings then the lower
and upper bound are  $\NCo$~\cite{MJT98,Buss97}).
Lindell's log-space result has been extended to partial 2-trees,
also known  as generalized series-parallel graphs~\cite{ADK08}.

In this paper we consider planar graph isomorphism and settle its complexity.
Note that trees and partial 2-trees are a special cases of planar graphs.
Planar Graph Isomorphism has been studied in its own right since the early days
 of computer science.
Weinberg~\cite{Wei66} presented an $O(n^2)$ algorithm for testing isomorphism of 
$3$-connected planar graphs. 
Hopcroft and Tarjan~\cite{HT74} extended this to general planar 
graphs, improving the time complexity to $O(n\log n)$. 
Hopcroft and Wong~\cite{HW74} further 
improved it to~$O(n)$. 
Recently Kukluk, Holder, and Cook \cite{KHC04}
gave an $O(n^2)$ algorithm for planar graph isomorphism, which is
suitable for practical applications.

The parallel complexity of  Planar Graph Isomorphism  was first considered by
Miller and Reif~\cite{MR91} and Ramachandran and Reif~\cite{RR94}. 
They showed that the upper bound is~$\ACo$, see also~\cite{Ver07}. 

Recent work has dealt with a further special case viz.\ 
3-connected planar graphs.
Thierauf and Wagner~\cite{TW08} presented a new upper bound of $\ULcoUL$, 
making use of the machinery developed for
the reachability problem~\cite{RA97} and specifically for
planar reachability~\cite{ADR05,BTV07}.
They also show that the problem is~$\Log$-hard.
Further progress, in the form of a log-space algorithm is made by 
Datta, Limaye, and Nimbhorkar~\cite{DLN08},
where the 3-connected planar case is settled, by building on ideas
from~\cite{TW08} and using Reingold's construction of universal exploration
sequences~\cite{Rei05}.

\par The current work is a natural culmination
of this series where we settle the complexity question for planar 
graph isomorphism by presenting the first log-space algorithm for the problem.
In fact, we give a log-space algorithm for the {\em graph canonization problem\/}, 
to which graph isomorphism reduces.
The canonization involves assigning to each graph an isomorphism invariant,
polynomial length string.
Our algorithm consists of the following steps.
\begin{enumerate}
\item Decompose the planar graph into its biconnected components and construct a 
	\emph{biconnected component tree\/} in log-space~\cite{ADK08} (Section~\ref{sec:cpg}).
\item Decompose biconnected planar components into their triconnected components  
	to obtain a \emph{triconnected component tree\/} in log-space. 
	This is essentially a parallel implementation of the sequential 
	algorithm of ~\cite{HT73} (Section~\ref{sec:dec}).
\item Invoke the algorithm of Datta, Limaye, and Nimbhorkar~\cite{DLN08} 
	to canonize the triconnected components of the graph.
\item Canonize biconnected planar graphs by applying tree canonization ideas 
	from~\cite{Lin92} to their triconnected component trees. Note that, 
	pairwise isomorphism of two trees labelled with the canons of their
	components does not imply isomorphism of the corresponding graphs. 
	Lindell's algorithm and complexity analysis had to be modified in a 
	non-trivial way for this step to work in log-space (Section~\ref{sec:cbp}).
\item Canonize planar graphs using their biconnected component trees, 
	and for biconnected components their triconnected component trees.
	For the canonization again, we use the basic structure of Lindell's algorithm.
	The new ingredients here are, an intricate case analysis, and
	a group theoretic lemma (Lemma~\ref{lem:aut}) to bound the number of 
	automorphisms of a coloured $3$-connected planar graph (Section~\ref{sec:cpg}). 
	It also requires a detailed analysis of the interferences of both tree structures.
\end{enumerate}

Our algorithm works recursively at various places.
The major challenge when developping a recursive log-space algorithm
is that very little can be stored at each level of the recursion.
But we must anyway be able to continue a computation
at the point where we made the recursive call,
when we return from the recursion.
We solve these problems by identifying in each case appropriate graph properties
that have a short description and which can be used to
recompute the point where we started from.

%% file: 2-prelim.tex
\section{Preliminaries}\label{sec:pre}
In this section, we recall some basic graph theoretic notions. 

\par A graph $G=(V,E)$ is \emph{connected\/} if there is a path between any 
two vertices in~$G$. 
For $U \subseteq V$ let $G(U)$ be the \emph{induced subgraph\/} of~$G$ on~$U$.
A vertex $v\in V$ is an \emph{articulation point\/} if 
$G(V\setminus \{v\})$ 
is not connected. A pair of vertices $u,v \in V$ is a \emph{separating
pair\/} if
$G(V\setminus\{u,v\})$ is not connected.
A \emph{biconnected graph\/} contains no articulation points. A
\emph{3-connected graph\/} contains no separating pairs.
A \emph{triconnected graph\/} is either a $3$-connected graph or a cycle
or a $3$-bond. A \emph{$k$-bond\/} is a graph consisting of two vertices
joined by $k$~edges.
A pair of vertices $(a,b)$ is said to be \emph{$3$-connected} if there are
three or more vertex-disjoint paths between them.

For a node~$v$ let $d(v)$ be the maximal distance that
$v$~has to any of the other nodes of~$G$.
Let~$C$ be the set of nodes~$v$ of~$G$ that have minimal value~$d(v)$.
The set~$C$ is called the {\em center of\/}~$G$.
In other words,
vertices in the center minimize the maximal distance from other vertices in the graph.
Note that if~$G$ is a tree such that every path from a leave to a leave has even length,
then the center consists of only one node,
namely the midpoint of a longest path in the  tree.

\par 
Let~$E_v$ be the set of edges incident to~$v$.
A permuatation~$\rho_v$ on~$E_v$ that has only one cycle is called a \emph{rotation\/}.
A \emph{rotation scheme\/} for a graph~$G$ is a set~$\rho$ of rotations,
\[\rho = \{\rho_v \mid v\in V \text{ and $\rho_v$ is a rotation on $E_v$} \}.\]
Let $\rho^{-1}$  be the set of inverse rotations,
$\rho^{-1} = \{\rho_v^{-1} \mid v\in V \}$.
A rotation scheme~$\rho$ describes an embedding of graph~$G$ in the plane.
If the embedding is planar, we call~$\rho$ a {\em planar rotation scheme\/}.
Note that in this case~$\rho^{-1}$ is a planar rotation scheme as well.
Allender and Mahajan~\cite{AM00} showed that a planar rotation scheme for a planar graph
can be computed in log-space.

\par Two graphs $G_1=(V_1,E_1)$ and $G_2=(V_2,E_2)$ are 
said to be \emph{isomorphic\/} ($G_1 \cong G_2$) 
if there is a bijection $\phi:V_1\rightarrow V_2$
such that $(u,v)\in E_1$ if and only if $(\phi(u),\phi(v))\in E_2$.
\emph{Graph isomorphism} (GI) is the problem of deciding whether two given
graphs are isomorphic.

\par A planar graph $G$, along with its planar embedding (given by~$\rho$) 
is called a \emph{plane graph\/} $\widehat{G} = (G,\rho)$. 
A plane graph divides the plane into regions. Each
such region is called a \emph{face}.
Let Planar-GI be the special case of
GI when the given graphs are planar.
The biconnected (respectively, $3$-connected) planar GI
is a special case of Planar-GI when the graphs are biconnected
($3$-connected) planar graphs.

\par Let~$\mathcal{G}$ be a class of graphs.
Let $f : \mathcal{G} \rightarrow \{0, 1\}^*$ be a function such that for all $G, H \in \mathcal{G}$
we have $G \cong H \Leftrightarrow f(G) = f(H)$. 
Then~$f$ computes a \emph{complete invariant} for $\mathcal{G}$. If~$f$ computes
for~$G$ a graph~$f(G)$ such that $G \cong f(G)$ then we call~$f(G)$ the
\emph{canon\/} for~$G$.

\par By \Log\ we denote the languages computable by a log-space bounded 
Turing machine. 

%% file: 3-tridecomp.tex
\section{Decomposition of Biconnected Planar Graphs}\label{sec:dec}

In this section, we prove the following theorem.

\begin{theorem}\label{thm:dec}
The decomposition of biconnected planar graphs into triconnected components is in log-space.
\end{theorem}

Hopcroft and Tarjan \cite{HT73} presented a sequential algorithm for the 
decomposition of a biconnected planar graph into its triconnected 
components. Their algorithm recursively removes separating pairs from
the graph and puts a copy of the separating pair in each of the
components so formed. The nodes in the separating pair are connected by
a virtual edge. If simple cycles are split at any intermediate steps then
they are combined later. This gives a decomposition which is unique \cite{M37}.
We describe a log-space algorithm for such a decomposition
of a biconnected planar graph.
We start with definitions and then prove some properties of separating pairs.

\begin{definition}
In a plane graph $\widehat{G}$,
a separating pair $\{a,b\}$ is said to \emph{span a face\/} $f$ if both its 
endpoints $a,b$ lie on the boundary of $f$.
Let $v_0,v_1,\ldots,v_k $ be a face boundary. Two separating pairs
$\{v_{i},v_{j}\}$, $\{v_{i'},v_{j'}\}$ are called \emph{intersecting\/} if $i < i' <j <j'$, and
\emph{non-intersecting\/} otherwise. 
\end{definition}

\begin{lemma}
Every separating pair spans some face.
\end{lemma}

To see this, note that 
in a plane graph $\widehat{G}$, 
a split component of a separating pair is embedded in some face.
This can be considered as the spanned face.
A separating pair $\{a,b\}$ that spans a face~$f$ is called 
$3$-{\em connected\/}
if there are at least three vertex-disjoint paths between $a,b$ i.e.
there is a path between $a,b$ in~$\widehat{G}$ which is vertex-disjoint 
from the boundary of~$f$. 
The following lemma enables us to remove all the $3$-connected separating pairs
simultaneously.

\begin{lemma}\label{lem:nint}
In a plane graph $\widehat{G}$,
$3$-connected separating pairs which span the same face are non-intersecting.
\end{lemma}
\begin{proof}
Suppose $\{a,c\}$ and $\{b,d\}$ are two $3$-connected
intersecting separating pairs on face $f$ in $\widehat{G}$
and let~$P$ be a path outside~$f$ from~$b$ to~$d$.
In particular,~$P$ does not pass through~$a$ or~$c$.

As the pair  $b,d$ is $3$-connected, it cannot be separated from the rest
of the graph by any other separating pair.
Let~$v$ be a vertex that gets separated from~$b$ and~$d$
when~$a$ and~$c$ are removed from the graph.
Since~$v$ lies outside~$f$, there is a path outside~$f$ from~$a$ via~$v$ to~$c$.
Since the graph is planar, this path must intersect~$P$.
Thus there is a path from~$v$ to~$b$ and~$d$ that does not pass through~$a$ or~$c$.
This contradicts the assumption that removal of $a$ and $c$ separated $v$
from $b$ and $d$.
\end{proof}

\begin{definition} 
Call a set of vertices $V' \subseteq V(\widehat{G})$ \emph{separable\/} 
if there exists a $3$-connected separating pair $\{a,b\}$ in $V(\widehat{G})$ 
such that the removal of $\{a,b\}$ divides $V'$ into different connected components.
Otherwise $V'$ is called \emph{inseparable}.
Given an inseparable triple $\tau = \{ u,v,w \}$, define 
$C_\tau = \{ x \mid \{u,v,w,x\} \text{ is inseparable} \}$.
\end{definition}

Note that the nodes of a simple cycle are trivially inseparable
because there are no $3$-connected separating pairs.
The following lemma states that  except for cycles,
all biconnected graphs have $3$-connected separating pairs and hence
the sets $C_{\tau}$ defined above are the 3-connected components of such a graph.

\begin{lemma}\label{lem:3con}
Let~$G$ be a biconnected planar graph.
If~$G$ is not $3$-connected and not a cycle
then~$G$ has a $3$-connected separating pair.
\end{lemma}

\begin{proof}
Let $G$ be neither $3$-connected nor a cycle
and let $a,b$ be a separating pair of~$G$.
If~$a$ and~$b$ are 3-connected then we are done.
So assume that~$a$ and~$b$ are not 3-connected.

Let~$f$ be a face spanned by~$a$ and~$b$.
Then~$a$ and~$b$ are connected by two vertex-disjoint paths,
say $P_1$ and $P_2$,
which form the boundary of~$f$,
and the removal of $(a,b)$ separates these two paths.
Since~$G$ is not a single cycle,
it has more faces apart from~$f$.
Therefore~$f$ shares some of its edges with another face, say $f'$.
Consider the common boundary between~$f$ and~$f'$.
The endpoints of this boundary, say $(u,v)$ have three
vertex-disjoint paths between them, and hence are $3$-connected.

Both~$u$ and~$v$ lie on~$P_1$ or both lie on~$P_2$,
since otherwise $P_1$ and $P_2$ will not be separated on the removal of $(a,b)$.
Without loss of generality,
assume that $u, v \in P_1$.
Let $P_1=\{a=v_1,v_2, \dots,v_k=b\}$ and
consider all $3$-connected pairs $(v_i,v_j)$ of vertices that lie on $P_1$.
Pick a pair, say  $(v_i,v_j)$, that is maximally apart on $P_1$.
We claim that $(v_i,v_j)$ is a separating pair:
if not,
there exists a path outside~$f$ from $v_{i'}$ to~$v_j$ for some $i' < i$,
or from $v_{j'}$ to~$v_i$ for some $j' > j$.
In the first case, $(v_{i'},v_j)$ is a $3$-connected pair that is further apart than
$(v_i,v_j)$,
in the second case the same holds for $(v_i,v_{j'})$.
But this contradicts the choice of $(v_i,v_j)$.
\end{proof}

Hence, with inseparable triples we can compute triconnected components.
If a triple of vertices is inseparable, then it is part of the same triconnected component.
For distinct $\tau_1,\tau_2$, the sets $C_{\tau_1}$ and $C_{\tau_2}$ are either disjoint or identical.
This allows us to identify any such $C_\tau$ with the lexicographical
smallest $\tau_0$ (considering the labels of vertices in $\tau$ lexicographically sorted) 
such that $C_\tau = C_{\tau_0}$.
This is the approach of Algorithm~\ref{alg:dec} below.
\begin{algorithm}[h]
{\bf Input:} Biconnected planar graph $G=(V,E)$. \\
\parbox{4in}{{\bf Output:} The triconnected components of $G$.}
\begin{algorithmic}[1]
\STATE fix a planar embedding $\widehat{G}$ of $G$.
\STATE {\bf for all} faces $f$ of $\widehat{G}$ {\bf do}
\STATE 		\mbox{\quad} $S_f \leftarrow \{(u,v)|(u,v)$ is a $3$-connected 
		separating pair that spans $f\}$
\STATE \mbox{\quad} $S \leftarrow \cup_{f\in \widehat{G}}S_f$ the set of $3$-connected separating pairs
\STATE {\bf for all} $(u,v)\in S$ {\bf do}
\STATE 		\mbox{\quad} {\bf if} $(u,v)\in E$ {\bf then} output a $3$-bond for $(u,v)$
\STATE compute the set of all inseparable triples $\tau_1,\ldots,\tau_k$
\STATE {\bf for} $i \leftarrow 1$ to $k$ {\bf do} ~~~~~~~~~~~~~~~~{\scriptsize \{compute $3$-connected components\}}
\STATE \mbox{\quad} {\bf if} $\forall h < i$ \; $\tau_i \cup \tau_h$ is a separable set {\bf then} 
\STATE		\mbox{\qquad} $C_i \leftarrow \tau_i$ ~~~~~~~~~~~~{\scriptsize \{create new $3$-connected component using 
inseparable triple $\tau_i$ \}}
\STATE		\mbox{\qquad} {\bf for} $j \leftarrow i+1$ to $k$ {\bf do}
\STATE		\mbox{\qquad\quad} {\bf if} $\tau_i \cup \tau_j$ is an inseparable set
		{\bf then} $C_i \leftarrow C_i \cup \tau_j$
\STATE 		\mbox{\qquad} output the induced subgraph on $C_i$ 
		without edges corresponding to $3$-bonds, 
		 \mbox{\qquad\quad}including virtual edges $\{s \in S \mid s \subseteq C_i\}$
\end{algorithmic}
\caption{Algorithm to decompose a graph into triconnected components.}\label{alg:dec}
\end{algorithm}

\begin{figure}[!ht]
\begin{center}
\scalebox{0.85}{\input{Figures/decomp.eepic}}
\end{center}
\caption{
The decomposition of a biconnected planar graph $\widehat{G}$.
Its triconnected components are $G_1,\dots,G_4$ and the
corresponding triconnected component tree is $T$.
In $\widehat{G}$, the pairs $(a,b)$ and $(c,d)$ are $3$-connected
separating pairs. The inseparable triples are
$\{a,b,c\}$, $\{b,c,d\}$, $\{a,c,d\}$, $\{a,b,d\}$, $\{a,b,f\}$, and $\{c,d,e\}$. Hence
the triconnected components are the induced graphs 
$G_1$ on $\{a,b,f\}$, 
$G_2$ on $\{a,b,c,d\}$, and
$G_4$ on $\{c,d,e\}$.
Since the 3-connected separating pair $(c,d)$ is connected by an edge in~$\widehat{G}$,
we also get $\{c,d\}$ as triple-bond~$G_3$.
The virtual edges corresponding to the 3-connected separating pairs are drawn with dashed lines.
}
\label{fig:decomp}
\end{figure}
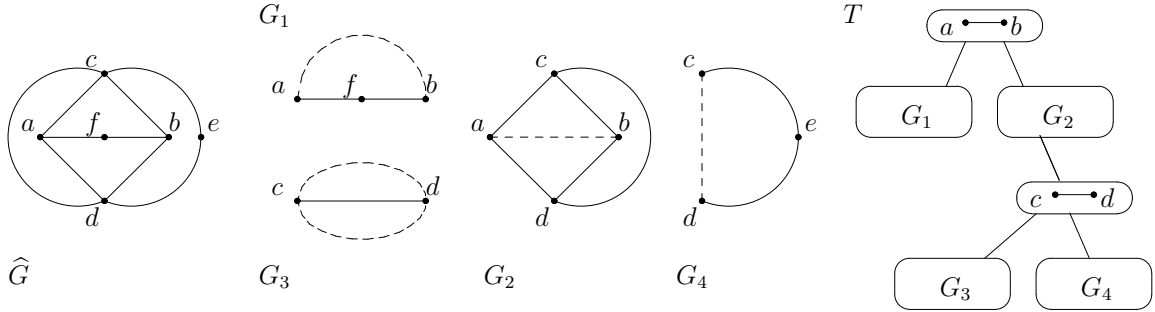

First, the algorithm computes all $3$-connected separating pairs in the set~$S$.
From these, we get all the  $3$-bonds.
The {\bf for}-loop from line~8 on computes the 3-connected components $C_{\tau}$:
In line~9, we search for the first inseparable triple $\tau \notin \{ C_{\tau_h} \mid 1 \leq h < i \}$
that can be separated from all previous ones.
In lines~11 and~12, we search for all $\tau_j \subseteq C_{\tau_i}$.
By Lemma~\ref{lem:3con}, it suffices to consider the pairs in~$S$
to check whether a set is separable or not.
The set $C_i$ finally equals $C_{\tau_i}$.
In line~13 we compute the triconnected component induced by $C_{\tau_i}$.
An example of a decomposition is provided in Figure~\ref{fig:decomp}.

\par 
Each step in the algorithm can be implemented in log-space.
For instance, a combinatorial embedding for
planar graphs can be computed in log-space \cite{AM00}. 
Separating pairs,
inseparable triples 
and  the triconnected components
can be computed in log-space, making oracle queries
to undirected reachability~\cite{Rei05}.

\paragraph{The triconnected component tree.}
Construct a graph~$T$ such that its nodes correspond to 
triconnected components and separating pairs, see Figure~\ref{fig:decomp}.
There is an edge between a {\em triconnected component node} and a {\em separating pair node} 
if the vertices of the separating pair are contained in the triconnected  component. Two triconnected component nodes or separating pair nodes do not share 
an edge.

It is easy to see that $T$ is a tree, referred to as the {\em triconnected component tree of\/}~$G$.
Conversely, given~$T$, we define  
$\graph(T) = G$, the graph which has the triconnected component tree~$T$.
We list some properties of~$T$.

\begin{lemma}\label{lem:propertiesOfTree}
The graph~$T$ defined above has the following properties:
\begin{enumerate}
\item $T$ is a tree and
all the leaves of~$T$ are triconnected components.
\item Each path in~$T$ is an alternating path of separating pairs and triconnected components.
	Hence, a path between two leaves always contains an odd number of nodes
	and therefore $T$~has a unique center node.
\item With an arbitrary separating pair node as root,  $T$~has odd depth. 
\item A $3$-bond is introduced as a child of a separating pair only as
	an indicator that the vertices of the separating pair have an edge between
	them in~$G$. Hence a $3$-bond is always a leaf node. In \cite{HT73}
	it is a $k$-bond, where~$k$ is the number of components formed by the removal of the separating
	pair. Observe, $k$ is the number of children of its parent separating pair and can be
	computed easily.
\end{enumerate}
\end{lemma}

\begin{proof}
We only show the first claim.
Suppose~$T$ has a cycle~$C$.
By definition,
$C$~is an alternating cycle of separating pairs and triconnected components,
$C = (p_1,c_1, p_2,c_2\ldots ,p_r, c_r, p_1)$.
Remove any separating pair $p_i$ from~$C$.
Then the triconnected components $c_{i-1}$ and $c_i$ 
remain connected through the other elements of the cycle, contradicting the 
assumption that $p_i$~separates them.
\end{proof}

%% file: Figures/decomp.eepic
\setlength{\unitlength}{0.00087489in}
\begingroup\makeatletter\ifx\SetFigFontNFSS\undefined%
\gdef\SetFigFontNFSS#1#2#3#4#5{%
  \reset@font\fontsize{#1}{#2pt}%
  \fontfamily{#3}\fontseries{#4}\fontshape{#5}%
  \selectfont}%
\fi\endgroup%
{\renewcommand{\dashlinestretch}{30}
\begin{picture}(8037,2199)(0,-10)
\put(690,777){\blacken\ellipse{36}{36}}
\put(690,777){\ellipse{36}{36}}
\put(1140,1227){\blacken\ellipse{36}{36}}
\put(1140,1227){\ellipse{36}{36}}
\put(240,1227){\blacken\ellipse{36}{36}}
\put(240,1227){\ellipse{36}{36}}
\put(690,1227){\blacken\ellipse{36}{36}}
\put(690,1227){\ellipse{36}{36}}
\put(690,1677){\blacken\ellipse{36}{36}}
\put(690,1677){\ellipse{36}{36}}
\put(7620,822){\blacken\ellipse{36}{36}}
\put(7620,822){\ellipse{36}{36}}
\put(7350,822){\blacken\ellipse{36}{36}}
\put(7350,822){\ellipse{36}{36}}
\put(6990,2037){\blacken\ellipse{36}{36}}
\put(6990,2037){\ellipse{36}{36}}
\put(6720,2037){\blacken\ellipse{36}{36}}
\put(6720,2037){\ellipse{36}{36}}
\put(1365,1227){\blacken\ellipse{36}{36}}
\put(1365,1227){\ellipse{36}{36}}
\put(2040,777){\blacken\ellipse{36}{36}}
\put(2040,777){\ellipse{36}{36}}
\put(2940,777){\blacken\ellipse{36}{36}}
\put(2940,777){\ellipse{36}{36}}
\put(2940,1497){\blacken\ellipse{36}{36}}
\put(2940,1497){\ellipse{36}{36}}
\put(2040,1497){\blacken\ellipse{36}{36}}
\put(2040,1497){\ellipse{36}{36}}
\put(2490,1497){\blacken\ellipse{36}{36}}
\put(2490,1497){\ellipse{36}{36}}
\put(3840,1677){\blacken\ellipse{36}{36}}
\put(3840,1677){\ellipse{36}{36}}
\put(3840,777){\blacken\ellipse{36}{36}}
\put(3840,777){\ellipse{36}{36}}
\put(4290,1227){\blacken\ellipse{36}{36}}
\put(4290,1227){\ellipse{36}{36}}
\put(3390,1227){\blacken\ellipse{36}{36}}
\put(3390,1227){\ellipse{36}{36}}
\put(4875,1677){\blacken\ellipse{36}{36}}
\put(4875,1677){\ellipse{36}{36}}
\put(5550,1227){\blacken\ellipse{36}{36}}
\put(5550,1227){\ellipse{36}{36}}
\put(4875,777){\blacken\ellipse{36}{36}}
\put(4875,777){\ellipse{36}{36}}
\dashline{60.000}(2040,1497)(2076,1668)(2127,1770)
	(2211,1851)(2307,1911)(2424,1944)
	(2535,1947)(2649,1920)(2757,1863)
	(2835,1785)(2901,1692)(2937,1584)(2940,1497)
\dashline{60.000}(2040,777)(2076,880)(2127,941)
	(2211,989)(2307,1025)(2424,1045)
	(2535,1047)(2649,1031)(2757,997)
	(2835,950)(2901,894)(2937,829)(2940,777)
\dashline{60.000}(2040,777)(2076,674)(2127,613)
	(2211,565)(2307,529)(2424,509)
	(2535,507)(2649,523)(2757,557)
	(2835,604)(2901,660)(2937,725)(2940,777)
\put(877.500,1227.000){\arc{975.000}{4.3176}{8.2488}}
\put(502.500,1227.000){\arc{975.000}{1.1760}{5.1072}}
\put(4027.500,1227.000){\arc{975.000}{4.3176}{8.2488}}
\put(5062.500,1227.000){\arc{975.000}{4.3176}{8.2488}}
\path(240,1227)(690,1677)(1140,1227)
	(690,777)(240,1227)(1140,1227)
\put(7050,1332){\arc{210}{1.5708}{3.1416}}
\put(7050,1482){\arc{210}{3.1416}{4.7124}}
\put(7650,1482){\arc{210}{4.7124}{6.2832}}
\put(7650,1332){\arc{210}{0}{1.5708}}
\path(6945,1332)(6945,1482)
\path(7050,1587)(7650,1587)
\path(7755,1482)(7755,1332)
\path(7650,1227)(7050,1227)
\path(6720,1902)(6585,1587)
\path(6990,1902)(7125,1587)
\path(7350,822)(7620,822)
\put(7185,792){\arc{210}{1.5708}{3.1416}}
\put(7185,807){\arc{210}{3.1416}{4.7124}}
\put(7785,807){\arc{210}{4.7124}{6.2832}}
\put(7785,792){\arc{210}{0}{1.5708}}
\path(7080,792)(7080,807)
\path(7185,912)(7785,912)
\path(7890,807)(7890,792)
\path(7785,687)(7185,687)
\path(7242,1239)(7377,924)
\path(7242,1239)(7377,924)
\put(7320,117){\arc{210}{1.5708}{3.1416}}
\put(7320,267){\arc{210}{3.1416}{4.7124}}
\put(7920,267){\arc{210}{4.7124}{6.2832}}
\put(7920,117){\arc{210}{0}{1.5708}}
\path(7215,117)(7215,267)
\path(7320,372)(7920,372)
\path(8025,267)(8025,117)
\path(7920,12)(7320,12)
\path(7451,692)(7586,377)
\path(6720,2037)(6990,2037)
\put(6555,2007){\arc{210}{1.5708}{3.1416}}
\put(6555,2022){\arc{210}{3.1416}{4.7124}}
\put(7155,2022){\arc{210}{4.7124}{6.2832}}
\put(7155,2007){\arc{210}{0}{1.5708}}
\path(6450,2007)(6450,2022)
\path(6555,2127)(7155,2127)
\path(7260,2022)(7260,2007)
\path(7155,1902)(6555,1902)
\put(6060,1332){\arc{210}{1.5708}{3.1416}}
\put(6060,1482){\arc{210}{3.1416}{4.7124}}
\put(6660,1482){\arc{210}{4.7124}{6.2832}}
\put(6660,1332){\arc{210}{0}{1.5708}}
\path(5955,1332)(5955,1482)
\path(6060,1587)(6660,1587)
\path(6765,1482)(6765,1332)
\path(6660,1227)(6060,1227)
\put(6330,117){\arc{210}{1.5708}{3.1416}}
\put(6330,267){\arc{210}{3.1416}{4.7124}}
\put(6930,267){\arc{210}{4.7124}{6.2832}}
\put(6930,117){\arc{210}{0}{1.5708}}
\path(6225,117)(6225,267)
\path(6330,372)(6930,372)
\path(7035,267)(7035,117)
\path(6930,12)(6330,12)
\path(7215,687)(6855,372)
\path(3840,777)(4290,1227)
\path(4290,1227)(3840,1677)
\path(3840,1677)(3390,1227)
\path(3390,1227)(3840,777)
\dashline{60.000}(3390,1227)(4290,1227)
\dashline{60.000}(4875,1677)(4875,777)
\path(2040,777)(2940,777)
\path(2940,1497)(2040,1497)
\put(555,1272){\makebox(0,0)[lb]{\smash{{\SetFigFontNFSS{12}{14.4}{\rmdefault}{\mddefault}{\updefault}$f$}}}}
\put(1140,1272){\makebox(0,0)[lb]{\smash{{\SetFigFontNFSS{12}{14.4}{\rmdefault}{\mddefault}{\updefault}$b$}}}}
\put(1410,1272){\makebox(0,0)[lb]{\smash{{\SetFigFontNFSS{12}{14.4}{\rmdefault}{\mddefault}{\updefault}$e$}}}}
\put(555,1722){\makebox(0,0)[lb]{\smash{{\SetFigFontNFSS{12}{14.4}{\rmdefault}{\mddefault}{\updefault}$c$}}}}
\put(555,597){\makebox(0,0)[lb]{\smash{{\SetFigFontNFSS{12}{14.4}{\rmdefault}{\mddefault}{\updefault}$d$}}}}
\put(7665,732){\makebox(0,0)[lb]{\smash{{\SetFigFontNFSS{12}{14.4}{\rmdefault}{\mddefault}{\updefault}$d$}}}}
\put(7035,1947){\makebox(0,0)[lb]{\smash{{\SetFigFontNFSS{12}{14.4}{\rmdefault}{\mddefault}{\updefault}$b$}}}}
\put(105,1272){\makebox(0,0)[lb]{\smash{{\SetFigFontNFSS{12}{14.4}{\rmdefault}{\mddefault}{\updefault}$a$}}}}
\put(6540,1947){\makebox(0,0)[lb]{\smash{{\SetFigFontNFSS{12}{14.4}{\rmdefault}{\mddefault}{\updefault}$a$}}}}
\put(7170,732){\makebox(0,0)[lb]{\smash{{\SetFigFontNFSS{12}{14.4}{\rmdefault}{\mddefault}{\updefault}$c$}}}}
\put(6270,1317){\makebox(0,0)[lb]{\smash{{\SetFigFontNFSS{12}{14.4}{\rmdefault}{\mddefault}{\updefault}$G_1$}}}}
\put(2940,1542){\makebox(0,0)[lb]{\smash{{\SetFigFontNFSS{12}{14.4}{\rmdefault}{\mddefault}{\updefault}$b$}}}}
\put(2355,1542){\makebox(0,0)[lb]{\smash{{\SetFigFontNFSS{12}{14.4}{\rmdefault}{\mddefault}{\updefault}$f$}}}}
\put(1860,1542){\makebox(0,0)[lb]{\smash{{\SetFigFontNFSS{12}{14.4}{\rmdefault}{\mddefault}{\updefault}$a$}}}}
\put(2940,822){\makebox(0,0)[lb]{\smash{{\SetFigFontNFSS{12}{14.4}{\rmdefault}{\mddefault}{\updefault}$d$}}}}
\put(7260,1317){\makebox(0,0)[lb]{\smash{{\SetFigFontNFSS{12}{14.4}{\rmdefault}{\mddefault}{\updefault}$G_2$}}}}
\put(7530,102){\makebox(0,0)[lb]{\smash{{\SetFigFontNFSS{12}{14.4}{\rmdefault}{\mddefault}{\updefault}$G_4$}}}}
\put(4290,1272){\makebox(0,0)[lb]{\smash{{\SetFigFontNFSS{12}{14.4}{\rmdefault}{\mddefault}{\updefault}$b$}}}}
\put(3705,1722){\makebox(0,0)[lb]{\smash{{\SetFigFontNFSS{12}{14.4}{\rmdefault}{\mddefault}{\updefault}$c$}}}}
\put(3705,597){\makebox(0,0)[lb]{\smash{{\SetFigFontNFSS{12}{14.4}{\rmdefault}{\mddefault}{\updefault}$d$}}}}
\put(3255,1272){\makebox(0,0)[lb]{\smash{{\SetFigFontNFSS{12}{14.4}{\rmdefault}{\mddefault}{\updefault}$a$}}}}
\put(5595,1272){\makebox(0,0)[lb]{\smash{{\SetFigFontNFSS{12}{14.4}{\rmdefault}{\mddefault}{\updefault}$e$}}}}
\put(4740,1722){\makebox(0,0)[lb]{\smash{{\SetFigFontNFSS{12}{14.4}{\rmdefault}{\mddefault}{\updefault}$c$}}}}
\put(4740,597){\makebox(0,0)[lb]{\smash{{\SetFigFontNFSS{12}{14.4}{\rmdefault}{\mddefault}{\updefault}$d$}}}}
\put(4695,192){\makebox(0,0)[lb]{\smash{{\SetFigFontNFSS{12}{14.4}{\rmdefault}{\mddefault}{\updefault}$G_4$}}}}
\put(3345,192){\makebox(0,0)[lb]{\smash{{\SetFigFontNFSS{12}{14.4}{\rmdefault}{\mddefault}{\updefault}$G_2$}}}}
\put(1770,192){\makebox(0,0)[lb]{\smash{{\SetFigFontNFSS{12}{14.4}{\rmdefault}{\mddefault}{\updefault}$G_3$}}}}
\put(15,192){\makebox(0,0)[lb]{\smash{{\SetFigFontNFSS{12}{14.4}{\rmdefault}{\mddefault}{\updefault}$\widehat{G}$}}}}
\put(1770,2037){\makebox(0,0)[lb]{\smash{{\SetFigFontNFSS{12}{14.4}{\rmdefault}{\mddefault}{\updefault}$G_1$}}}}
\put(5865,2037){\makebox(0,0)[lb]{\smash{{\SetFigFontNFSS{12}{14.4}{\rmdefault}{\mddefault}{\updefault}$T$}}}}
\put(6540,102){\makebox(0,0)[lb]{\smash{{\SetFigFontNFSS{12}{14.4}{\rmdefault}{\mddefault}{\updefault}$G_3$}}}}
\put(1860,822){\makebox(0,0)[lb]{\smash{{\SetFigFontNFSS{12}{14.4}{\rmdefault}{\mddefault}{\updefault}$c$}}}}
\end{picture}
}

%% file: 4-bicanon.tex
\section{Canonization of Biconnected Planar Graphs}\label{sec:cbp}
In this section, we give a log-space algorithm to canonize biconnected planar graphs. 
For this, we define an isomorphism ordering on triconnected component trees which is similar to that of Lindell's tree isomorphism ordering.
We first give a brief overview of Lindell's algorithm and
then describe our canonization procedure.

\subsection{Overview of Lindell's Algorithm}

Lindell~\cite{Lin92} gave a log-space algorithm for tree canonization.
The algorithm is based on an order relation~$\leq$ on trees defined below.
The order relation has the property that 
two trees~$S$ and~$T$ are isomorphic if and only if~$S = T$.
Because of this property it is called a {\em canonical order\/}.
Clearly,
an algorithm that decides the order can be used as an isomorphism test.
Lindell showed how to extend such an algorithm to compute a canon for a tree
in log-space.
Let ~$S$ and~$T$ be two trees with root~$s$ and~$t$, respectively.
The canonical order is defined as follows.~$S< T$ if
\begin{enumerate}
\item \label{itm:size} 
	$|S|<|T|$, or
\item \label{itm:chi} 
	$|S|=|T|$ but $\#s<\#t$, 
	where $\#s$ and $\#t$ are the number of children of~$s$ and~$t$, respectively, 
	or
\item \label{itm:ind} 
	$|S|=|T|$ and $\#s=\#t = k$, 
	but $(S_1, \ldots, S_k) < (T_1, \ldots, T_k)$ lexicographically, 
	where it is inductively assumed that
	$S_1\leq \ldots \leq S_k$ and $T_1\leq \ldots \leq T_k$ are the ordered
	subtrees of $S$ and $T$ rooted at the~$k$ children of~$s$ and~$t$, respectively.
\end{enumerate}

The comparisons in steps~\ref{itm:size} and~\ref{itm:chi} can be made in log-space. 
Lindell proved that
even the third step can be performed in log-space using 
\emph{two-pronged\/} depth-first search, and \emph{cross-comparing\/} 
only a child of~$S$ with a child of~$T$. This is briefly described below:
\begin{itemize}
\item Find the number of minimal sized children of~$s$ and~$t$. If these
	numbers are different then the tree with a larger number of minimal children
	is declared to be smaller. If equality is found then remember the
	minimal size and check for the next size. This process is continued till
	an inequality in the sizes is detected or all the children of~$s$ and~$t$
	are exhausted.
\item   If~$s$ and~$t$ have the same number of children of each size then
	assume that the children of~$s$ and~$t$ are partitioned into
	\emph{size-classes\/} (referred to as \emph{blocks\/} in \cite{Lin92})
	in the increasing order of the the sizes of the subtrees rooted at them.
	That is,
	the~$k$ children of~$s$ and~$t$ are partitioned into groups, 
	such that the $i$-th group is of cardinality~$k_i$
	and the subtrees in the $i$-th group all have size~$N_i$,
	where $N_1 < N_2 < \cdots$.
	It follows that $\sum_i k_i =k$ and $\sum_i k_i N_i = n-1$.
	Then compare the children in each size-class recursively as follows:

\vspace{\topsep}
\noindent{\bf Case 1,~$k=0$.}
Hence~$s$ and~$t$ have no children. 
They are isomorphic as all one-node trees are isomorphic.
We conclude that $S = T$.

\vspace{\itemsep}
\noindent{\bf Case 2,~$k=1$.}
	Recursively consider the grand-children of~$s$ and~$t$.
	No space is needed for the recursive call.

\vspace{\itemsep}
\noindent{\bf Case 3,~$k\geq 2$.}
	For each of the subtrees~$S_j$  compute its order profile.
	The order profile consists of three counters, $c_<$, $c_>$ and $c_=$. 
	These counters indicate the number of subtrees in the size-class of~$S_j$
	that are respectively smaller than, greater than, or equal to~$S_j$.
	The counters are computed by making cross-comparisons.

	\par Note, that isomorphic subtrees in the same size-class have the same order profile.
	Therefore, it suffices to check that each such order profile occurs the same number of times 
	in each size-class in~$S$ and~$T$.
	To perform this check, compare the different order profiles of every size class
	in lexicographic order.
	The subtrees in the size-class~$i$ of~$S$ and~$T$, which is currently being considered, 
	with a count $c_<=0$ form the first
	isomorphism class. The size of this isomorphism class is compared across
	the trees by comparing the values of the $c_=$ variables. If these values match
	then both trees have the same number of minimal children. 
	Note that the lexicographical next larger order profile has the current value of $c_< + c_=$ 
	as its value for the $c_<$-counter.
	
	\par This way, one can loop through all the order profiles.
	If a difference in the order profiles of the subtrees of~$S$ and~$T$ is found
	then the lexicographical smaller order profile defines the smaller tree.
	
	\par The last order profile considered is the one with $c_< + c_=  = k$  for the current counters.
	If this point is passed without uncovering an inequality
	then the trees must be isomorphic and it follows that $S = T$.
\end{itemize}

Since $\sum_i k_i N_i  \leq n$, the following recursion equation for the space complexity holds.
For each new size class, the work-tape allocated for the former computations can be reused.

$$ \mathcal{S}(n) = \max_i\{\mathcal{S}(N_i) + O(\log k_i)\}   \leq  
	\max_i\{~\mathcal{S}\!\left( \frac{n}{k_i} \right) + O(\log k_i) \}, $$
where $k_i \geq 2$ for all~$i$.
It is not hard to see that  $\mathcal{S}(n) = O(\log n)$.

\subsection{Isomorphism Order of Triconnected Component Trees}\label{subsec:ord}

\begin{figure}[!ht]
\begin{center}
\scalebox{0.82}{\input{Figures/trees3.eepic}}
\end{center}
\caption{Triconnected component trees.}\label{fig:trees}
\end{figure}
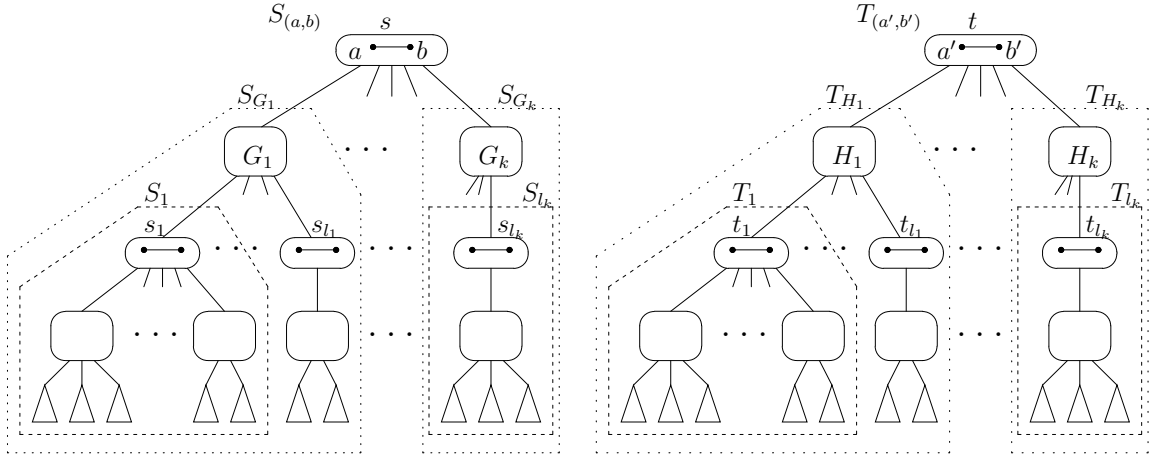

We now describe an isomorphism order procedure for two triconnected
component trees~$S$ and~$T$, corresponding to two biconnected planar
graphs~$G$ and~$H$, respectively.
Both, $S$ and $T$ are rooted at separating
pair nodes, as described in Section~\ref{sec:dec}, say $s=(a,b)$ and $t=(a',b')$. 
Therefore we also write~$S_{(a,b)}$ and~$T_{(a',b')}$.
They have separating pair nodes at odd levels and triconnected component nodes
at even levels.
Figure~\ref{fig:trees} shows two trees to be compared.
Our canonical order procedure is more complex than Lindell's algorithm, because
each node of the tree is a separating pair or a triconnected component.
Thus in particular, unlike in the case of Lindell's algorithm, two leaves in a
triconnected component tree are not always isomorphic. 
In the easiest case the components to these leaves are not of the same size.
We start by defining the size of a triconnected component tree.

\begin{definition} \label{def:TriSize}
For a triconnected component tree~$T$,
the {\em size of an individual component node\/}~$C$ of~$T$ is the number~$n_C$ of nodes in~$C$.
Note that the separating pair nodes are counted in in every component where they occur.
The {\em size of the tree\/}~$T$, denoted by~$|T|$,
is the sum of the sizes of its component nodes.
\end{definition}

Note that the size of~$T$ is at least as large as the number of 
vertices in $\graph(T)$, the graph corresponding to the triconnected component tree $T$.

We define the isomorphism order~$<_{\tt{T}}$ for~$S_{(a,b)}$ and~$T_{(a',b')}$ by
first comparing their sizes, 
then the number of children of~$s$ and~$t$.
These two steps are exactly the same as in Lindell's algorithm. 
If equality is found in these two steps, then in the third step we make 
recursive comparisons of the subtrees of~$S_{(a,b)}$ and~$T_{(a',b')}$.
However,
here it does not suffice to compare the order profiles of the subtrees in the
different size classes as in Lindell's algorithm explained above.
We need a further comparison step to
ensure that~$G$ and~$H$ are indeed isomorphic. 

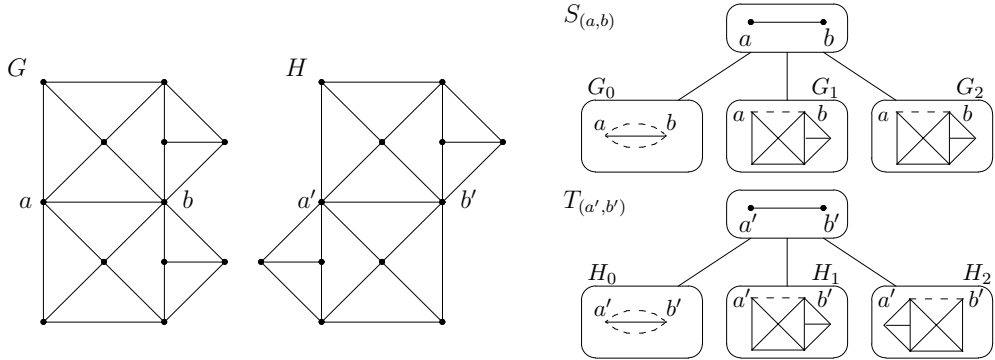
\begin{figure}[!ht]
\begin{center}
\scalebox{0.8}{\input{Figures/orientation.eepic}}
\end{center}
\caption{
The graphs $G$ and $H$ have the same triconnected component trees but are not isomorphic.
In $S_{(a,b)}$, the $3$-bonds form one isomorphism class~$I_1$ and the other two
components form the second isomorphism class~$I_2$, as they all are pairwise isomorphic. 
The non-isomorphism is detected by comparing the directions given to the parent separating pair.
We have $p=2$ isomorphism classes
and  for the orientation counters we have $O_1=O_1'=(0,0)$, 
whereas $O_2=(2,0)$ and $O_2'=(1,1)$ and hence $O'_2$ is lexicographically smaller than~$O_2$.
Therefore we have $T_{(a',b')} <_{\tt{T}} S_{(a,b)}$.
}
\label{fig:orientation}
\end{figure}

To see this assume that~$s$ and~$t$ 
have two children each,~$G_1$,~$G_2$ and~$H_1$,~$H_2$
such that~$G_1 \cong H_1$ and~$G_2 \cong H_2$.
Still we cannot conclude that~$G$ and~$H$ are isomorphic
because it is possible that
the isomorphism between~$G_1$ and~$H_1$ maps~$a$ to~$a'$ and~$b$ to~$b'$,
but the isomorphism between~$G_2$ and~$H_2$ maps~$a$ to~$b'$ and~$b$ to~$a'$.
Then these two isomorphisms cannot be
extended to an isomorphism between~$G$ and~$H$.
For an example see Figure~\ref{fig:orientation} on page~\pageref{fig:orientation}.

To handle this, we introduce
the notion of an \emph{orientation of a separating pair\/}. 
A separating pair gets an orientation from subtrees
rooted at its children. Also, every subtree rooted at a 
triconnected component node gives an orientation to the parent separating pair.
If the orientation is consistent,
then we define $S_{(a,b)} =_{\tt{T}} T_{(a',b')}$ and we will show 
that~$G$ and~$H$ are isomorphic in this case.

\paragraph{Isomorphism order of two subtrees rooted at triconnected components.}
We consider the isomorphism order of two subtrees~$S_{G_i}$ and~$T_{H_j}$ rooted
at triconnected component nodes~$G_i$ and~$H_j$, respectively.
We distinguish the following cases.

\vspace{\topsep}\noindent
{\bf Case~1,} {\em $G_i$ and $H_j$ are of different types\/}.
	$G_i$ and $H_j$ can be either $3$-bonds or cycles or $3$-connected components. 
	If the types of~$G_i$ and~$H_j$ are different, 
	we immediately detect an inequality, as it suffices to check whether 
	each of them is a cycle or a 3-bond or neither of them.
	We define a canonical order among subtrees rooted at triconnected components
	in this ascending order: 3-bond, cycle, 3-connected component, such that
	e.g. $S_{G_i} <_{\tt{T}} T_{H_j}$ if $G_i$ is a 3-bond and $H_j$ is a cycle.

\vspace{\itemsep}\noindent
{\bf Case~2,} {\em $G_i$ and~$H_j$ are $3$-bonds\/}.
	In this case, $S_{G_i}$ and~$T_{H_j}$  are leaves,
	immediately define $S_{G_i} =_{\tt{T}} T_{H_j}$.
 	Clearly, $G_i\cong H_j$ as all $3$-bonds are isomorphic.

\vspace{\itemsep}\noindent
{\bf Case~3,} {\em $G_i$ and~$H_j$ are cycles or $3$-connected components\/}.
	We construct the canons of~$G_i$ and~$H_j$ and compare
	them bit-by-bit. To canonize a cycle, we traverse it starting from the
	virtual edge that corresponds to its parent, and then traversing the
	entire cycle along the edges encountered. There are two possible
	traversals depending on which direction of the starting edge is chosen.
	Thus, a cycle has two possible canons. 
	
	\par To canonize a $3$-connected component~$G_i$, we use the log-space
	algorithm from Datta, Limaye, and Nimbhorkar~\cite{DLN08}.
 	Besides~$G_i$, the algorithm gets as input a starting edge
	and a combinatorial embedding~$\rho$ of~$G_i$.
	We always take the
	virtual edge $(a,b)$ corresponding to~$G_i$'s parent as the starting edge.
	Then there are two choices for the direction of this edge, $(a,b)$ or $(b,a)$. 
	Further, a $3$-connected graph has two planar combinatorial
	embeddings~\cite{Whi33}.
	Hence, there are four possible ways to canonize~$G_i$. 
	
	\par We start the canonization of~$G_i$ and~$H_j$ in all the possible ways (two if
	they are cycles and four if they are $3$-connected components),
	and compare these canons bit-by-bit.
	Let $C_g$ and $C_h$ be two canons to be compared.
	The base case is that~$G_i$ and~$H_j$ are leaf nodes
	and therefore contain no further virtual edges.
	In this case we use the lexicographic order between~$C_g$ and~$C_h$.
	If~$G_i$ and~$H_j$ contain further virtual edges then these edges are specially treated
	in the bitwise comparison of~$C_g$ and~$C_h$:

\begin{enumerate}
\item If a virtual edge is traversed in the construction of one of the canons~$C_g$ or~$C_h$ 
	but not in the other,
	then we define the one without the virtual edge to be the {\em smaller canon.\/}
\item If~$C_g$ and~$C_h$ encounter virtual edges $(u,v)$ and
	$(u',v')$ corresponding to a child of~$G_i$ and~$H_j$, respectively, we need to
	recursively compare the subtrees rooted at $(u,v)$ and $(u',v')$. 
	If we find in the recursion that one of the subtrees is smaller than the other,
	then the canon with the smaller subtree is defined to be the {\em smaller canon.\/}
\item If we find that the subtrees rooted at $(u,v)$ and
	$(u',v')$ are equal then we look at the orientations given to $(u,v)$ and $(u',v')$ 
	by their children.
	This orientation, called the {\em reference orientation\/}, is defined below.
	If one of the canons traverses the virtual edge in the direction of its reference orientation
	but the other one not,
	then the one with the same direction is defined to be the {\em smaller canon.\/}
\end{enumerate}

We eliminate the canons which were found to be the larger canons in 
at least one of the comparisons.
In the end, the canons that are not eliminated are the {\em minimum canons\/}.
If we have minimum canons for both~$G_i$ and~$H_j$ then we
define $S_{G_i} =_{\tt{T}} T_{H_j}$.
The construction of the canons also defines an isomorphism 
between the subgraphs described by~$S_{G_i}$ and~$T_{H_j}$,
i.e. $\graph(S_{G_i}) \cong \graph(T_{H_j})$.
For a single triconnected component this follows from Datta, Limaye, and Nimbhorkar~\cite{DLN08}.
If the trees contain several components,
then our definition of $S_{G_i} =_{\tt{T}} T_{H_j}$ guarantees that we can combine
the isomorphisms of the components to an isomorphism between $\graph(S_{G_i})$ and $\graph(T_{H_j})$.

Finally, we define the {\em orientation given to the parent separating pair of\/}~$G_i$ and~$H_j$
as the direction in which the minimum canon traverses this edge.
If the minimum canons are obtained for both choices of directions of
the edge, we say that~$S_{G_i}$ and~$T_{H_j}$ are 
{\em symmetric about their parent separating pair\/}, 
and thus do not give an orientation.
This finishes the description of the order for the case of subtrees rooted at triconnected components.

\vspace{\topsep}

Observe, that we do not need to
compare the sizes and the degree of the root nodes of~$S_{G_i}$ and~$T_{H_j}$ 
in an intermediate step, as it is done in Lindell's algorithm for subtrees.
That is, because the degree of the root node~$G_i$ is encoded as the number of
virtual edges in~$G_i$. The size of~$S_{G_i}$ is checked by the length of the 
minimal canons for~$G_i$ and when we compare the sizes of the children of the root node~$G_i$
with those of~$H_j$.

\paragraph{Isomorphism order of two subtrees rooted at separating pairs.}
The first three steps of the isomorphism ordering are performed similar
to that of \cite{Lin92} maintaining the order profiles. Now we assume that
the subtrees are partitioned into isomorphism classes. The additional step 
involves comparison of {\em orientations} given by the corresponding isomorphism
classes defined as follows: 

\par Let $(G_1,\ldots, G_k)$ be the children of the root $(a,b)$ of~$S_{(a,b)}$,
and $(S_{G_1},\ldots, S_{G_k})$ be the subtrees rooted at $(G_1,\ldots, G_k)$.
Similarly let $(H_1,\ldots, H_k)$ be the children of the root $(a',b')$ of~$T_{(a',b')}$
and $(T_{H_1},\ldots, T_{H_k})$ be the subtrees rooted at $(H_1,\ldots, H_k)$.
We first order the subtrees,
say $S_{G_1} \leq_{\tt{T}} \cdots \leq_{\tt{T}} S_{G_k}$ and $T_{H_1} \leq_{\tt{T}} \cdots \leq_{\tt{T}} T_{H_k}$,
and verify that $S_{G_i} =_{\tt{T}} T_{H_i}$ for all~$i$.
If we find an inequality then
the one with the smallest index $i$ defines the order between~$S_{(a,b)}$ and~$T_{(a',b')}$.
Now assume that $S_{G_i} =_{\tt{T}} T_{H_i}$ for all~$i$.
Inductively, the corresponding split components are isomorphic, i.e.\
$\graph(S_{G_i}) \cong \graph(T_{H_i})$ for all~$i$.

\par The next comparison concerns the orientation of $(a,b)$ and $(a',b')$.
We already explained above the orientation given by each of the~$S_{G_i}$'s to $(a,b)$.
We define a \emph{reference orientation\/} for the root nodes $(a,b)$ and $(a',b')$
which is given by their children. This is done as follows.
We partition  $(S_{G_1},\ldots, S_{G_k})$ into  classes of isomorphic subtrees,
say $I_1 <_{\tt{T}} \ldots <_{\tt{T}} I_p$ for some $p \leq k$, and similar
$(T_{H_1}, \ldots, T_{H_k})$ into $I'_1 <_{\tt{T}} \ldots <_{\tt{T}} I'_p$.
It follows that~$I_j$ and~$I'_j$ contain the same number of subtrees for every~$j$.

\begin{itemize}
\item 
Consider the orientation given to $(a,b)$ by an isomorphism class~$I_j$:
For each isomorphism class~$I_j$
	we compute an {\em orientation counter\/}, which is a pair $O_j=(c^{\rightarrow}_j,c^{\leftarrow}_j)$,
	where
	$c^{\rightarrow}_j$  is the number of subtrees of~$I_j$  which give one orientation, say $(a,b)$,
	and $c^{\leftarrow}_j$ is the number of subtrees from $I_j$  which give the other orientation, $(b,a)$.
	The  larger number decides the orientation given to $(a,b)$.
	If these numbers are equal, or if each component in this class is symmetric about $(a,b)$ 
	then no orientation is given to $(a,b)$ by this class, and the class is
	said to be {\em symmetric about\/} $(a,b)$.
	Note that in an isomorphism class,
	either all or none of the components are symmetric about the parent. 

\item
The \emph{reference orientation of\/} $(a,b)$ is defined as the 
	orientation given to $(a,b)$ by the smallest non-symmetric isomorphism class.
	If all isomorphism classes are  symmetric about $(a,b)$, 
	then we say that $(a,b)$ has \emph{no reference orientation\/}.

We order all the orientation counters $O_j=(c^{\rightarrow}_j,c^{\leftarrow}_j)$ such that the first component $c^{\rightarrow}_j$ 
is the counter  for the reference orientation of $(a,b)$.
\end{itemize}

Let $O'_j= (d^{\rightarrow}_j,d^{\leftarrow}_j)$ be the corresponding orientation counters for
the isomorphism classes~$I'_j$.
Now we compare the orientation counters~$O_j$ and~$O'_j$ for $j = 1, \dots,p$.
If they are all pairwise equal, 
then the graphs~$G$ and~$H$ are isomorphic and we define $S_{(a,b)} =_{\tt{T}} T_{(a',b')}$.
Otherwise, let~$j$ be the smallest index such that $O_j \not= O'_j$.
Then we define $S_{(a,b)}<_{\tt{T}} T_{(a',b')}$ if 
$O_j$ is lexicographically smaller than~$ O'_j$,
and $T_{(a',b')} <_{\tt{T}} S_{(a,b)}$ otherwise.
This finishes the definition of the order.
For an example, see Figure~\ref{fig:orientation}. 

\paragraph{Summary of the steps in the isomorphism order.} \label{summary-tri}
The isomorphism order of two triconnected component trees $S$ and $T$ rooted
at separating pairs $s=(a,b)$ and $t=(a',b')$ is defined $S_{(a,b)}<_{\tt{T}} T_{(a',b')}$ if:

\begin{enumerate}
\item \label{itm:1} $|S_{(a,b)}|<|T_{(a',b')}|$ or

\item  \label{itm:2}$|S_{(a,b)}|=|T_{(a',b')}|$ but $\#s<\#t$ or

\item \label{itm:3} $|S_{(a,b)}|=|T_{(a',b')}|$, $\#s=\#t=k$, 
	but $(S_{G_1}, \ldots , S_{G_k}) <_{\tt{T}}  (T_{H_1} ,\ldots , T_{H_k})$ lexicographically,
	where we assume that $S_{G_1} \leq_{\tt{T}} \ldots \leq_{\tt{T}} S_{G_k}$
	and $T_{H_1} \leq_{\tt{T}} \ldots \leq_{\tt{T}} T_{H_k}$ are the ordered subtrees of~$S_{(a,b)}$
	and~$T_{(a',b')}$, respectively.
	To compute the order between the subtrees~$S_{G_i}$ and~$T_{H_i}$
	we compare lexicographically the canons of~$G_i$ and~$H_i$
	and {\em recursively\/} the subtrees rooted at the children of~$G_i$ and~$H_i$.
	Note, that these children are again separating pair nodes.

\item  \label{itm:4} $|S_{(a,b)}|=|T_{(a',b')}|$, $\#s=\#t=k$, 
	$(S_{G_1} \leq_{\tt{T}} \ldots \leq_{\tt{T}} S_{G_k})=_{\tt{T}} (T_{H_1}\leq_{\tt{T}} \ldots \leq_{\tt{T}} T_{H_k})$, 
	but $(O_1,\ldots, O_p)<(O'_1,\ldots ,O'_p)$ lexicographically, 
	where~$O_j$ and~$O_j'$ are the orientation counters
	of the $j^{th}$ isomorphism classes~$I_j$ and~$I'_j$ of all the~$S_{G_i}$'s and the~$T_{H_i}$'s.
\end{enumerate}
We say that two triconnected component trees~$S_{e}$ and~$T_{e'}$ are \emph{equal
according to the isomorphism order\/}, denoted by $S_{e}=_{\tt{T}} T_{e'}$, if
neither $S_{e}<_{\tt{T}} T_{e'}$ nor $T_{e'}<_{\tt{T}} S_{e}$ holds.
The following theorem states that
two trees are $=_{\tt{T}}$-equal, precisely when the underlying graphs are isomorphic.

\begin{theorem}
\label{thm:bcor}
The biconnected planar graphs~$G$ and~$H$ are isomorphic if and only if 
there is a choice of 
separating pairs $e,e'$ in~$G$ and~$H$ 
such that $S_e =_{\tt{T}}  T_{e'}$ when rooted at~$e$ and~$e'$, respectively.
\end{theorem}

\begin{proof}
Assume that $S_e =_{\tt{T}} T_{e'}$.
The argument is an induction on the depth of the trees
that follows the inductive definition of the isomorphism order.
The induction goes from depth~$d$ to~$d+2$. If the grandchildren 
of separating pairs, say~$s$ and~$t$, are $=_{\tt{T}}$-equal up to step~4, 
then we compare the children of~$s$ and~$t$.
If they are equal then
we can extend the $=_{\tt{T}}$-equality to the separating pairs~$s$ and~$t$.

When subtrees are rooted at separating pair nodes,
the comparison describes an order on the subtrees which correspond to 
split components of the separating pairs. The order describes an isomorphism 
among the split components.

When subtrees are rooted at triconnected component nodes, say~$G_i$ and~$H_j$,
the comparison states equality if the components have the same canon, 
i.e. are isomorphic.
By the  induction hypothesis
we know that the children rooted at virtual edges of~$G_i$ and~$H_j$
are isomorphic.
The equality in the comparisons inductively describes 
an isomorphism between the vertices in the children of the root nodes.

Hence, the isomorphism between the children at any level
can be extended to an isomorphism between 
the corresponding subgraphs in~$G$ and~$H$ and 
therefore to~$G$ and~$H$ itself.

\vspace{\topsep}
The reverse direction holds obviously as well.
Namely, if~$G$ and~$H$ are isomorphic and
there is an isomorphism 
that maps the separating pair $(a,b)$ of~$G$
to the separating pair $(a',b')$ of~$H$,
then the triconnected component trees~$S_{(a,b)}$ of~$G$
and~$T_{(a',b')}$ of~$H$
rooted respectively at $(a,b)$ and $(a',b')$ will clearly be equal.
Hence, such an isomorphism mapps separating pairs of $G$ onto separating pairs of $H$.
This isomorphism describes a permutation on the split components of separating pairs,
which means we have a permutation on triconnected components, the children of the separating pairs.
By induction hypothesis, the children (at depth $d+2$) of two such 
triconnected components are isomorphic and equal according to $=_{\tt{T}}$.
More formally,
one can argue inductively on the depth of~$S_{(a,b)}$ and~$T_{(a',b')}$.
\end{proof}


\subsection{Complexity of the Isomorphism Order Algorithm}\label{sec:ComplTriIsoOrdAlg}

We analyse the space complexity of the isomorphism order algorithm.
The first two steps of the isomorphism order algorithm 
can be computed in log-space as in Lindell's algorithm~\cite{Lin92}.
We show that steps~$3$ and~$4$ can also be performed in log-space. 
We use the algorithm of Datta, Limaye, and Nimbhorkar~\cite{DLN08} to canonize
a triconnected component~$G_i$ of size~$n_{G_i}$ in space $O(\log{n_{G_i}})$.

\paragraph{Comparing two subtrees rooted at triconnected components.}
For this, we consider two subtrees~$S_{G_i}$ and~$T_{H_j}$ 
with~$\size{S_{G_i}} = \size{T_{H_j}} = N$ rooted
at triconnected component nodes~$G_i$ and~$H_j$, respectively.
The cases that~$G_i$ and~$H_j$ are of different types or are both 3-bonds
are easy to handle.
Assume now that both are cycles or 3-connected components.
Then we start constructing and comparing all the possible canons of~$G_i$ and~$H_j$.
We eliminate the larger ones and make recursive comparisons
whenever the canons encounter virtual edges simultaneously.
We can keep track of the canons, which are not eliminated, in constant space.

\par Suppose we construct and compare two canons~$C_g$ and~$C_h$
and consider the moment when we encounter virtual edges $(a,b)$ and $(a',b')$
in~$C_g$ and~$C_h$, respectively.
Now we recursively compare the subtrees rooted at the separating pair nodes $(a,b)$ and $(a',b')$.
Note, that we cannot afford to store the entire work-tape content.
It suffices to store the information of
\begin{itemize}
\item	the canons which are not eliminated,
\item	which canons encountered the virtual edges corresponding to $(a,b)$ and $(a',b')$, and
\item	the direction in which the virtual edges $(a,b)$ and $(a',b')$ were encountered. 
\end{itemize}
This takes altogether $O(1)$ space.

\par When a recursive call is completed, 
we look at the work-tape and compute the canons $C_G$ and $C_h$.
Therefore, recompute the parent separating pair of the component, 
where the virtual edge $(a,b)$ is contained. With a look on the 
bits stored on the work-tape, 
we can recompute the canons $C_g$ and $C_h$.
Recompute for them, where $(a,b)$ and $(a',b')$ are encountered
in the correct direction of the edges
and resume the computation from that point.

Although we only need $O(1)$ space per recursion level,
we cannot guarantee yet, that the implementation of the algorithm described so far 
works in log-space. The problem is,
that the subtrees where we go into recursion might be of size~$> N/2$
and in this case the recursion depth can get too large.
To get around this problem, we check
whether~$G_i$ and~$H_j$ have a large child, before starting the
construction and comparison of their canons.
A \emph{large child} is a child which has  size~$> N/2$.
If we find a large child of $G_i$ and $H_j$
then we compare them a priori and store the result of their recursive comparison.
Because~$G_i$ and~$H_j$ can have at most one large child each,
this needs only $O(1)$ additional bits.
Now, whenever the virtual edges
corresponding to the large children from~$S_{G_i}$ and~$T_{H_j}$ are encountered
simultaneously in a canon of~$G_i$ and~$H_j$, the stored result can be
used, thus avoiding a recursive call.

\paragraph{Comparing two subtrees rooted at separating pairs.}
Consider two subtrees~$S_{(a,b)}$ and~$T_{(a',b')}$ of  size~$N$, 
rooted at separating pair nodes $(a,b)$ and $(a',b')$,
respectively.
We start comparing  all the subtrees~$S_{G_i}$ and~$T_{H_j}$ 
of~$S_{(a,b)}$ and~$T_{(a',b')}$, respectively.
These subtrees are rooted at triconnected components
and we can use the implementation described above.
Therefore, we store on the work-tape the counters $c_<,c_=,c_>$.
If they turn out to be pairwise equal,
we compute the orientation counters~$O_j$ and~$O'_j$ 
of the isomorphism classes~$I_j$ and~$I'_j$, for all~$j$.
The isomorphism classes are computed via the order profiles of the subtrees,
as in Lindell's algorithm.

\par When we return from recursion,
it is an easy task to find $(a,b)$ and $(a',b')$ again,
since a triconnected component has a unique parent, which 
always is a separating pair node. Since we have the counters $c_<,c_=,c_>$
and the orientation counters on the work-tape, 
we can proceed with the next comparison.

\par Let~$k_j$ be the number of subtrees in~$I_j$.
The counters $c_<,c_=,c_>$ and the orientation counters need altogether 
at most $O(\log{k_j})$ space.
From the orientation counters we also get the reference orientation of $(a,b)$.
Let~$N_j$ be the  size of the subtrees in~$I_j$.
Then we have $N_j \leq N/k_j$.
This would lead to a log-space implementation as in Lindell's algorithm
except for the case that~$N_j$ is large, i.e.\ $N_j > N/2$.

We handle the case of large children as above:
we recurse on large children a priori and store the result in $O(1)$ bits. 
Then we process the other subtrees of~$S_{(a,b)}$ and~$T_{(a',b')}$. When
we reach the  size-class of the large child, we know the reference
orientation, if any. Now we use the stored result to compare the
orientations given by the large children to their respective parent, and
return the result accordingly.

As seen above, while comparing two trees of size~$N$, 
the algorithm uses  no space for making a recursive call for a subtree of 
size~larger than~$N/2$, and it uses $O(\log{k_j})$ space 
if the subtrees are of size~at most $N/k_j$, where $k_j \geq 2$. 
Hence we get the same recurrence for the space $\mathcal{S}(N)$ as Lindell:
	\begin{displaymath}\label{eq:space-bic}
	\mathcal{S}(N) \leq \max_j~\mathcal{S}\!\left(\frac{N}{k_j} \right)+O(\log{k_j}),
	\end{displaymath}
where $k_j \geq 2$ for all~$j$.
Thus $\mathcal{S}(N)=O(\log{N})$.
Note that the number~$n$ of nodes of~$G$ is in general smaller than~$N$,
because the separating pair nodes occur in all components split off by this pair.
But we certainly have $n < N \leq O(n^2)$ \cite{HT73}. 
This proves the following theorem.

\begin{theorem}
The isomorphism order between two triconnected component trees of 
biconnected planar graphs can be computed in log-space.
\end{theorem}

\subsection{The Canon of a Biconnected Planar Graph} \label{sec:bicanon}

Once we know the ordering among the subtrees, 
it is straight forward to output the canon of the triconnected
component tree $T$. 
We traverse $T$ in the tree isomorphism order as in Lindell \cite{Lin92}, 
outputting the canon of each of the nodes along with virtual edges and delimiters. 
That is, we output a `[' while going down a subtree, and `]' while going up a subtree. 

We need to choose a separating pair as root for the tree.
Since there is no distinguished separating pair,
we simply cycle through all of them.
Since there are less than~$n^2$ many separating pairs, 
a log-space transducer can cycle through all of them
and can determine the separating pair which, when chosen as the root,
leads to the lexicographically minimum canon of~$S$. 
We describe the canonization procedure for a fixed root, say $(a,b)$.

The canonization procedure has two steps.
In the first step we compute what we call a {\em canonical list\/} for~$S_{(a,b)}$.
This is a list of the edges of~$G$, also including virtual edges.
In the second step
we compute the final canon from the canonical list.

\paragraph{Canonical list of a subtree rooted at a separating pair.}
Consider a subtree~$S_{(a,b)}$  rooted at the separating pair node~$(a,b)$.
We start with computing the reference orientation of~$(a,b)$
and output the edge in this direction.
This can be done by comparing the children of the separating pair node~$(a,b)$ 
according to their isomorphism order with the help of the oracle.
Then we recursively output the canonical lists of the subtrees of~$(a,b)$ 
according to the increasing isomorphism order.
Among isomorphic siblings, those which give the reference orientation to
the parent are considered before those which give the reverse orientation.
We denote this canonical list of edges $l(S,a,b)$.
If the subtree rooted at $(a,b)$ does not give any
orientation to $(a,b)$, then take that orientation for $(a,b)$, in which
it is encountered during the construction of the above canon
of its parent.

\par Assume now, the parent of $S_{(a,b)}$ is a triconnected component.
In the symmetric case, $S_{(a,b)}$ does not give an orientation $(a,b)$ to its parent.
Then take the reference orientation which is given to the parent of all siblings.

\paragraph{Canonical list  of a  subtree rooted at a triconnected component.}
Consider the subtree~$S_{G_i}$ rooted at the triconnected component node~$G_i$.
Let~$(a,b)$ be the parent separating pair of~$S_{G_i}$ with reference orientation~$(a,b)$.
If~$G_i$ is a $3$-bond then output its canonical list $l(G_i,a,b)$ as~$(a,b)$.
If~$G_i$ is a cycle then it has a unique canonical list with respect to the orientation~$(a,b)$,
that is $l(G_i,a,b)$.

Now we consider the case that~$G_i$ is a $3$-connected component. 
Then~$G_i$ has two possible canons with respect to the orientation~$(a,b)$, 
one for each of the two embeddings. 
Query the oracle
for the embedding that leads to the lexicographically smaller canonical list and output it
as $l(G_i,a,b)$. 
If we encounter a virtual edge~$(c,d)$ during the construction,
we determine its reference orientation with the help of the oracle
and output it in this direction.
If the children of the virtual edge do not give an orientation,
we output~$(c,d)$ in the direction in which
it is encountered during the construction of the canon for~$G_i$.
Finally, the children rooted at separating pair node $(c,d)$ are
ordered with the canonical order procedure.

\par We give now an example.
Consider the canonical list~$l(S,a,b)$ of edges for the tree~$S_{(a,b)}$ 
of Figure~\ref{fig:trees}.
Let $s_i$ be the edge connecting the vertices $a_i$ with $b_i$.
We also write for short $l'(S_i,s_i)$ which is one of $l(S_i,a_i,b_i)$ or $l(S_i, b_i,a_i)$.
The direction of $s_i$ is as described above.
\begin{eqnarray*}
l(S,a,b) &=& [\; (a,b)\; l(S_{G_1},a,b) \; \dots \; l(S_{G_k},a,b)\; ], \text{ where}\\
l(S_{G_1},a,b) &=& [\; l(G_1,a,b)\; [l'(S_1,s_1)]\; \dots\;  [l'(S_{l_1},s_{l_1})]\; ]\\
&\vdots& \\
l(S_{G_k},a,b) &=& [\; l(G_k,a,b)\; [l'(S_{l_k},s_{l_k})]\; ]
\end{eqnarray*}

\paragraph{Canon for the biconnected planar graph.}
This list is now almost the canon,
except that the names of the nodes are still the ones they have in~$G$.
Clearly,
a canon must be independent of the original names of the nodes.
The final canon for $S_{(a,b)}$ can be obtained by a log-space transducer which
relabels the vertices in the order of their first occurrence in this
canonical list and outputs the list using these new labels.

\par Note that the canonical list of edges contains virtual edges as
well, which are not a part of~$G$. However, this is not a problem as the
virtual edges can be distinguished from real edges because of the
presence of $3$-bonds.
To get the canon for $G$, remove these virtual edges and the delimiters `['
and `]' in the canon for $S_{(a,b)}$.
This is sufficient, because we describe here a bijective 
function $f$ which transforms an automorphism $\phi$ of $S_{(a,b)}$ into an automorphism
$f(\phi)$ for $G$ with $(a,b)$ fixed.
We get the following result.

\begin{theorem}
A biconnected planar graph can be canonized in log-space.
\end{theorem}

%% file: Figures/trees3.eepic
\setlength{\unitlength}{0.00087489in}
\begingroup\makeatletter\ifx\SetFigFontNFSS\undefined%
\gdef\SetFigFontNFSS#1#2#3#4#5{%
  \reset@font\fontsize{#1}{#2pt}%
  \fontfamily{#3}\fontseries{#4}\fontshape{#5}%
  \selectfont}%
\fi\endgroup%
{\renewcommand{\dashlinestretch}{30}
\begin{picture}(8304,3324)(0,-10)
\put(2937,2982){\blacken\ellipse{36}{36}}
\put(2937,2982){\ellipse{36}{36}}
\put(2667,2982){\blacken\ellipse{36}{36}}
\put(2667,2982){\ellipse{36}{36}}
\path(2667,2982)(2937,2982)
\put(2502,2952){\arc{210}{1.5708}{3.1416}}
\put(2502,2967){\arc{210}{3.1416}{4.7124}}
\put(3102,2967){\arc{210}{4.7124}{6.2832}}
\put(3102,2952){\arc{210}{0}{1.5708}}
\path(2397,2952)(2397,2967)
\path(2502,3072)(3102,3072)
\path(3207,2967)(3207,2952)
\path(3102,2847)(2502,2847)
\put(1272,1497){\blacken\ellipse{36}{36}}
\put(1272,1497){\ellipse{36}{36}}
\put(1002,1497){\blacken\ellipse{36}{36}}
\put(1002,1497){\ellipse{36}{36}}
\path(552,687)(552,507)
\path(642,687)(822,507)
\path(552,507)(462,237)(642,237)(552,507)
\path(822,507)(732,237)(912,237)(822,507)
\path(462,687)(282,507)
\path(282,507)(192,237)(372,237)(282,507)
\put(2397,1497){\blacken\ellipse{36}{36}}
\put(2397,1497){\ellipse{36}{36}}
\put(2127,1497){\blacken\ellipse{36}{36}}
\put(2127,1497){\ellipse{36}{36}}
\path(2127,1497)(2397,1497)
\put(2097,1467){\arc{210}{1.5708}{3.1416}}
\put(2097,1482){\arc{210}{3.1416}{4.7124}}
\put(2427,1482){\arc{210}{4.7124}{6.2832}}
\put(2427,1467){\arc{210}{0}{1.5708}}
\path(1992,1467)(1992,1482)
\path(2097,1587)(2427,1587)
\path(2532,1482)(2532,1467)
\path(2427,1362)(2097,1362)
\path(3522,687)(3522,507)
\path(3612,687)(3792,507)
\path(3522,507)(3432,237)(3612,237)(3522,507)
\path(3792,507)(3702,237)(3882,237)(3792,507)
\path(3432,687)(3252,507)
\path(3252,507)(3162,237)(3342,237)(3252,507)
\path(1452,507)(1362,237)(1542,237)(1452,507)
\path(1722,507)(1632,237)(1812,237)(1722,507)
\path(1452,507)(1542,687)
\path(1722,507)(1632,687)
\path(2127,507)(2037,237)(2217,237)(2127,507)
\path(2397,507)(2307,237)(2487,237)(2397,507)
\path(2127,507)(2217,687)
\path(2397,507)(2307,687)
\put(3657,1497){\blacken\ellipse{36}{36}}
\put(3657,1497){\ellipse{36}{36}}
\put(3387,1497){\blacken\ellipse{36}{36}}
\put(3387,1497){\ellipse{36}{36}}
\put(5547,1497){\blacken\ellipse{36}{36}}
\put(5547,1497){\ellipse{36}{36}}
\put(5277,1497){\blacken\ellipse{36}{36}}
\put(5277,1497){\ellipse{36}{36}}
\path(4827,687)(4827,507)
\path(4917,687)(5097,507)
\path(4827,507)(4737,237)(4917,237)(4827,507)
\path(5097,507)(5007,237)(5187,237)(5097,507)
\path(4737,687)(4557,507)
\path(4557,507)(4467,237)(4647,237)(4557,507)
\put(6672,1497){\blacken\ellipse{36}{36}}
\put(6672,1497){\ellipse{36}{36}}
\put(6402,1497){\blacken\ellipse{36}{36}}
\put(6402,1497){\ellipse{36}{36}}
\path(6402,1497)(6672,1497)
\put(6372,1467){\arc{210}{1.5708}{3.1416}}
\put(6372,1482){\arc{210}{3.1416}{4.7124}}
\put(6702,1482){\arc{210}{4.7124}{6.2832}}
\put(6702,1467){\arc{210}{0}{1.5708}}
\path(6267,1467)(6267,1482)
\path(6372,1587)(6702,1587)
\path(6807,1482)(6807,1467)
\path(6702,1362)(6372,1362)
\path(7797,687)(7797,507)
\path(7887,687)(8067,507)
\path(7797,507)(7707,237)(7887,237)(7797,507)
\path(8067,507)(7977,237)(8157,237)(8067,507)
\path(7707,687)(7527,507)
\path(7527,507)(7437,237)(7617,237)(7527,507)
\path(5727,507)(5637,237)(5817,237)(5727,507)
\path(5997,507)(5907,237)(6087,237)(5997,507)
\path(5727,507)(5817,687)
\path(5997,507)(5907,687)
\path(6402,507)(6312,237)(6492,237)(6402,507)
\path(6672,507)(6582,237)(6762,237)(6672,507)
\path(6402,507)(6492,687)
\path(6672,507)(6582,687)
\put(7932,1497){\blacken\ellipse{36}{36}}
\put(7932,1497){\ellipse{36}{36}}
\put(7662,1497){\blacken\ellipse{36}{36}}
\put(7662,1497){\ellipse{36}{36}}
\put(7212,2982){\blacken\ellipse{36}{36}}
\put(7212,2982){\ellipse{36}{36}}
\put(6942,2982){\blacken\ellipse{36}{36}}
\put(6942,2982){\ellipse{36}{36}}
\path(2577,2847)(1857,2397)
\path(2712,2847)(2622,2622)
\path(3027,2847)(3522,2397)
\path(2892,2847)(2982,2622)
\path(2802,2847)(2802,2622)
\path(1677,2037)(1137,1587)
\path(1002,1497)(1272,1497)
\path(1047,1362)(1002,1227)
\path(1137,1362)(1137,1227)
\path(1227,1362)(1272,1227)
\put(432,792){\arc{210}{1.5708}{3.1416}}
\put(432,942){\arc{210}{3.1416}{4.7124}}
\put(672,942){\arc{210}{4.7124}{6.2832}}
\put(672,792){\arc{210}{0}{1.5708}}
\path(327,792)(327,942)
\path(432,1047)(672,1047)
\path(777,942)(777,792)
\path(672,687)(432,687)
\put(972,1467){\arc{210}{1.5708}{3.1416}}
\put(972,1482){\arc{210}{3.1416}{4.7124}}
\put(1302,1482){\arc{210}{4.7124}{6.2832}}
\put(1302,1467){\arc{210}{0}{1.5708}}
\path(867,1467)(867,1482)
\path(972,1587)(1302,1587)
\path(1407,1482)(1407,1467)
\path(1302,1362)(972,1362)
\path(1857,2037)(1902,1902)
\path(1767,2037)(1722,1902)
\path(1947,2037)(2217,1587)
\path(2262,1362)(2262,1047)
\put(1692,2142){\arc{210}{1.5708}{3.1416}}
\put(1692,2292){\arc{210}{3.1416}{4.7124}}
\put(1932,2292){\arc{210}{4.7124}{6.2832}}
\put(1932,2142){\arc{210}{0}{1.5708}}
\path(1587,2142)(1587,2292)
\path(1692,2397)(1932,2397)
\path(2037,2292)(2037,2142)
\path(1932,2037)(1692,2037)
\path(957,1362)(552,1047)
\put(1467,792){\arc{210}{1.5708}{3.1416}}
\put(1467,942){\arc{210}{3.1416}{4.7124}}
\put(1707,942){\arc{210}{4.7124}{6.2832}}
\put(1707,792){\arc{210}{0}{1.5708}}
\path(1362,792)(1362,942)
\path(1467,1047)(1707,1047)
\path(1812,942)(1812,792)
\path(1707,687)(1467,687)
\path(1317,1362)(1587,1047)
\put(2142,792){\arc{210}{1.5708}{3.1416}}
\put(2142,942){\arc{210}{3.1416}{4.7124}}
\put(2382,942){\arc{210}{4.7124}{6.2832}}
\put(2382,792){\arc{210}{0}{1.5708}}
\path(2037,792)(2037,942)
\path(2142,1047)(2382,1047)
\path(2487,942)(2487,792)
\path(2382,687)(2142,687)
\path(3387,1497)(3657,1497)
\path(3522,2037)(3522,1587)
\path(3522,1362)(3522,1047)
\put(3402,2142){\arc{210}{1.5708}{3.1416}}
\put(3402,2292){\arc{210}{3.1416}{4.7124}}
\put(3642,2292){\arc{210}{4.7124}{6.2832}}
\put(3642,2142){\arc{210}{0}{1.5708}}
\path(3297,2142)(3297,2292)
\path(3402,2397)(3642,2397)
\path(3747,2292)(3747,2142)
\path(3642,2037)(3402,2037)
\put(3402,792){\arc{210}{1.5708}{3.1416}}
\put(3402,942){\arc{210}{3.1416}{4.7124}}
\put(3642,942){\arc{210}{4.7124}{6.2832}}
\put(3642,792){\arc{210}{0}{1.5708}}
\path(3297,792)(3297,942)
\path(3402,1047)(3642,1047)
\path(3747,942)(3747,792)
\path(3642,687)(3402,687)
\put(3357,1467){\arc{210}{1.5708}{3.1416}}
\put(3357,1482){\arc{210}{3.1416}{4.7124}}
\put(3687,1482){\arc{210}{4.7124}{6.2832}}
\put(3687,1467){\arc{210}{0}{1.5708}}
\path(3252,1467)(3252,1482)
\path(3357,1587)(3687,1587)
\path(3792,1482)(3792,1467)
\path(3687,1362)(3357,1362)
\dashline{30.000}(3072,1812)(3972,1812)(3972,147)
	(3072,147)(3072,1812)
\dashline{30.000}(1542,1812)(1902,1227)(1902,147)
	(102,147)(102,1227)(957,1812)(1542,1812)
\dashline{30.000}(5817,1812)(6177,1227)(6177,147)
	(4377,147)(4377,1227)(5232,1812)(5817,1812)
\path(6852,2847)(6132,2397)
\path(6987,2847)(6897,2622)
\path(7302,2847)(7797,2397)
\path(7167,2847)(7257,2622)
\path(7077,2847)(7077,2622)
\path(5952,2037)(5412,1587)
\path(5277,1497)(5547,1497)
\path(5322,1362)(5277,1227)
\path(5412,1362)(5412,1227)
\path(5502,1362)(5547,1227)
\put(4707,792){\arc{210}{1.5708}{3.1416}}
\put(4707,942){\arc{210}{3.1416}{4.7124}}
\put(4947,942){\arc{210}{4.7124}{6.2832}}
\put(4947,792){\arc{210}{0}{1.5708}}
\path(4602,792)(4602,942)
\path(4707,1047)(4947,1047)
\path(5052,942)(5052,792)
\path(4947,687)(4707,687)
\put(5247,1467){\arc{210}{1.5708}{3.1416}}
\put(5247,1482){\arc{210}{3.1416}{4.7124}}
\put(5577,1482){\arc{210}{4.7124}{6.2832}}
\put(5577,1467){\arc{210}{0}{1.5708}}
\path(5142,1467)(5142,1482)
\path(5247,1587)(5577,1587)
\path(5682,1482)(5682,1467)
\path(5577,1362)(5247,1362)
\path(6132,2037)(6177,1902)
\path(6042,2037)(5997,1902)
\path(6222,2037)(6492,1587)
\path(6537,1362)(6537,1047)
\put(5967,2142){\arc{210}{1.5708}{3.1416}}
\put(5967,2292){\arc{210}{3.1416}{4.7124}}
\put(6207,2292){\arc{210}{4.7124}{6.2832}}
\put(6207,2142){\arc{210}{0}{1.5708}}
\path(5862,2142)(5862,2292)
\path(5967,2397)(6207,2397)
\path(6312,2292)(6312,2142)
\path(6207,2037)(5967,2037)
\path(5232,1362)(4827,1047)
\put(5742,792){\arc{210}{1.5708}{3.1416}}
\put(5742,942){\arc{210}{3.1416}{4.7124}}
\put(5982,942){\arc{210}{4.7124}{6.2832}}
\put(5982,792){\arc{210}{0}{1.5708}}
\path(5637,792)(5637,942)
\path(5742,1047)(5982,1047)
\path(6087,942)(6087,792)
\path(5982,687)(5742,687)
\path(5592,1362)(5862,1047)
\put(6417,792){\arc{210}{1.5708}{3.1416}}
\put(6417,942){\arc{210}{3.1416}{4.7124}}
\put(6657,942){\arc{210}{4.7124}{6.2832}}
\put(6657,792){\arc{210}{0}{1.5708}}
\path(6312,792)(6312,942)
\path(6417,1047)(6657,1047)
\path(6762,942)(6762,792)
\path(6657,687)(6417,687)
\path(7662,1497)(7932,1497)
\path(7797,2037)(7797,1587)
\path(7797,1362)(7797,1047)
\put(7677,2142){\arc{210}{1.5708}{3.1416}}
\put(7677,2292){\arc{210}{3.1416}{4.7124}}
\put(7917,2292){\arc{210}{4.7124}{6.2832}}
\put(7917,2142){\arc{210}{0}{1.5708}}
\path(7572,2142)(7572,2292)
\path(7677,2397)(7917,2397)
\path(8022,2292)(8022,2142)
\path(7917,2037)(7677,2037)
\put(7677,792){\arc{210}{1.5708}{3.1416}}
\put(7677,942){\arc{210}{3.1416}{4.7124}}
\put(7917,942){\arc{210}{4.7124}{6.2832}}
\put(7917,792){\arc{210}{0}{1.5708}}
\path(7572,792)(7572,942)
\path(7677,1047)(7917,1047)
\path(8022,942)(8022,792)
\path(7917,687)(7677,687)
\put(7632,1467){\arc{210}{1.5708}{3.1416}}
\put(7632,1482){\arc{210}{3.1416}{4.7124}}
\put(7962,1482){\arc{210}{4.7124}{6.2832}}
\put(7962,1467){\arc{210}{0}{1.5708}}
\path(7527,1467)(7527,1482)
\path(7632,1587)(7962,1587)
\path(8067,1482)(8067,1467)
\path(7962,1362)(7632,1362)
\path(6942,2982)(7212,2982)
\put(6777,2952){\arc{210}{1.5708}{3.1416}}
\put(6777,2967){\arc{210}{3.1416}{4.7124}}
\put(7377,2967){\arc{210}{4.7124}{6.2832}}
\put(7377,2952){\arc{210}{0}{1.5708}}
\path(6672,2952)(6672,2967)
\path(6777,3072)(7377,3072)
\path(7482,2967)(7482,2952)
\path(7377,2847)(6777,2847)
\dottedline{60}(12,12)(12,1452)(1632,2532)
	(2217,2532)(2577,1902)(2577,12)(12,12)
\dottedline{75}(4287,12)(4287,1452)(5907,2532)
	(6492,2532)(6852,1902)(6852,12)(4287,12)
\dottedline{75}(7302,12)(8292,12)(8292,2532)
	(7302,2532)(7302,12)
\dashline{30.000}(7347,1812)(8247,1812)(8247,147)
	(7347,147)(7347,1812)
\dottedline{60}(3027,12)(4017,12)(4017,2532)
	(3027,2532)(3027,12)
\path(3466,2040)(3432,1902)
\path(3432,2037)(3342,1902)
\path(7741,2040)(7707,1902)
\path(7707,2037)(7617,1902)
\put(2982,2892){\makebox(0,0)[lb]{\smash{{\SetFigFontNFSS{12}{14.4}{\rmdefault}{\mddefault}{\updefault}$b$}}}}
\put(2487,2892){\makebox(0,0)[lb]{\smash{{\SetFigFontNFSS{12}{14.4}{\rmdefault}{\mddefault}{\updefault}$a$}}}}
\put(2712,3117){\makebox(0,0)[lb]{\smash{{\SetFigFontNFSS{12}{14.4}{\rmdefault}{\mddefault}{\updefault}$s$}}}}
\put(1722,2127){\makebox(0,0)[lb]{\smash{{\SetFigFontNFSS{12}{14.4}{\rmdefault}{\mddefault}{\updefault}$G_1$}}}}
\put(2622,867){\makebox(0,0)[lb]{\smash{{\SetFigFontNFSS{12}{14.4}{\rmdefault}{\mddefault}{\updefault}\huge{\dots}}}}}
\put(2622,1497){\makebox(0,0)[lb]{\smash{{\SetFigFontNFSS{12}{14.4}{\rmdefault}{\mddefault}{\updefault}\huge{\dots}}}}}
\put(1497,1497){\makebox(0,0)[lb]{\smash{{\SetFigFontNFSS{12}{14.4}{\rmdefault}{\mddefault}{\updefault}\huge{\dots}}}}}
\put(2442,2217){\makebox(0,0)[lb]{\smash{{\SetFigFontNFSS{12}{14.4}{\rmdefault}{\mddefault}{\updefault}\huge{\dots}}}}}
\put(3432,2127){\makebox(0,0)[lb]{\smash{{\SetFigFontNFSS{12}{14.4}{\rmdefault}{\mddefault}{\updefault}$G_k$}}}}
\put(1002,1632){\makebox(0,0)[lb]{\smash{{\SetFigFontNFSS{12}{14.4}{\rmdefault}{\mddefault}{\updefault}$s_1$}}}}
\put(912,867){\makebox(0,0)[lb]{\smash{{\SetFigFontNFSS{12}{14.4}{\rmdefault}{\mddefault}{\updefault}\huge{\dots}}}}}
\put(6897,867){\makebox(0,0)[lb]{\smash{{\SetFigFontNFSS{12}{14.4}{\rmdefault}{\mddefault}{\updefault}\huge{\dots}}}}}
\put(6897,1497){\makebox(0,0)[lb]{\smash{{\SetFigFontNFSS{12}{14.4}{\rmdefault}{\mddefault}{\updefault}\huge{\dots}}}}}
\put(5772,1497){\makebox(0,0)[lb]{\smash{{\SetFigFontNFSS{12}{14.4}{\rmdefault}{\mddefault}{\updefault}\huge{\dots}}}}}
\put(6717,2217){\makebox(0,0)[lb]{\smash{{\SetFigFontNFSS{12}{14.4}{\rmdefault}{\mddefault}{\updefault}\huge{\dots}}}}}
\put(5187,867){\makebox(0,0)[lb]{\smash{{\SetFigFontNFSS{12}{14.4}{\rmdefault}{\mddefault}{\updefault}\huge{\dots}}}}}
\put(6987,3117){\makebox(0,0)[lb]{\smash{{\SetFigFontNFSS{12}{14.4}{\rmdefault}{\mddefault}{\updefault}$t$}}}}
\put(6762,2892){\makebox(0,0)[lb]{\smash{{\SetFigFontNFSS{12}{14.4}{\rmdefault}{\mddefault}{\updefault}$a'$}}}}
\put(7257,2892){\makebox(0,0)[lb]{\smash{{\SetFigFontNFSS{12}{14.4}{\rmdefault}{\mddefault}{\updefault}$b'$}}}}
\put(7707,2127){\makebox(0,0)[lb]{\smash{{\SetFigFontNFSS{12}{14.4}{\rmdefault}{\mddefault}{\updefault}$H_k$}}}}
\put(5997,2127){\makebox(0,0)[lb]{\smash{{\SetFigFontNFSS{12}{14.4}{\rmdefault}{\mddefault}{\updefault}$H_1$}}}}
\put(5277,1632){\makebox(0,0)[lb]{\smash{{\SetFigFontNFSS{12}{14.4}{\rmdefault}{\mddefault}{\updefault}$t_1$}}}}
\put(3567,1632){\makebox(0,0)[lb]{\smash{{\SetFigFontNFSS{12}{14.4}{\rmdefault}{\mddefault}{\updefault}$s_{l_k}$}}}}
\put(7842,1632){\makebox(0,0)[lb]{\smash{{\SetFigFontNFSS{12}{14.4}{\rmdefault}{\mddefault}{\updefault}$t_{l_k}$}}}}
\put(2217,1632){\makebox(0,0)[lb]{\smash{{\SetFigFontNFSS{12}{14.4}{\rmdefault}{\mddefault}{\updefault}$s_{l_1}$}}}}
\put(6492,1632){\makebox(0,0)[lb]{\smash{{\SetFigFontNFSS{12}{14.4}{\rmdefault}{\mddefault}{\updefault}$t_{l_1}$}}}}
\put(1902,3162){\makebox(0,0)[lb]{\smash{{\SetFigFontNFSS{12}{14.4}{\rmdefault}{\mddefault}{\updefault}$S_{(a,b)}$}}}}
\put(6177,3162){\makebox(0,0)[lb]{\smash{{\SetFigFontNFSS{12}{14.4}{\rmdefault}{\mddefault}{\updefault}$T_{(a',b')}$}}}}
\put(1002,1857){\makebox(0,0)[lb]{\smash{{\SetFigFontNFSS{12}{14.4}{\rmdefault}{\mddefault}{\updefault}$S_1$}}}}
\put(3747,1857){\makebox(0,0)[lb]{\smash{{\SetFigFontNFSS{12}{14.4}{\rmdefault}{\mddefault}{\updefault}$S_{l_k}$}}}}
\put(5277,1857){\makebox(0,0)[lb]{\smash{{\SetFigFontNFSS{12}{14.4}{\rmdefault}{\mddefault}{\updefault}$T_1$}}}}
\put(8022,1857){\makebox(0,0)[lb]{\smash{{\SetFigFontNFSS{12}{14.4}{\rmdefault}{\mddefault}{\updefault}$T_{l_k}$}}}}
\put(1677,2577){\makebox(0,0)[lb]{\smash{{\SetFigFontNFSS{12}{14.4}{\rmdefault}{\mddefault}{\updefault}$S_{G_1}$}}}}
\put(3567,2577){\makebox(0,0)[lb]{\smash{{\SetFigFontNFSS{12}{14.4}{\rmdefault}{\mddefault}{\updefault}$S_{G_k}$}}}}
\put(7842,2577){\makebox(0,0)[lb]{\smash{{\SetFigFontNFSS{12}{14.4}{\rmdefault}{\mddefault}{\updefault}$T_{H_k}$}}}}
\put(5952,2577){\makebox(0,0)[lb]{\smash{{\SetFigFontNFSS{12}{14.4}{\rmdefault}{\mddefault}{\updefault}$T_{H_1}$}}}}
\end{picture}
}

%% file: Figures/orientation.eepic
\setlength{\unitlength}{0.00087489in}
\begingroup\makeatletter\ifx\SetFigFontNFSS\undefined%
\gdef\SetFigFontNFSS#1#2#3#4#5{%
  \reset@font\fontsize{#1}{#2pt}%
  \fontfamily{#3}\fontseries{#4}\fontshape{#5}%
  \selectfont}%
\fi\endgroup%
{\renewcommand{\dashlinestretch}{30}
\begin{picture}(7362,2694)(0,-10)
\path(1185,1182)(1635,732)(1185,282)
	(285,282)(285,1182)(1185,282)
	(1185,1182)(285,282)
\path(2355,1182)(1905,732)(2355,282)
	(3255,282)(3255,1182)(2355,282)
	(2355,1182)(3255,282)
\put(5550,1137){\blacken\ellipse{36}{36}}
\put(5550,1137){\ellipse{36}{36}}
\put(6090,1137){\blacken\ellipse{36}{36}}
\put(6090,1137){\ellipse{36}{36}}
\put(285,2082){\blacken\ellipse{36}{36}}
\put(285,2082){\ellipse{36}{36}}
\put(1185,2082){\blacken\ellipse{36}{36}}
\put(1185,2082){\ellipse{36}{36}}
\put(1635,1632){\blacken\ellipse{36}{36}}
\put(1635,1632){\ellipse{36}{36}}
\put(735,1632){\blacken\ellipse{36}{36}}
\put(735,1632){\ellipse{36}{36}}
\put(285,282){\blacken\ellipse{36}{36}}
\put(285,282){\ellipse{36}{36}}
\put(1185,282){\blacken\ellipse{36}{36}}
\put(1185,282){\ellipse{36}{36}}
\put(735,732){\blacken\ellipse{36}{36}}
\put(735,732){\ellipse{36}{36}}
\put(1185,1632){\blacken\ellipse{36}{36}}
\put(1185,1632){\ellipse{36}{36}}
\put(1185,732){\blacken\ellipse{36}{36}}
\put(1185,732){\ellipse{36}{36}}
\put(1635,732){\blacken\ellipse{36}{36}}
\put(1635,732){\ellipse{36}{36}}
\put(2355,2082){\blacken\ellipse{36}{36}}
\put(2355,2082){\ellipse{36}{36}}
\put(3255,2082){\blacken\ellipse{36}{36}}
\put(3255,2082){\ellipse{36}{36}}
\put(3705,1632){\blacken\ellipse{36}{36}}
\put(3705,1632){\ellipse{36}{36}}
\put(2805,1632){\blacken\ellipse{36}{36}}
\put(2805,1632){\ellipse{36}{36}}
\put(3255,1182){\blacken\ellipse{36}{36}}
\put(3255,1182){\ellipse{36}{36}}
\put(2355,1182){\blacken\ellipse{36}{36}}
\put(2355,1182){\ellipse{36}{36}}
\put(2355,1182){\blacken\ellipse{36}{36}}
\put(2355,1182){\ellipse{36}{36}}
\put(2355,282){\blacken\ellipse{36}{36}}
\put(2355,282){\ellipse{36}{36}}
\put(3255,282){\blacken\ellipse{36}{36}}
\put(3255,282){\ellipse{36}{36}}
\put(3255,1182){\blacken\ellipse{36}{36}}
\put(3255,1182){\ellipse{36}{36}}
\put(2805,732){\blacken\ellipse{36}{36}}
\put(2805,732){\ellipse{36}{36}}
\put(2355,732){\blacken\ellipse{36}{36}}
\put(2355,732){\ellipse{36}{36}}
\put(3255,1632){\blacken\ellipse{36}{36}}
\put(3255,1632){\ellipse{36}{36}}
\put(1905,732){\blacken\ellipse{36}{36}}
\put(1905,732){\ellipse{36}{36}}
\put(5550,2532){\blacken\ellipse{36}{36}}
\put(5550,2532){\ellipse{36}{36}}
\put(6090,2532){\blacken\ellipse{36}{36}}
\put(6090,2532){\ellipse{36}{36}}
\path(5948,1857)(6143,1662)(5948,1467)
	(5948,1857)(5558,1467)(5948,1467)
	(5558,1857)(5558,1467)
\dashline{60.000}(5550,1857)(5955,1857)
\path(6735,454)(6540,259)(6735,64)
	(6735,454)(7125,64)(6735,64)
	(7125,454)(7125,64)
\dashline{60.000}(7133,454)(6728,454)
\path(7028,1857)(7223,1662)(7028,1467)
	(7028,1857)(6638,1467)(7028,1467)
	(6638,1857)(6638,1467)
\dashline{60.000}(6630,1857)(7035,1857)
\put(285,1182){\blacken\ellipse{36}{36}}
\put(285,1182){\ellipse{36}{36}}
\put(1185,1182){\blacken\ellipse{36}{36}}
\put(1185,1182){\ellipse{36}{36}}
\put(5475,1017){\arc{210}{1.5708}{3.1416}}
\put(5475,1167){\arc{210}{3.1416}{4.7124}}
\put(6165,1167){\arc{210}{4.7124}{6.2832}}
\put(6165,1017){\arc{210}{0}{1.5708}}
\path(5370,1017)(5370,1167)
\path(5475,1272)(6165,1272)
\path(6270,1167)(6270,1017)
\path(6165,912)(5475,912)
\path(5550,1137)(6090,1137)
\put(4395,117){\arc{210}{1.5708}{3.1416}}
\put(4395,447){\arc{210}{3.1416}{4.7124}}
\put(5085,447){\arc{210}{4.7124}{6.2832}}
\put(5085,117){\arc{210}{0}{1.5708}}
\path(4290,117)(4290,447)
\path(4395,552)(5085,552)
\path(5190,447)(5190,117)
\path(5085,12)(4395,12)
\put(5475,117){\arc{210}{1.5708}{3.1416}}
\put(5475,447){\arc{210}{3.1416}{4.7124}}
\put(6165,447){\arc{210}{4.7124}{6.2832}}
\put(6165,117){\arc{210}{0}{1.5708}}
\path(5370,117)(5370,447)
\path(5475,552)(6165,552)
\path(6270,447)(6270,117)
\path(6165,12)(5475,12)
\path(5550,912)(5010,552)
\path(5820,912)(5820,552)
\path(285,2082)(1185,2082)(1635,1632)
	(1185,1182)(1185,2082)(285,1182)
	(1185,1182)(285,2082)(285,1182)
\path(1185,1632)(1635,1632)
\path(1185,732)(1635,732)
\path(2355,2082)(3255,2082)(3705,1632)
	(3255,1182)(3255,2082)(2355,1182)
	(3255,1182)(2355,2082)(2355,1182)
\path(3255,1632)(3705,1632)
\path(1905,732)(2355,732)
\put(5475,2412){\arc{210}{1.5708}{3.1416}}
\put(5475,2562){\arc{210}{3.1416}{4.7124}}
\put(6165,2562){\arc{210}{4.7124}{6.2832}}
\put(6165,2412){\arc{210}{0}{1.5708}}
\path(5370,2412)(5370,2562)
\path(5475,2667)(6165,2667)
\path(6270,2562)(6270,2412)
\path(6165,2307)(5475,2307)
\path(5550,2532)(6090,2532)
\put(4395,1512){\arc{210}{1.5708}{3.1416}}
\put(4395,1842){\arc{210}{3.1416}{4.7124}}
\put(5085,1842){\arc{210}{4.7124}{6.2832}}
\put(5085,1512){\arc{210}{0}{1.5708}}
\path(4290,1512)(4290,1842)
\path(4395,1947)(5085,1947)
\path(5190,1842)(5190,1512)
\path(5085,1407)(4395,1407)
\put(5475,1512){\arc{210}{1.5708}{3.1416}}
\put(5475,1842){\arc{210}{3.1416}{4.7124}}
\put(6165,1842){\arc{210}{4.7124}{6.2832}}
\put(6165,1512){\arc{210}{0}{1.5708}}
\path(5370,1512)(5370,1842)
\path(5475,1947)(6165,1947)
\path(6270,1842)(6270,1512)
\path(6165,1407)(5475,1407)
\path(5550,2307)(5010,1947)
\path(5820,2307)(5820,1947)
\path(6090,2307)(6630,1947)
\put(6555,117){\arc{210}{1.5708}{3.1416}}
\put(6555,447){\arc{210}{3.1416}{4.7124}}
\put(7245,447){\arc{210}{4.7124}{6.2832}}
\put(7245,117){\arc{210}{0}{1.5708}}
\path(6450,117)(6450,447)
\path(6555,552)(7245,552)
\path(7350,447)(7350,117)
\path(7245,12)(6555,12)
\put(6555,1512){\arc{210}{1.5708}{3.1416}}
\put(6555,1842){\arc{210}{3.1416}{4.7124}}
\put(7245,1842){\arc{210}{4.7124}{6.2832}}
\put(7245,1512){\arc{210}{0}{1.5708}}
\path(6450,1512)(6450,1842)
\path(6555,1947)(7245,1947)
\path(7350,1842)(7350,1512)
\path(7245,1407)(6555,1407)
\path(5948,462)(6143,267)(5948,72)
	(5948,462)(5558,72)(5948,72)
	(5558,462)(5558,72)
\dashline{60.000}(5550,462)(5955,462)
\path(4470,282)(4920,282)
\dashline{30.000}(4470,282)(4539,333)(4626,366)
	(4743,366)(4836,342)(4902,297)(4920,279)
\dashline{30.000}(4470,282)(4539,231)(4626,198)
	(4743,198)(4836,222)(4902,267)(4920,285)
\path(4470,1677)(4920,1677)
\dashline{30.000}(4470,1677)(4539,1626)(4626,1593)
	(4743,1593)(4836,1617)(4902,1662)(4920,1680)
\dashline{30.000}(4467,1683)(4536,1734)(4623,1767)
	(4740,1767)(4833,1743)(4899,1698)(4917,1680)
\path(7212,1662)(7035,1662)
\path(6732,258)(6558,258)
\path(6135,267)(5955,267)
\path(6132,1662)(5961,1662)
\path(6090,912)(6630,552)
\put(6090,957){\makebox(0,0)[lb]{\smash{{\SetFigFontNFSS{12}{14.4}{\rmdefault}{\mddefault}{\updefault}$b'$}}}}
\put(105,1137){\makebox(0,0)[lb]{\smash{{\SetFigFontNFSS{12}{14.4}{\rmdefault}{\mddefault}{\updefault}$a$}}}}
\put(1320,1137){\makebox(0,0)[lb]{\smash{{\SetFigFontNFSS{12}{14.4}{\rmdefault}{\mddefault}{\updefault}$b$}}}}
\put(15,2127){\makebox(0,0)[lb]{\smash{{\SetFigFontNFSS{12}{14.4}{\rmdefault}{\mddefault}{\updefault}$G$}}}}
\put(3390,1137){\makebox(0,0)[lb]{\smash{{\SetFigFontNFSS{12}{14.4}{\rmdefault}{\mddefault}{\updefault}$b'$}}}}
\put(2175,1137){\makebox(0,0)[lb]{\smash{{\SetFigFontNFSS{12}{14.4}{\rmdefault}{\mddefault}{\updefault}$a'$}}}}
\put(2085,2127){\makebox(0,0)[lb]{\smash{{\SetFigFontNFSS{12}{14.4}{\rmdefault}{\mddefault}{\updefault}$H$}}}}
\put(5460,2352){\makebox(0,0)[lb]{\smash{{\SetFigFontNFSS{12}{14.4}{\rmdefault}{\mddefault}{\updefault}$a$}}}}
\put(6090,2352){\makebox(0,0)[lb]{\smash{{\SetFigFontNFSS{12}{14.4}{\rmdefault}{\mddefault}{\updefault}$b$}}}}
\put(4920,1722){\makebox(0,0)[lb]{\smash{{\SetFigFontNFSS{11}{13.2}{\rmdefault}{\mddefault}{\updefault}$b$}}}}
\put(4920,327){\makebox(0,0)[lb]{\smash{{\SetFigFontNFSS{11}{13.2}{\rmdefault}{\mddefault}{\updefault}$b'$}}}}
\put(5460,957){\makebox(0,0)[lb]{\smash{{\SetFigFontNFSS{12}{14.4}{\rmdefault}{\mddefault}{\updefault}$a'$}}}}
\put(4380,327){\makebox(0,0)[lb]{\smash{{\SetFigFontNFSS{11}{13.2}{\rmdefault}{\mddefault}{\updefault}$a'$}}}}
\put(5415,1792){\makebox(0,0)[lb]{\smash{{\SetFigFontNFSS{11}{13.2}{\rmdefault}{\mddefault}{\updefault}$a$}}}}
\put(6045,397){\makebox(0,0)[lb]{\smash{{\SetFigFontNFSS{11}{13.2}{\rmdefault}{\mddefault}{\updefault}$b'$}}}}
\put(7170,397){\makebox(0,0)[lb]{\smash{{\SetFigFontNFSS{11}{13.2}{\rmdefault}{\mddefault}{\updefault}$b'$}}}}
\put(6495,1792){\makebox(0,0)[lb]{\smash{{\SetFigFontNFSS{10}{12.0}{\rmdefault}{\mddefault}{\updefault}$a$}}}}
\put(7125,1792){\makebox(0,0)[lb]{\smash{{\SetFigFontNFSS{11}{13.2}{\rmdefault}{\mddefault}{\updefault}$b$}}}}
\put(5415,397){\makebox(0,0)[lb]{\smash{{\SetFigFontNFSS{11}{13.2}{\rmdefault}{\mddefault}{\updefault}$a'$}}}}
\put(6495,397){\makebox(0,0)[lb]{\smash{{\SetFigFontNFSS{11}{13.2}{\rmdefault}{\mddefault}{\updefault}$a'$}}}}
\put(4380,1722){\makebox(0,0)[lb]{\smash{{\SetFigFontNFSS{11}{13.2}{\rmdefault}{\mddefault}{\updefault}$a$}}}}
\put(6000,1992){\makebox(0,0)[lb]{\smash{{\SetFigFontNFSS{11}{13.2}{\rmdefault}{\mddefault}{\updefault}$G_1$}}}}
\put(7080,1992){\makebox(0,0)[lb]{\smash{{\SetFigFontNFSS{11}{13.2}{\rmdefault}{\mddefault}{\updefault}$G_2$}}}}
\put(6000,597){\makebox(0,0)[lb]{\smash{{\SetFigFontNFSS{11}{13.2}{\rmdefault}{\mddefault}{\updefault}$H_1$}}}}
\put(7125,597){\makebox(0,0)[lb]{\smash{{\SetFigFontNFSS{11}{13.2}{\rmdefault}{\mddefault}{\updefault}$H_2$}}}}
\put(6045,1792){\makebox(0,0)[lb]{\smash{{\SetFigFontNFSS{11}{13.2}{\rmdefault}{\mddefault}{\updefault}$b$}}}}
\put(4335,597){\makebox(0,0)[lb]{\smash{{\SetFigFontNFSS{11}{13.2}{\rmdefault}{\mddefault}{\updefault}$H_0$}}}}
\put(4335,1992){\makebox(0,0)[lb]{\smash{{\SetFigFontNFSS{11}{13.2}{\rmdefault}{\mddefault}{\updefault}$G_0$}}}}
\put(4155,2532){\makebox(0,0)[lb]{\smash{{\SetFigFontNFSS{12}{14.4}{\rmdefault}{\mddefault}{\updefault}$S_{(a,b)}$}}}}
\put(4155,1137){\makebox(0,0)[lb]{\smash{{\SetFigFontNFSS{12}{14.4}{\rmdefault}{\mddefault}{\updefault}$T_{(a',b')}$}}}}
\end{picture}
}

%% file: 5-plcanon.tex
\section{Canonization of Planar Graphs}\label{sec:cpg}

In this section, we give a log-space algorithm for the canonization of planar graphs.
The main part is to show how to canonize {\em connected\/} planar graphs.
Then,
if a given graph is not connected,
we compute its connected components in log-space
and canonize each of these components.
The canons of the connected components are output in lexicographical increasing order.
Hence,
from now on we assume that the given planar graph is connected.

We decompose a planar graph
into its biconnected components and then
construct a tree on these biconnected components and articulation points.
We refer to this tree as the \emph{biconnected component tree\/}.
We also refer to the components as \emph{biconnected component nodes\/} 
and \emph{articulation point nodes\/}.
This tree is unique and can be constructed in log-space~\cite{ADK08}.

Similar to triconnected component trees,
we put a copy of an articulation point~$a$ into each of the components formed by the 
removal of~$a$.
An articulation point~$a$ has a copy in each of the biconnected components 
obtained by its removal.
In the discussion below,
we refer to a copy of an articulation point in a biconnected component~$B$ as an {\em 
articulation point in\/}~$B$.
Although an articulation point has at most one copy in each of the biconnected components,
the corresponding triconnected component trees can have many copies of 
the same articulation point,
if it belongs to a separating pair in the biconnected component.

\par Given a planar graph~$G$, we root its biconnected component tree  at an articulation 
point.
During the isomorphism ordering of two such trees~$S$ and~$T$, we can
fix the root of~$S$ arbitrarily
and make an equality test for all choices of roots for~$T$.
As there are ~$ \leq n$~articulation points, a
log-space transducer can cycle through all of them for the choice of the root for~$T$.
We state some properties of articulation points. 

\begin{lemma}\label{le:bitree}
Let~$B$ be a biconnected component in~$S$ and~$\T(B)$ be its triconnected component tree. 
Then the following holds.
\begin{enumerate}
\item~$S$ has a unique center, similar to a triconnected component tree.
\item If an articulation point~$a$ of~$S$ appears in a separating pair node~$s$ in~$\T(B)$,
	then it appears in all the triconnected component nodes which are adjacent to~$s$ in~$\T(B)$.
\item If an articulation point~$a$ appears in two nodes~$C$ and~$D$ in~$\T(B)$,
	it appears in all the nodes that lie on the path between~$C$ and~$D$ in~$\T(B)$.
	Hence, there is a unique node~$A$ in~$\T(B)$ that contains~$a$
	which is nearest to the center of~$\T(B)$.
	We call~$A$ the triconnected component \emph{associated with\/}~$a$.
	Thus we can uniquely associate each articulation point contained in~$B$ with a 
	triconnected component in~$\T(B)$.
\end{enumerate}
\end{lemma}

\subsection{Isomorphism Order for Biconnected Component Trees}

The isomorphism order for biconnected component trees rooted at articulation points
is defined in three steps that correspond to the first three steps of the isomorphism 
order for triconnected component trees
in Section~\ref{subsec:ord} on page~\pageref{summary-tri}.
We mention the main differences in the isomorphism ordering for biconnected component trees
from that of triconnected component trees.

\begin{enumerate}
\item The biconnected component nodes are connected by articulation point nodes. 
	The resulting graph is a tree similar to the tree of
	triconnected component nodes and separating pair nodes.
	For articulation points, we do not need the notion of orientation.
	Instead,
	we color the copy of the parent articulation
	point in a biconnected component with a distinct color and
	then the pairwise isomorphism among the subtrees of~$S$ and~$T$ 
	can be extended to the isomorphism between the corresponding 
	planar graphs~$G$ and~$H$ in a straight forward way.
\item In the triconnected component trees
	we have a separating pair as root, i.e. an edge.
	Now,
	we only have an articulation point, i.e., one vertex.
	Hence, when we compare biconnected components~$B$ and~$B'$,
	then we do not have an obvious, uniquely defined edge as root 
	for the corresponding component trees~$\T(B)$ and~$\T(B')$.
	The naive approach would be to cycle through all separating pairs and 
	finally define the one as root that leads to a minimal canon.
	However,
	that way we cannot guarantee that the algorithm works in log-space.
	Let~$n_B$ be the size of~$B$.
	Note that there can be upto~$O(n_B)$ separating pairs.
	When we go into a recursion at some point,
	we need to store the edge that is currently the root.
	That is, we need~$O(\log n_B)$ space at one level of recursion
	and this is too much for an overall log-space bound.
	Hence our major task will be to limit the number of possible choices 
	of roots appropriately
	so that the algorithm runs in log-space.
\item There are some more nontrivial tasks, to guarantee the log-space bound.
	It is not obvious, what to store on the work-tape when we 
	go into recursion at some node in~$S$ or some node in~$\T(B)$ and,
	what can be recomputed. 
	We also need a new definition of the size of a subtree, to correctly
	descide, which child is a large child and must be considered a priori 
	by the comparison algorithm.
\end{enumerate}

The size of a triconnected component tree is defined in Definition~\ref{def:TriSize} 
on page~\pageref{def:TriSize}.
Here we extend the definition to biconnected component trees.

\begin{definition} \label{def:BiSize}
Let~$B$ be a biconnected component node in a biconnected component tree~$S$,
and let~$\T(B)$ be the triconnected component tree of~$B$.
The {\em size of\/}~$B$ is defined as~$\size{\T(B)}$ as in Definition~\ref{def:TriSize}.
The size of an articulation point node in~$S$ is defined as~$1$.
Note that the articulation points maybe counted  several times, namely in every component they occur.
The {\em size of\/}~$S$, denoted by~$\size{S}$, is the sum of the sizes of its components.
\end{definition}

\par
\begin{figure}[!ht]
\begin{center}
\scalebox{0.82}{\input{Figures/biConCompTrees.eepic}}
\end{center}
\caption{Biconnected component trees.}\label{fig:biConCompTrees}
\end{figure}
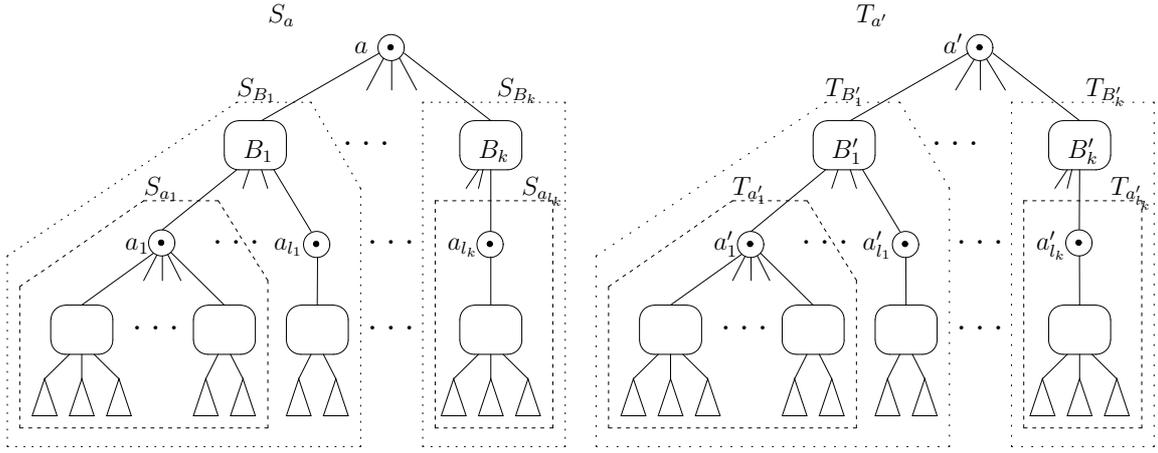

 We define the isomorphism order for two biconnected component trees~$S_a$ and~$T_{a'}$
rooted at nodes~$s$ and~$t$ corresponding to articulation points~$a$ and~$a'$, 
respectively
(see Figure~\ref{fig:biConCompTrees}).
Define~$S_a <_{\tt{B}} T_{a'}$ if

\begin{enumerate}
\item~$|S_a| < |T_{a'}|$ or
\item~$|S_a| = |T_{a'}|$ but~$\#s < \#t$ or
\item~$|S_a| = |T_{a'}|$,~$\#s = \#t = k$,
	but~$(S_{B_1}, \ldots , S_{B_k}) <_{\tt{B}}  (T_{B'_1} ,\ldots , T_{B'_k})$ lexicographically,
	where we assume that~$S_{B_1} \leq_{\tt{B}} \cdots \leq_{\tt{B}} S_{B_k}$
	and~$T_{B'_1} \leq_{\tt{B}} \cdots \leq_{\tt{B}} T_{B'_k}$ are the ordered subtrees
	of~$S_a$ and~$T_{a'}$, respectively.
	To compare the order between the subtrees~$S_{B_i}$ and~$T_{B'_j}$
	we compare the triconnected component trees~$\T(B_i)$ of~$B_i$ and~$\T(B'_j)$ of~$B'_j$.
	When we reach the first occurences of some articulation points 
	in $\T(B_i)$ and $\T(B_{j'})$
	(i.e. the \emph{reference copies\/} of these articulation points as described later)
	then we compare \emph{recursively\/} the corresponding 
	subtrees rooted at the children of~$B_i$ and~$B'_j$.
	Note, that these children are again articulation point nodes.
\end{enumerate}

We say that two biconnected component trees are \emph{equal\/},
denoted by~$S_a =_{\tt{B}}  T_{a'}$,
if neither of~$S_a <_{\tt{B}} T_{a'}$ and~$T_{a'} <_{\tt{B}} S_a$ holds.
The inductive ordering of the subtrees of~$S_a$ and~$T_{a'}$ proceeds exactly
as in Lindell's algorithm, by partitioning them into
size-classes and comparing the children in the same size-class recursively. The
book-keeping required (e.g. the order profile of a node, the number of nodes
in a size-class that have been compared so far) is similar to that in Lindell's
algorithm. We discuss how to compare two such subtrees~$S_{B}$ and~$T_{B'}$, rooted at
biconnected component nodes~$B$ and~$B'$, respectively.

\paragraph{Isomorphism order of two subtrees rooted at biconnected components.}
We consider the isomorphism order of two subtrees~$S_{B_i}$ and~$T_{B'_j}$ rooted
at biconnected component nodes~$B_i$ and~$B'_j$, and let~$a$ and~$a'$ be their parent
articulation points, respectively.
We start by constructing and comparing the canons of the triconnected
component trees of~$B_i$ and~$B'_j$.
To do so, we choose a separating pair as root for each of them.
As explained above,
we cannot afford to simply try all possible choices as root.
We will show below that we can compute in log-space a sufficiently small
number of separating pairs as roots for~$B_i$ and~$B'_j$
which suffices for our purpose.
That is, we make pairwise cross-comparisons of the canons obtained for these separating
pairs as roots and determine the minimum canons.

The base case is that~$B_i$ and~$B'_j$ are leaf nodes
and therefore contain no articulation points other than the parent articulation point.
In this case, we can cycle through all the edges as roots
and find the lexicographically smallest canon.

If~$B_i$ and~$B'_j$ contain articulation points,
we go into recursion.
Note that if an articulation point is part of a separating pair,
it can occur several times in the triconnected component tree.
To avoid recursion on the same pair of articulation points multiple times,
we need to additionally keep track of whether a pair of 
articulation points is encountered for the first time in the comparison.

Also, while canonizing the triconnected component trees of~$B_i$ and~$B'_j$,
we give a separate color to the copy of~$a$ and~$a'$ in these trees,
to ensure that the parent articulation points are always mapped to each other.

\paragraph{Limiting the number of possible choices for the root.}
Let~$S_a$ be a biconnected component tree rooted at articulation point~$a$.
Let~$B$ be a child of~$a$ in~$S_a$ and
$\T(B)$ be the triconnected component tree of the biconnected component~$B$.
We show how to limit the number of potential root nodes for~$\T(B)$.

Besides the parent~$a$, 
let~$B$ have articulation points~$a_1, \dots, a_l$ for some integer~$l \geq 0$,
such that~$a_j$ is the root node of the subtree~$S_{a_j}$ of~$S_a$ 
(see Figure~\ref{fig:biConCompTrees}).
We partition the subtrees~$S_{a_1}, \dots, S_{a_l}$ into classes  
$E_1, \dots, E_p$ of equal size subtrees 
(i.e.\ size according to Definition~\ref{def:BiSize}).
Let~$k_j$ be the number of subtrees in~$E_j$.
Let the order of the size classes be such that~$k_1 \leq k_2 \leq \cdots \leq k_p$.
All articulation points  with their subtrees
in size class~$E_j$ are colored with color~$j$.

To limit the number of potential root nodes for~$\T(B)$,
we distinguish several cases below.
The center of~$\T(B)$, denoted by~$C$, will play an important role thereby.
In some of the cases
we will show that the number of automorphisms of the center~$C$ is small.
This already suffices for our purpose:
in this case, we canonize the component~$C$ separately for all edges in~$C$ as starting edge
and with colored articulation points.
We determine the edges that lead to the minimum canon.
The number of such edges is linear in the number of automorphisms of~$C$,
and hence this number is small, too.
Now,
if~$C$ contains no separating pairs,
we directly take these edges as roots for~$T_i$.
Otherwise
we take the separating pairs that occur first in the canons that start with these edges.
Hence,
in either case the number of potential root nodes for~$\T(B)$ is small.

We start our case analysis by considering properties of the center~$C$ of~$\T(B)$. 

\begin{itemize}
\item \label{itm:sep} 
{\bf The center~$C$ of~$\T(B)$ is a separating pair:}
	We choose this separating pair as the root of~$\T(B)$. Thus we have only one choice for
	the root, and the subtree rooted at~$B$ can be canonized in a unique way. 
\item \label{itm:atri} 
{\bf~$C$ is a triconnected component and~$a$ is not
	associated with~$C$:} 
Let~$a$ be associated with a triconnected component~$R$. 
We find the path from~$R$ to~$C$
	in~$\T(B)$ and find the separating pair closest to~$C$ on this path. This
	serves as the unique choice for the root of~$\T(B)$. 
\item \label{itm:cyc} 
{\bf~$a$ is associated with~$C$ and~$C$ is a cycle:}
	We canonize~$C$ for the two edges
	incident on~$a$ as starting edges, and~$a$ as the starting vertex. We
	construct these canons till a virtual edge is encountered in one or both
	of them. We choose the separating pairs corresponding to the first virtual
	edges encountered in these canons as the roots of~$\T(B)$. Thus we get
	at most two choices for the root of~$\T(B)$. 
\end{itemize}

For the following cases, we assume that the center~$C$ is a~$3$-connected component
and~$a$~is associated with~$C$.
We proceed with the case analysis according to the number~$l$ of articulation points in~$B$ besides~$a$.

\vspace{\topsep}
\par\noindent{\bf Case I:~$l=0$.}
$B$ is a leaf node in~$S_a$, it contains no articulation points besides~$a$.
We color~$a$ with a distinct color.
In this case we can cycle through all edges as root for~$\T(B)$.

\vspace{\itemsep}
\par\noindent{\bf Case II:~$l=1$.}
If~$B$ has exactly one articulation point besides~$a$,
then we process this child a priori and store the result.
We color~$a$ and~$a_1$ with distinct colors
and proceed with~$B$ as it would be a leaf node.

\vspace{\itemsep}
\par\noindent{\bf Case III:~$l\geq 2$.}
We distinguish two subcases.
\begin{enumerate}
\item \label{itm:aout}
	{\bf Some articulation point~$a_j$ in~$E_1$ is not associated with~$C$.}
	Let~$a_j$ be associated with a triconnected component~$D \not= C$.
	Find the path from~$D$ to~$C$ in~$\T(B)$ and
	select the separating pair node closest to~$C$ on this path.
	Thus~$a_j$ uniquely defines a separating pair.
	In the worst case, every component in~$E_1$ contains an 
	articulation point that is not associated with~$C$.
	Therefore, we get up to~$k_1$ separating pairs as candidates for the root.

\item{\bf All articulation points in~$E_1$ are associated with~$C$.}
	We distinguish three further subcases.
\begin{enumerate}
\item \label{itm:itm:two}
	{\bf~$k_1 = k_2 = 1$.}
	$C$~has at least three vertices that are fixed by all its automorphisms
	(i.e.\ $a$ and the articulation point with its subtree in~$E_1$ and that in~$E_2$).
	We will show  in Corollary~\ref{co:3-fixed} below
	that~$C$ has at most one non-trivial automorphism in this case.
	Thus, we have at most two ways of choosing the root of~$\T(B)$.
\item \label{itm:itm:one}
	{\bf~$k_1 = 1$ and~$k_2 \geq 2$.}
	We process the child in~$E_1$ a priori and store the result.
	We prove in Lemma~\ref{lem:aut} below
	that~$C$ can have at most~$4k_2$ automorphisms in this case. 
	Thus, we have at most~$4k_2$ ways of choosing the root of~$\T(B)$.
\item \label{itm:itm:zero}
	{\bf~$k_1 \geq 2$.}
	Again by Lemma~\ref{lem:aut} below,
	$C$~can have at most~$4k_1$ automorphisms.
	Thus, we have at most~$4k_1$ ways of choosing the root of~$\T(B)$.
\end{enumerate}
\end{enumerate}

Let~$N = \size{S_B}$.
The subtrees in the size class~$E_m$ clearly have size~$\leq N/k_m$.
Since the size classes are ordered according to increasing~$k_j$'s,
the subtrees in~$E_j$ also have size~$\leq N/k_m$ for all~$j \geq m$.
Therefore we have:
in the subcases~\ref{itm:aout}) and~\ref{itm:itm:zero}) 
of case~III above we use ~$O(\log{k_1})$ space to keep track of
which of the potential root edges we are currently using,
and all subtrees are of size~$\leq N/k_1$.
The same holds with respect to~$k_2$ in subcase~\ref{itm:itm:one}).
This will suffice to bound the total space used for the subtree rooted at~$B$ by~$O(\log{N})$.

The following lemma gives a relation between the size of the smallest color
class and the number of automorphisms for a~$3$-connected graph, which
has one distinctly colored vertex.

\begin{lemma}\label{lem:aut}
Let~$G$ be a~$3$-connected planar graph with colors on its vertices such that
one vertex~$a$ is colored distinctly, and let~$k \geq 2$ be the size of the
smallest color class apart from the one which contains~$a$.
$G$~has~$\leq 4k$ automorphisms.
\end{lemma}

To prove Lemma \ref{lem:aut}, we refer to the following results.
\begin{lemma}{\rm  \cite{Babai95}(P.\ Mani)}
Every triconnected planar graph~$G$ can be embedded on the~$2$-sphere as
a convex polytope~$P$ such that the automorphism group of~$G$ coincides
with the automorphism group of the convex polytope~$P$ formed by the
embedding.
\end{lemma}

\begin{lemma}{\rm \cite{AD04,Babai95, Artin96}} \label{lem:prod}
For any convex polytope other than tetrahedron, octahedron, cube, icosahedron,
dodecahedron, the automorphism group is the product of its rotation group and
$(1, \tau )$, where~$\tau$ is a
reflection. The rotation group is either~$C_k$ or~$D_k$, where~$C_k$ is the
cyclic group of order~$k$ and~$D_k$ is the dihedral group of order~$2k$.
\end{lemma}

\begin{proofof}{Lemma~\ref{lem:aut}}
Let~$H$ be the subgroup of the rotation group, which permutes the vertices of the
smallest color class among themselves. Then~$H$ is cyclic since the rotation group
is cyclic. Let~$H$ be generated by a permutation~$\pi$.

Notice that a non-trivial rotation of the sphere fixes exactly two
points of the sphere viz.\  the end-points of the axis of rotation.
Then, the following claim holds.
\begin{claim}
\label{clm:bounded}
In the cycle decomposition of~$\pi$ each non-trivial cycle
has the same length.
\end{claim}

\begin{proofof}{Claim~\ref{clm:bounded}}
Suppose~$\pi_1,\pi_2$ are two non-trivial cycles of lengths~$p_1 < p_2$ respectively
in the cycle decomposition of~$\pi$. Then~$\pi^{p_1}$ fixes all elements of~$\pi_1$
but not all elements of~$\pi_2$. Thus~$\pi^{p_1} \in H$ cannot be a rotation of the 
sphere which contradicts the definition of~$H$.
\end{proofof}

As a consequence, the order of~$H$ is bounded by~$k$,
since the length of any cycle containing one of the~$k$ colored points is at most~$k$.
\end{proofof}

This leads to the following corollary, which justifies subcase~\ref{itm:itm:two}) of case~III.
\begin{corollary}\label{co:3-fixed}
Let~$G$ be a~$3$-connected planar graph with at least~$3$ colored
vertices, each having a distinct color. Then~$G$ has at most one
non-trivial automorphism.
\end{corollary}
\begin{proof}
An automorphism of~$G$ has to fix all the colored vertices. Consider
the embedding of~$G$ on a~$2$-sphere. The only possible symmetry is a
reflection about the plane containing the colored vertices, which leads
to exactly one non-trivial automorphism.
\end{proof}

Note, if the triconnected
component~$C$ is one of the exceptions stated in Lemma~\ref{lem:prod},
it implies that~$C$ has~$O(1)$ size. Thus, we do not have to limit its
number of possible minimum canons. 
The preceding discussion implies that 
if two biconnected component trees are equal for the isomorphism order for some choice of the root,
then the corresponding graphs are isomorphic.
The reverse direction clearly holds as well.

\begin{theorem}
\label{thm:cor}
Given two connected planar graphs~$G$ and~$H$, and their biconnected
component trees~$S$ and~$T$, 
then~$G\cong H$ if and only if
there is a choice of articulation points~$a,a'$ in~$G$ and~$H$
such that~$S_a =_{\tt{B}} T_{a'}$.
\end{theorem}

\begin{proof}
Assume that~$S_a =_{\tt{B}} T_{a'}$.
The argument is an induction on the depth of the trees
that follows the inductive definition of the isomorphism order.
The induction goes from depth~$d$ to~$d+2$. If the grandchildren 
of articulation points, say~$s$ and~$t$, are~$=_{\tt{B}}$-equal up to step~3, 
then we compare the children of~$s$ and~$t$.
If they are equal,
we can extend the~$=_{\tt{B}}$-equality to the articulation points~$s$ and~$t$.

When subtrees are rooted at articulation point nodes,
the comparison describes an order on the subgraphs which correspond to 
split components of the articulation points. The order describes an isomorphism 
among the split components.

When subtrees are rooted at biconnected component nodes, say~$B_i$ and~$B'_j$,
the comparison states equality if the components have the same canon, 
i.e. are isomorphic (cf. Theorem~\ref{thm:bcor}) and by induction hypothesis 
we know that the children rooted at articulation points of~$B_i$ and~$B'_j$
are isomorphic.
The equality in the comparisons inductively describes 
an isomorphism between the vertices in the children of the root nodes.

Hence, the isomorphism between the children at any level
can be extended to an isomorphism between 
the corresponding subgraphs in~$G$ and~$H$ and 
therefore to~$G$ and~$H$ itself.

\vspace{\topsep}
The reverse direction holds obviously as well.
Namely, if~$G$ and~$H$ are isomorphic and
there is an isomorphism between~$G$ and~$H$
that maps the articulation point~$a$ of~$G$
to the articulation point~$a'$ of~$H$,
then the biconnected component trees~$S_a$ of~$G$
and~$T_{a'}$ of~$H$ rooted respectively at~$a$ and~$a'$ will clearly be equal.
Hence, such an isomorphism maps articulation points of~$G$ to articulation points of~$H$.
This isomorphism describes a permutation of the split components of the articulation points.
By induction hypothesis, the children at depth~$d+2$ of two such 
biconnected components are isomorphic and equal according to~$=_{\tt{B}}$.
More formally,
one can argue inductively on the depth of~$S_a$ and~$T_{a'}$.
\end{proof}


\subsection{Complexity of the Isomorphism Order Algorithm}

The space analysis of the isomorphism order algorithm is similar to that of 
Lindell's algorithm. We highlight the differences needed 
in the analysis first.

\par 
When we compare biconnected components~$B$ and~$B'$ in the biconnected component tree
then a typical query is of the form~$(s,r)$,
where~$s$ is the chosen root of the triconnected
component tree and~$r$ is the index of the edge in the canon, which is to be retrieved. 
If there are~$k$~choices for the  root for the triconnected component trees of~$B$ and~$B'$, 
the base machine cycles through all of them one by one, keeping track of the minimum canon.
This takes~$O(\log k)$ space. From the discussion above,
we know that the possible choices for the root can be restricted to~$O(k)$,
and that the subtrees rooted at the children of~$B$
have size~$ \leq \size{S_B}/k$, when~$k \geq 2$.
Hence the comparison of~$B$ and~$B'$ can be done in log-space in this case.

We compare the triconnected component trees~$\T(B)$ and~$\T(B')$ according to~$B$ and~$B'$.
When we compare triconnected components in~$\T(B)$ and~$\T(B')$ then 
the algorithm asks oracle queries to the triconnected planar graph canonization algorithm.
The base machine retrieves edges in these canons one by one
from the oracle and compares them. Two edges~$(a,b)$ and~$(a',b')$ are 
compared by first comparing~$a$ and~$a'$. 
If both are articulation points,
we check whether we reach them for the first time.
In this case,
we compare the biconnected subtrees~$S_a$ and~$S_{a'}$ rooted at~$a$ and~$a'$.
If these are equal then we look, whether~$(a,b)$ and~$(a',b')$ are separating pairs.
If so, then we compare their triconnected subtrees.
If these are equal then we proceed with the next edge, 
e.g.\ $(b,c)$,
and continue in the same way.

We now describe in detail, 
how to find out whether
articulation points~$a$ and~$a'$ occur for the first time in our traversal,
and what is stored on the work-tape when we go into recursion.

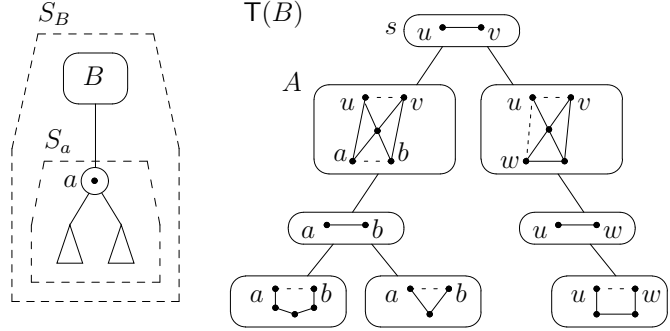
\begin{figure}[!ht]
\begin{center}
\scalebox{0.85}{\input{Figures/biConTriConTree.eepic}}
\end{center}
\caption{A biconnected component tree~$S_{B}$ rooted at
	biconnected component~$B$ which has
	an articulation point~$a$ as child, 
	which occurs in the triconnected component tree~$\T(B)$ of~$B$. 
	In~$A$ and the other triconnected components the dashed edges are separating pairs.
	}\label{fig:biConTriConTree}
\end{figure}

\paragraph{Limiting the number of recursive calls for articulation points.}
When we compare the triconnected component trees~$\T(B)$ and~$\T(B')$, 
respectively (see Figure~\ref{fig:biConTriConTree}), then
we might find several copies of articulation points~$a$ and~$a'$.
That is,~$a$ may occur in several components in~$\T(B)$,
because~$a$ can be part of a separating pair.
We want to go into recursion on~$a$ to the subtree~$S_a$ only once.
This will be either directly when we reach~$\T(B)$, 
in the case that~$S_a$ is a large child of~$B$,
or at a uniquely defined point in~$T(B)$.
The first case will be described in detail below on page~\pageref{large-child}.
Otherwise
we will define a unique component node~$A$ of~$\T(B)$ that contains~$a$,
and we go into recursion on~$a$ only in this component.
Note that $a$~can occur several times in the canon of the triconnected component~$A$, 
once for every edge connected to~$a$.
We go into recursion at the first edge where~$a$ occurs, when we examine~$A$.
We call this occurance of~$a$ the {\em reference copy of\/}~$a$,  \label{ref-point}
and similar for~$a'$ in~$A'$ which is a node in~$\T(B')$.
Note, that the reference copy of~$a$ depends on the chosen root for~$\T(B)$.
We will show that the position of the reference copy 
(i.e.\ the component~$A$ and the position in the canon for~$A$) 
can be found again after recursion without storing any extra information on the work-tape.

\begin{lemma}\label{lem:FirstOccOfArtPoint}
The reference copy of an articulation point~$a$ in~$\T(B)$ and~$a'$ in~$\T(B')$ for
the comparison of triconnected component trees~$\T(B)$ with~$\T(B')$ can be found in log-space.
\end{lemma}

\begin{proof} 
To prove the lemma, we distinguish three cases for~$a$ in~$\T(B)$.
Assume, that we have the same situation for~$a'$ in~$\T(B')$.
If not, then we found an inequality.
We define now a unique component~$A$, where~$a$ is contained.
We distinguish the following cases.
\begin{itemize}
\item Articulation point~$a$ occurs in the root separating pair of~$\T(B)$.
	That is,~$a$ occurs already at the beginning of the comparisons for~$\T(B)$.
	Then we define~$A$ as the  root separating pair.
\item Articulation point~$a$ occurs in  separating pairs other than the root of~$\T(B)$.
	Then~$a$ occurs in all the component nodes, which contain such a separating pair.
	By Lemma~\ref{lem:propertiesOfTree} these nodes form a connected subtree of~$\T(B)$.
	Hence, one of these component nodes is the closest to the root of~$\T(B)$. This 
	component is always a triconnected component node. Let~$A$ be this component. 
	Note, that the comparison first compares~$a$ with~$a'$ before comparing 
	the biconnected or triconnected subtrees, so we reach these copies
	first in the comparison.
\item Articulation point~$a$ does not occur in a separating pair.
	Then,~$a$ occurs in only one triconnected component node in~$\T(B)$.
	Let~$A$ be this component.
\end{itemize}

In all except the first case, we find~$a$ in a triconnected component node~$A$ first. 
Let~$a'$ be found first in component node~$A'$, accordingly.
Assume, we start the comparison of~$A$ and~$A'$.
More precisely, we start to compare the canons~$C$ of~$A$ and~$C'$ of~$A'$ bit for bit.
We go into recursion if and only if we reach the first edge in the canons which contain~$a$ and~$a'$.
Note, that~$C$ can contain more than one edge with endpoint~$a$. 
On all the other edges in~$C$ and~$C'$ we do not go again into recursion.
It is easy to see, that we can recompute the first occurence of~$A$ and~$A'$.
\end{proof}

\paragraph{Comparing two subtrees rooted at separating pairs or triconnected components.}
We go into recursion at separating pairs and triconnected components in~$\T(B)$ and~$\T(B')$.
When we reach a reference copy of an articulation point in both trees, then we interrupt
the comparison of~$B$ with~$B'$ and go into recurison as described before,
i.e.\ we compare the corresponding articulation point nodes, the children of~$B$ and~$B'$.
When we return from recursion, we proceed with the comparison of~$\T(B)$ and~$\T(B')$.

In this part we concentrate on the comparison of~$\T(B)$ and~$\T(B')$.
We give an overview of what is stored on the work-tape when we go into recursion
at separating pairs and triconnected components.
Basically, the comparison is similar to that in Section~\ref{sec:ComplTriIsoOrdAlg}.
We summarize the changes.

\begin{itemize}
\item We use the size function according to Definition \ref{def:BiSize}.
	That is, the size of a triconnected subtree rooted
	at a node $C$ in $\T(B)$
	also includes the sizes of the biconnected subtrees rooted at 
	the reference articulation points which appear
	in the subtree of $\T(B)$ rooted at $C$.

\item For a root separating pair node, 
	we store at most~$O(\log k)$ bits on the work-tape, 
	when we have~$k$~candidates as root separating pairs for~$\T(B)$.
	Hence, whenever we make recomputations in~$\T(B)$, we have
	to find the root separating pair node first.
	For this, we compute~$\T(B)$ in log-space and with the rules described above,
	we find the candidate edges in log-space. With the bits on the work-tape, we know
	which of these candidate edges is the current root separating pair.
	We proceed as in the case of non-root separating pair nodes described next.

\item For a non-root separating pair node and triconnected component nodes,
	we store the same  on the work-tape as described in Section~\ref{sec:ComplTriIsoOrdAlg},
	i.e.\ the counters~$c_<,c_=,c_>$, orientation counters for separating pair nodes,
	and the information of the current canon for triconnected component nodes.
	First, recompute the root separating pair node, 
	then we can determine the parent component node.
	With the information on the work-tape, 
	we can proceed with the computations as described in Section~\ref{sec:ComplTriIsoOrdAlg}.
\end{itemize}

For the triconnected component trees~$\T(B)$ and~$\T(B')$,
we get the same space-bounds as in the previous section on page~\pageref{eq:space-bic}.
That is, for the cross-comparison of the children of 
separating pair nodes~$s$ of~$\T(B)$ and~$t$ of~$\T(B')$
we use~$O(\log k_j)$ space when we go into recursion on subtrees of size~$\leq N/k_j$,
where~$N$ is the size of the subtree rooted at~$s$
and~$k_j$ is the cardinality of the~$j$-th isomorphism class.
For each such child (a triconnected component node), 
we use~$O(1)$ bits, when we go into recursion.
In the case we have large children (of size $\geq N/2$), we treat them a priori.
We will discuss this below.

\paragraph{Comparing two subtrees rooted at articulation points.}

When we consider the trees~$S_a$ and~$S_{a'}$ 
rooted at articulation points~$a$ and~$a'$
then we have for the cross comparison of their children, say~$B_1,\dots,B_k$
and~$B'_1,\dots,B'_k$ respectively,
a similar space analysis as in the case of separating pair nodes.
That is, 
we use~$O(\log k_j)$ space when we go into recursion on subtrees of size~$\leq N/k_j$,
where~$N = \size{S_a}$ 
and~$k_j$ is the cardinality of the~$j$-th isomorphism class.
Large children (of size $\geq N/2$) are treated a priori.
We will discuss this below.

When we compare biconnected components~$B_i$ and~$B'_i$,
then we compute~$\T(B_i)$ and~$\T(B'_i)$.
We have a set of separating pairs as candidates for the root of~$\T(B_i)$.
Recall, that for~$B_i$,	its children are partitioned into size classes.
Let~$k_i$ be the number of elements of the smallest size class with~$k_i \geq 2$,
there are~$O(k_i)$ separating pairs as roots for~$\T(B_i)$.
Except for the trivial cases,
the algorithm uses~$O(\log k_i)$ 
space when it starts to compare the trees~$\T(B_i)$ and~$\T(B'_i)$.

Assume now that we compare~$\T(B_i)$ and~$\T(B'_i)$.
In particular, assume we compare triconnected components~$A$ and~$A'$ of these trees.
We follow the canons of~$A$ and~$A'$ as described above, until 
we reach articulation points, say~$a$ and~$a'$.
First, we recompute whether~$a$ and~$a'$ already occured in the parent node.
If not, then we recompute the canons of~$A$ and~$A'$ and check, 
whether~$a$ and~$a'$ occur for the first time.
If so, then we store nothing and go into recursion.

\par When we return from recursion, 
we recompute the components~$A$ and~$A'$ in~$\T(B)$ and~$\T(B')$.
On the stack there is information about which are the current and the unerased canons.
We run through the current canons and find the first occurence of~$a$ and~$a'$.

\paragraph{Large children.}\label{large-child}
As in the case of biconnected graphs in Section~\ref{sec:cbp},
we deviate from the algorithm described so far in the case 
that the recursion would lead to a large child.
Large subtrees are again treated a priori. 

However,
the notion of a large child is somewhat subtle here.
We already defined the size of biconnected component trees~$S_a$ and~$S_B$
with an articulation point~$a$ or a biconnected component~$B$ as root.
A {\em large child\/} of such a tree of size~$N$ is a child of size~$\geq N/2$.

Now consider $\T(B)$, the triconnected component tree of~$B$.
Let~$A$ be a triconnected component and~$(u,v)$ be a separating pair in~$\T(B)$.
We have not yet defined the subtrees~$S_A$ and~$S_{(u,v)}$ rooted at~$A$ and~$(u,v)$, respectively,
and this has to be done quite carefully.

We already described above that an articulation point~$a$ may occur in several components of a triconnected component tree.
We said that we go into recursion to the biconnected component tree~$S_a$ only once,
namely either when we reach the reference copy of~$a$ (as defined on page~\pageref{ref-point})
or even before in the following case:
let~$a$ be an articulation point in the biconnected component~$B$.
Let~$\T(B)$ be the trinconnected component tree of~$B$,
and let~$C$ be the node in~$\T(B)$ that contains the reference copy of~$a$.
Then it might be the case that
$S_a$ is a large child of~$S_B$ {\em and\/} of~$S_C$.
In this case we visit~$S_a$ when we reach~$B$,
i.e.\ before we start to compute the root for~$\T(B)$.
Then, when we reach the reference copy of~$a$ in~$C$,
we first check whether we already visited~$S_a$.
In this case the comparison result (with some large child~$S_{a'}$ of~$B'$)
is already stored on the work-tape 
and we don not visit~$S_a$ a second time.
Note, if we would go into recursion at the reference copy a second time then
we cannot guarantee the log-space bound of the transducer, because we already
have written bits on the work-tape for $B$
when we traverse the child, the biconnected subtree $S_a$ for the second time.
Otherwise, we visit~$S_a$ at the reference copy of~$a$.

Consequently, we consider~$S_a$ as a subtree only at the place
where we go into recursion to~$S_a$.
Recall, that this is not a static property,
because for example the position of the reference copy depends on the chosen root of the tree,
and we try several possibilities for the root.
Figure~\ref{fig:biConTriConTree2} shows an example.

\begin{figure}[!ht]
\begin{center}
\scalebox{0.85}{\input{Figures/biConTriConTree2.eepic}}
\end{center}
\caption{
The triconnected component tree~$\T(B)$ of the biconnected component~$B$. 
The triconnected component~$A$ contains the reference copy of articulation point~$a$.
If~$S_a$ is not a large child of~$B$,
then the subtree~$S_A$ consists of the subtree of~$T(B)$ rooted at~$A$ 
{\bf and} the subtree~$S_a$.
In contrast, $S_a$ is not part of the subtree~$S_{(a,b)}$
because it does not contain the reference copy of~$a$.
}\label{fig:biConTriConTree2}
\end{figure}

\begin{definition}
Let~$B$ be a biconnected component and~$\T(B)$ its triconnected component tree.
Let~$C$ be a node in~$\T(B)$, i.e.\ a triconnected component node or a separating pair node.
The {\em tree~$S_C$  rooted at\/}~$C$ consists of the subtree of~$\T(B)$ rooted 
at~$C$ 
(with respect to the root of~$\T(B)$) and of the subtrees~$S_a$ for all articulation points~$a$
that have a reference copy in the subtree of~$\T(B)$ rooted at~$C$,
with exception of those~$S_a$ that are a large child of~$S_B$.
The {\em size of\/}~$S_C$ is the sum of the sizes of its components.

Let~$N$ be the size of $S_C$. 
A {\em large child of\/}~$S_C$ is a subtree of the root of~$S_C$ of size~$\geq N/2$.
\end{definition}

Whenever the algorithm reaches a component~$a$,~$B$ or~$C$ as above,
it first checks whether the corresponding tree~$S_a$,~$S_B$, or~$S_C$ has a large child
and treats it a priori.
The result is stored with~$O(1)$ bits.
In the case of triconnected components, we also store the orientation.
We distinguish large children as follows.

\begin{itemize} 
\item Large children with respect to the biconnected component tree.
	These are children of node~$a$ in~$S_a$ or~$B$ in~$S_B$.
	These children are biconnected component nodes or articulation point nodes.
	When comparing~$S_B$ with~$S_{B'}$, then we go for large children into recursion 
	before computing the trees~$\T(B)$ and~$\T(B')$.
\item Large children with respect to the triconnected component tree.
	These are children of node~$C$ in~$S_C$.
	These children are separating pair nodes, 
	triconnected component nodes 
	or reference copies of articulation point nodes in~$C$.
\end{itemize}

\paragraph{Analysis of the space requirement.}
We analyze the 
comparison algorithm when it compares subtrees rooted at 
separating pairs and subtrees rooted at articulation points. 
For the analysis, the recursion goes here from depth~$d$ to~$d+2$ of the trees.
Observe, that large children are handled a priori at any level of the trees.
We set up the following recursion equation for the space requirement of our algorithm.
	\begin{displaymath}
	\mathcal{S}(N)= \max_j~\mathcal{S}\!\left( \frac{N}{k_j}\right) + O(\log{k_j}),
	\end{displaymath}
where~$k_j \geq 2$ (for all~$j$) are the values 
mentioned above in the corresponding cases.
Hence,~$\mathcal{S}(N)= O( \log{N} )$.

For the explanation of the  recursion equation
it is helpful to imagine that we have two work-tapes.
We use the first work-tape  when we go into recursion at articulation point nodes,
and the second work-tape when we go into recursion at separating pair nodes.
The total space needed is the sum of the space of the two work-tapes.

\begin{itemize}
\item At an articulation point node,
	the value~$k_j$ is  the number of elements in the~$j$-th size class among the 
	children~$B_1,\dots,B_k$ of the articulation point node. 
	We store~$O(\log k_j)$ bits and recursively consider subtrees of size~$\leq N / k_j$.

\item At a separating pair node
	the value~$k_j$ is  the number of elements in the~$j$-th isomorphism class 
	among the children~$G_1,\dots,G_k$
	of the separating pair node.
	We store~$O(\log k_j)$ bits and recursively consider subtrees of size~$\leq N / k_j$.
\end{itemize}

This finishes the complexity analysis. We get the following theorem.

\begin{theorem}\label{thm:plc}
The isomorphism order between two planar graphs can be computed in log-space.
\end{theorem}

\subsection{The Canon of a Planar Graph}
From Theorem~\ref{thm:plc}, we know that the isomorphism order of biconnected
component trees can be computed in log-space. Using this algorithm, we show
that the canon of a planar graph can be output 
in log-space.
\par 
The canonization of planar graphs proceeds exactly as in the case of biconnected planar graphs.
A log-space procedure traverses the biconnected
component tree and makes oracle queries
to the isomorphism order algorithm and outputs a canonical list of edges,
along with delimiters to separate the lists for siblings. 

For an example, consider the canonical list~$l(S,a)$ of edges 
for the tree~$S_a$ of Figure \ref{fig:biConCompTrees}.
Let $l(B_i,a)$ be the canonical list edges of the biconnected component $B_i$ 
(i.e. the canonical list of $\T(B_i)$ with $a$ the parent articulation point).
Let $a_1, \dots, a_{l_1}$ be the order of the reference copies of articulation points
as they occur in the canon of $\T(B_i)$.
Then we get the following canonical list for $S_a$.

\begin{eqnarray*}
l(S,a) &=& [\; (a)\; l(S_{B_1},a) \; \dots \; l(S_{B_k},a)\; ], \text{ where}\\
l(S_{B_1},a) &=& [\; l(B_1,a)\; [l(S_{a_1},a_1)]\; \dots\;  [l(S_{a_{l_1}},a_{l_1})]\; ]\\
&\vdots& \\
l(S_{B_k},a) &=& [\; l(B_k,a)\; [l(S_{a_{l_k}},a_{l_k})]\; ]
\end{eqnarray*}

A log-space transducer then renames the vertices according to their first occurrence
in this list, to get the final canon for the biconnected component tree.
This canon depends upon the choice
of the root of the biconnected component tree. Further log-space
transducers cycle through all the articulation points as roots to find 
the minimum canon among them, then rename the vertices according to their first 
occurrence in the canon 
and finally, remove the virtual edges and delimiters to obtain a canon for the planar graph.
This proves the main theorem.

\begin{theorem}
A planar graph can be canonized in log-space.
\end{theorem}

%% file: Figures/biConCompTrees.eepic
\setlength{\unitlength}{0.00087489in}
\begingroup\makeatletter\ifx\SetFigFontNFSS\undefined%
\gdef\SetFigFontNFSS#1#2#3#4#5{%
  \reset@font\fontsize{#1}{#2pt}%
  \fontfamily{#3}\fontseries{#4}\fontshape{#5}%
  \selectfont}%
\fi\endgroup%
{\renewcommand{\dashlinestretch}{30}
\begin{picture}(8349,3279)(0,-10)
\path(552,687)(552,507)
\path(642,687)(822,507)
\path(552,507)(462,237)(642,237)(552,507)
\path(822,507)(732,237)(912,237)(822,507)
\path(462,687)(282,507)
\path(282,507)(192,237)(372,237)(282,507)
\path(3522,687)(3522,507)
\path(3612,687)(3792,507)
\path(3522,507)(3432,237)(3612,237)(3522,507)
\path(3792,507)(3702,237)(3882,237)(3792,507)
\path(3432,687)(3252,507)
\path(3252,507)(3162,237)(3342,237)(3252,507)
\path(1452,507)(1362,237)(1542,237)(1452,507)
\path(1722,507)(1632,237)(1812,237)(1722,507)
\path(1452,507)(1542,687)
\path(1722,507)(1632,687)
\path(2127,507)(2037,237)(2217,237)(2127,507)
\path(2397,507)(2307,237)(2487,237)(2397,507)
\path(2127,507)(2217,687)
\path(2397,507)(2307,687)
\path(4827,687)(4827,507)
\path(4917,687)(5097,507)
\path(4827,507)(4737,237)(4917,237)(4827,507)
\path(5097,507)(5007,237)(5187,237)(5097,507)
\path(4737,687)(4557,507)
\path(4557,507)(4467,237)(4647,237)(4557,507)
\path(7797,687)(7797,507)
\path(7887,687)(8067,507)
\path(7797,507)(7707,237)(7887,237)(7797,507)
\path(8067,507)(7977,237)(8157,237)(8067,507)
\path(7707,687)(7527,507)
\path(7527,507)(7437,237)(7617,237)(7527,507)
\path(5727,507)(5637,237)(5817,237)(5727,507)
\path(5997,507)(5907,237)(6087,237)(5997,507)
\path(5727,507)(5817,687)
\path(5997,507)(5907,687)
\path(6402,507)(6312,237)(6492,237)(6402,507)
\path(6672,507)(6582,237)(6762,237)(6672,507)
\path(6402,507)(6492,687)
\path(6672,507)(6582,687)
\put(7069,2935){\blacken\ellipse{36}{36}}
\put(7069,2935){\ellipse{36}{36}}
\put(7072,2932){\ellipse{190}{190}}
\put(5404,1495){\blacken\ellipse{36}{36}}
\put(5404,1495){\ellipse{36}{36}}
\put(6529,1495){\blacken\ellipse{36}{36}}
\put(6529,1495){\ellipse{36}{36}}
\put(7789,1505){\blacken\ellipse{36}{36}}
\put(7789,1505){\ellipse{36}{36}}
\put(3514,1495){\blacken\ellipse{36}{36}}
\put(3514,1495){\ellipse{36}{36}}
\put(2254,1495){\blacken\ellipse{36}{36}}
\put(2254,1495){\ellipse{36}{36}}
\put(1129,1505){\blacken\ellipse{36}{36}}
\put(1129,1505){\ellipse{36}{36}}
\put(2794,2935){\blacken\ellipse{36}{36}}
\put(2794,2935){\ellipse{36}{36}}
\put(5407,1492){\ellipse{190}{190}}
\put(6532,1492){\ellipse{190}{190}}
\put(7792,1502){\ellipse{190}{190}}
\put(3517,1492){\ellipse{190}{190}}
\put(2257,1492){\ellipse{190}{190}}
\put(1132,1502){\ellipse{190}{190}}
\put(2797,2932){\ellipse{190}{190}}
\path(2712,2892)(1857,2397)
\path(2742,2851)(2622,2622)
\path(2892,2892)(3522,2397)
\path(2858,2855)(2982,2622)
\path(2802,2834)(2802,2622)
\path(1677,2037)(1137,1598)
\path(1098,1414)(1002,1227)
\path(1137,1407)(1137,1227)
\path(1176,1418)(1272,1227)
\put(432,792){\arc{210}{1.5708}{3.1416}}
\put(432,942){\arc{210}{3.1416}{4.7124}}
\put(672,942){\arc{210}{4.7124}{6.2832}}
\put(672,792){\arc{210}{0}{1.5708}}
\path(327,792)(327,942)
\path(432,1047)(672,1047)
\path(777,942)(777,792)
\path(672,687)(432,687)
\path(1857,2037)(1902,1902)
\path(1767,2037)(1722,1902)
\path(1947,2037)(2217,1587)
\path(2264,1397)(2262,1047)
\put(1692,2142){\arc{210}{1.5708}{3.1416}}
\put(1692,2292){\arc{210}{3.1416}{4.7124}}
\put(1932,2292){\arc{210}{4.7124}{6.2832}}
\put(1932,2142){\arc{210}{0}{1.5708}}
\path(1587,2142)(1587,2292)
\path(1692,2397)(1932,2397)
\path(2037,2292)(2037,2142)
\path(1932,2037)(1692,2037)
\path(1073,1427)(552,1047)
\put(1467,792){\arc{210}{1.5708}{3.1416}}
\put(1467,942){\arc{210}{3.1416}{4.7124}}
\put(1707,942){\arc{210}{4.7124}{6.2832}}
\put(1707,792){\arc{210}{0}{1.5708}}
\path(1362,792)(1362,942)
\path(1467,1047)(1707,1047)
\path(1812,942)(1812,792)
\path(1707,687)(1467,687)
\path(1197,1435)(1587,1047)
\put(2142,792){\arc{210}{1.5708}{3.1416}}
\put(2142,942){\arc{210}{3.1416}{4.7124}}
\put(2382,942){\arc{210}{4.7124}{6.2832}}
\put(2382,792){\arc{210}{0}{1.5708}}
\path(2037,792)(2037,942)
\path(2142,1047)(2382,1047)
\path(2487,942)(2487,792)
\path(2382,687)(2142,687)
\path(3522,2037)(3522,1587)
\path(3522,1401)(3522,1047)
\put(3402,2142){\arc{210}{1.5708}{3.1416}}
\put(3402,2292){\arc{210}{3.1416}{4.7124}}
\put(3642,2292){\arc{210}{4.7124}{6.2832}}
\put(3642,2142){\arc{210}{0}{1.5708}}
\path(3297,2142)(3297,2292)
\path(3402,2397)(3642,2397)
\path(3747,2292)(3747,2142)
\path(3642,2037)(3402,2037)
\put(3402,792){\arc{210}{1.5708}{3.1416}}
\put(3402,942){\arc{210}{3.1416}{4.7124}}
\put(3642,942){\arc{210}{4.7124}{6.2832}}
\put(3642,792){\arc{210}{0}{1.5708}}
\path(3297,792)(3297,942)
\path(3402,1047)(3642,1047)
\path(3747,942)(3747,792)
\path(3642,687)(3402,687)
\dashline{30.000}(3117,1812)(3972,1812)(3972,147)
	(3117,147)(3117,1812)
\dashline{30.000}(1542,1812)(1902,1227)(1902,147)
	(102,147)(102,1182)(1002,1812)(1542,1812)
\path(6987,2892)(6132,2397)
\path(7032,2847)(6897,2622)
\path(7167,2892)(7797,2397)
\path(7122,2847)(7257,2622)
\path(7076,2838)(7077,2622)
\path(5952,2037)(5412,1587)
\path(5368,1405)(5277,1227)
\path(5411,1397)(5412,1227)
\path(5457,1407)(5547,1227)
\put(4707,792){\arc{210}{1.5708}{3.1416}}
\put(4707,942){\arc{210}{3.1416}{4.7124}}
\put(4947,942){\arc{210}{4.7124}{6.2832}}
\put(4947,792){\arc{210}{0}{1.5708}}
\path(4602,792)(4602,942)
\path(4707,1047)(4947,1047)
\path(5052,942)(5052,792)
\path(4947,687)(4707,687)
\path(6132,2037)(6177,1902)
\path(6042,2037)(5997,1902)
\path(6222,2037)(6492,1587)
\path(6536,1397)(6537,1047)
\put(5967,2142){\arc{210}{1.5708}{3.1416}}
\put(5967,2292){\arc{210}{3.1416}{4.7124}}
\put(6207,2292){\arc{210}{4.7124}{6.2832}}
\put(6207,2142){\arc{210}{0}{1.5708}}
\path(5862,2142)(5862,2292)
\path(5967,2397)(6207,2397)
\path(6312,2292)(6312,2142)
\path(6207,2037)(5967,2037)
\path(5342,1418)(4827,1047)
\put(5742,792){\arc{210}{1.5708}{3.1416}}
\put(5742,942){\arc{210}{3.1416}{4.7124}}
\put(5982,942){\arc{210}{4.7124}{6.2832}}
\put(5982,792){\arc{210}{0}{1.5708}}
\path(5637,792)(5637,942)
\path(5742,1047)(5982,1047)
\path(6087,942)(6087,792)
\path(5982,687)(5742,687)
\path(5480,1427)(5862,1047)
\put(6417,792){\arc{210}{1.5708}{3.1416}}
\put(6417,942){\arc{210}{3.1416}{4.7124}}
\put(6657,942){\arc{210}{4.7124}{6.2832}}
\put(6657,792){\arc{210}{0}{1.5708}}
\path(6312,792)(6312,942)
\path(6417,1047)(6657,1047)
\path(6762,942)(6762,792)
\path(6657,687)(6417,687)
\path(7797,2037)(7794,1602)
\path(7796,1409)(7797,1047)
\put(7677,2142){\arc{210}{1.5708}{3.1416}}
\put(7677,2292){\arc{210}{3.1416}{4.7124}}
\put(7917,2292){\arc{210}{4.7124}{6.2832}}
\put(7917,2142){\arc{210}{0}{1.5708}}
\path(7572,2142)(7572,2292)
\path(7677,2397)(7917,2397)
\path(8022,2292)(8022,2142)
\path(7917,2037)(7677,2037)
\put(7677,792){\arc{210}{1.5708}{3.1416}}
\put(7677,942){\arc{210}{3.1416}{4.7124}}
\put(7917,942){\arc{210}{4.7124}{6.2832}}
\put(7917,792){\arc{210}{0}{1.5708}}
\path(7572,792)(7572,942)
\path(7677,1047)(7917,1047)
\path(8022,942)(8022,792)
\path(7917,687)(7677,687)
\dottedline{60}(12,12)(12,1407)(1677,2532)
	(2217,2532)(2577,1902)(2577,12)(12,12)
\dottedline{75}(4287,12)(4287,1407)(5952,2532)
	(6492,2532)(6852,1902)(6852,12)(4287,12)
\dottedline{75}(7302,12)(8337,12)(8337,2532)
	(7302,2532)(7302,12)
\dashline{30.000}(7392,1812)(8247,1812)(8247,147)
	(7392,147)(7392,1812)
\dottedline{60}(3027,12)(4062,12)(4062,2532)
	(3027,2532)(3027,12)
\path(3466,2040)(3432,1902)
\path(3432,2037)(3342,1902)
\path(7741,2040)(7707,1902)
\path(7707,2037)(7617,1902)
\dashline{30.000}(5817,1812)(6177,1227)(6177,147)
	(4377,147)(4377,1182)(5277,1812)(5817,1812)
\put(2622,867){\makebox(0,0)[lb]{\smash{{\SetFigFontNFSS{12}{14.4}{\rmdefault}{\mddefault}{\updefault}\huge{\dots}}}}}
\put(1497,1497){\makebox(0,0)[lb]{\smash{{\SetFigFontNFSS{12}{14.4}{\rmdefault}{\mddefault}{\updefault}\huge{\dots}}}}}
\put(2442,2217){\makebox(0,0)[lb]{\smash{{\SetFigFontNFSS{12}{14.4}{\rmdefault}{\mddefault}{\updefault}\huge{\dots}}}}}
\put(912,867){\makebox(0,0)[lb]{\smash{{\SetFigFontNFSS{12}{14.4}{\rmdefault}{\mddefault}{\updefault}\huge{\dots}}}}}
\put(6897,867){\makebox(0,0)[lb]{\smash{{\SetFigFontNFSS{12}{14.4}{\rmdefault}{\mddefault}{\updefault}\huge{\dots}}}}}
\put(5772,1497){\makebox(0,0)[lb]{\smash{{\SetFigFontNFSS{12}{14.4}{\rmdefault}{\mddefault}{\updefault}\huge{\dots}}}}}
\put(6717,2217){\makebox(0,0)[lb]{\smash{{\SetFigFontNFSS{12}{14.4}{\rmdefault}{\mddefault}{\updefault}\huge{\dots}}}}}
\put(5187,867){\makebox(0,0)[lb]{\smash{{\SetFigFontNFSS{12}{14.4}{\rmdefault}{\mddefault}{\updefault}\huge{\dots}}}}}
\put(2622,1497){\makebox(0,0)[lb]{\smash{{\SetFigFontNFSS{12}{14.4}{\rmdefault}{\mddefault}{\updefault}\huge{\dots}}}}}
\put(6897,1497){\makebox(0,0)[lb]{\smash{{\SetFigFontNFSS{12}{14.4}{\rmdefault}{\mddefault}{\updefault}\huge{\dots}}}}}
\put(1722,2127){\makebox(0,0)[lb]{\smash{{\SetFigFontNFSS{12}{14.4}{\rmdefault}{\mddefault}{\updefault}$B_1$}}}}
\put(3432,2127){\makebox(0,0)[lb]{\smash{{\SetFigFontNFSS{12}{14.4}{\rmdefault}{\mddefault}{\updefault}$B_k$}}}}
\put(5997,2127){\makebox(0,0)[lb]{\smash{{\SetFigFontNFSS{12}{14.4}{\rmdefault}{\mddefault}{\updefault}$B'_1$}}}}
\put(7707,2127){\makebox(0,0)[lb]{\smash{{\SetFigFontNFSS{12}{14.4}{\rmdefault}{\mddefault}{\updefault}$B'_k$}}}}
\put(6807,2892){\makebox(0,0)[lb]{\smash{{\SetFigFontNFSS{12}{14.4}{\rmdefault}{\mddefault}{\updefault}$a'$}}}}
\put(6222,1452){\makebox(0,0)[lb]{\smash{{\SetFigFontNFSS{12}{14.4}{\rmdefault}{\mddefault}{\updefault}$a'_{l_1}$}}}}
\put(7482,1452){\makebox(0,0)[lb]{\smash{{\SetFigFontNFSS{12}{14.4}{\rmdefault}{\mddefault}{\updefault}$a'_{l_k}$}}}}
\put(5142,1452){\makebox(0,0)[lb]{\smash{{\SetFigFontNFSS{12}{14.4}{\rmdefault}{\mddefault}{\updefault}$a'_1$}}}}
\put(1947,1452){\makebox(0,0)[lb]{\smash{{\SetFigFontNFSS{12}{14.4}{\rmdefault}{\mddefault}{\updefault}$a_{l_1}$}}}}
\put(867,1452){\makebox(0,0)[lb]{\smash{{\SetFigFontNFSS{12}{14.4}{\rmdefault}{\mddefault}{\updefault}$a_1$}}}}
\put(3207,1452){\makebox(0,0)[lb]{\smash{{\SetFigFontNFSS{12}{14.4}{\rmdefault}{\mddefault}{\updefault}$a_{l_k}$}}}}
\put(2532,2892){\makebox(0,0)[lb]{\smash{{\SetFigFontNFSS{12}{14.4}{\rmdefault}{\mddefault}{\updefault}$a$}}}}
\put(6177,3117){\makebox(0,0)[lb]{\smash{{\SetFigFontNFSS{12}{14.4}{\rmdefault}{\mddefault}{\updefault}$T_{a'}$}}}}
\put(1902,3117){\makebox(0,0)[lb]{\smash{{\SetFigFontNFSS{12}{14.4}{\rmdefault}{\mddefault}{\updefault}$S_a$}}}}
\put(1002,1857){\makebox(0,0)[lb]{\smash{{\SetFigFontNFSS{12}{14.4}{\rmdefault}{\mddefault}{\updefault}$S_{a_1}$}}}}
\put(5277,1857){\makebox(0,0)[lb]{\smash{{\SetFigFontNFSS{12}{14.4}{\rmdefault}{\mddefault}{\updefault}$T_{a'_1}$}}}}
\put(3747,1857){\makebox(0,0)[lb]{\smash{{\SetFigFontNFSS{12}{14.4}{\rmdefault}{\mddefault}{\updefault}$S_{a_{l_k}}$}}}}
\put(8022,1857){\makebox(0,0)[lb]{\smash{{\SetFigFontNFSS{12}{14.4}{\rmdefault}{\mddefault}{\updefault}$T_{a'_{l_k}}$}}}}
\put(5952,2577){\makebox(0,0)[lb]{\smash{{\SetFigFontNFSS{12}{14.4}{\rmdefault}{\mddefault}{\updefault}$T_{B'_1}$}}}}
\put(7842,2577){\makebox(0,0)[lb]{\smash{{\SetFigFontNFSS{12}{14.4}{\rmdefault}{\mddefault}{\updefault}$T_{B'_k}$}}}}
\put(3567,2577){\makebox(0,0)[lb]{\smash{{\SetFigFontNFSS{12}{14.4}{\rmdefault}{\mddefault}{\updefault}$S_{B_k}$}}}}
\put(1677,2577){\makebox(0,0)[lb]{\smash{{\SetFigFontNFSS{12}{14.4}{\rmdefault}{\mddefault}{\updefault}$S_{B_1}$}}}}
\end{picture}
}

%% file: Figures/biConTriConTree.eepic
\setlength{\unitlength}{0.00087489in}
\begingroup\makeatletter\ifx\SetFigFontNFSS\undefined%
\gdef\SetFigFontNFSS#1#2#3#4#5{%
  \reset@font\fontsize{#1}{#2pt}%
  \fontfamily{#3}\fontseries{#4}\fontshape{#5}%
  \selectfont}%
\fi\endgroup%
{\renewcommand{\dashlinestretch}{30}
\begin{picture}(4614,2334)(0,-10)
\path(2487,1632)(2397,1182)(2757,1632)
	(2667,1182)(2487,1587)
\path(4107,282)(4107,102)(4377,102)(4377,282)
\path(2802,282)(2937,102)(3072,282)
\path(3657,1632)(3882,1182)(3612,1182)
	(3927,1632)(3882,1182)
\dashline{30.000}(3657,1632)(3612,1182)
\path(1857,282)(1857,147)(1992,102)
	(2127,147)(2127,282)
\put(589,1045){\blacken\ellipse{36}{36}}
\put(589,1045){\ellipse{36}{36}}
\put(3297,2127){\blacken\ellipse{36}{36}}
\put(3297,2127){\ellipse{36}{36}}
\put(3027,2127){\blacken\ellipse{36}{36}}
\put(3027,2127){\ellipse{36}{36}}
\path(3027,2127)(3297,2127)
\put(2487,732){\blacken\ellipse{36}{36}}
\put(2487,732){\ellipse{36}{36}}
\put(2217,732){\blacken\ellipse{36}{36}}
\put(2217,732){\ellipse{36}{36}}
\path(2217,732)(2487,732)
\put(2052,702){\arc{210}{1.5708}{3.1416}}
\put(2052,717){\arc{210}{3.1416}{4.7124}}
\put(2652,717){\arc{210}{4.7124}{6.2832}}
\put(2652,702){\arc{210}{0}{1.5708}}
\path(1947,702)(1947,717)
\path(2052,822)(2652,822)
\path(2757,717)(2757,702)
\path(2652,597)(2052,597)
\put(4107,732){\blacken\ellipse{36}{36}}
\put(4107,732){\ellipse{36}{36}}
\put(3837,732){\blacken\ellipse{36}{36}}
\put(3837,732){\ellipse{36}{36}}
\path(3837,732)(4107,732)
\put(3612,1182){\blacken\ellipse{36}{36}}
\put(3612,1182){\ellipse{36}{36}}
\put(2127,282){\blacken\ellipse{36}{36}}
\put(2127,282){\ellipse{36}{36}}
\put(1857,282){\blacken\ellipse{36}{36}}
\put(1857,282){\ellipse{36}{36}}
\put(4377,282){\blacken\ellipse{36}{36}}
\put(4377,282){\ellipse{36}{36}}
\put(4107,282){\blacken\ellipse{36}{36}}
\put(4107,282){\ellipse{36}{36}}
\put(3072,282){\blacken\ellipse{36}{36}}
\put(3072,282){\ellipse{36}{36}}
\put(2802,282){\blacken\ellipse{36}{36}}
\put(2802,282){\ellipse{36}{36}}
\put(3927,1632){\blacken\ellipse{36}{36}}
\put(3927,1632){\ellipse{36}{36}}
\put(3657,1632){\blacken\ellipse{36}{36}}
\put(3657,1632){\ellipse{36}{36}}
\put(2667,1182){\blacken\ellipse{36}{36}}
\put(2667,1182){\ellipse{36}{36}}
\put(2397,1182){\blacken\ellipse{36}{36}}
\put(2397,1182){\ellipse{36}{36}}
\put(2757,1632){\blacken\ellipse{36}{36}}
\put(2757,1632){\ellipse{36}{36}}
\put(2487,1632){\blacken\ellipse{36}{36}}
\put(2487,1632){\ellipse{36}{36}}
\put(1857,147){\blacken\ellipse{36}{36}}
\put(1857,147){\ellipse{36}{36}}
\put(1992,102){\blacken\ellipse{36}{36}}
\put(1992,102){\ellipse{36}{36}}
\put(2127,147){\blacken\ellipse{36}{36}}
\put(2127,147){\ellipse{36}{36}}
\put(2937,102){\blacken\ellipse{36}{36}}
\put(2937,102){\ellipse{36}{36}}
\put(4107,102){\blacken\ellipse{36}{36}}
\put(4107,102){\ellipse{36}{36}}
\put(4377,102){\blacken\ellipse{36}{36}}
\put(4377,102){\ellipse{36}{36}}
\put(3882,1182){\blacken\ellipse{36}{36}}
\put(3882,1182){\ellipse{36}{36}}
\put(2572,1397){\blacken\ellipse{36}{36}}
\put(2572,1397){\ellipse{36}{36}}
\put(3772,1407){\blacken\ellipse{36}{36}}
\put(3772,1407){\ellipse{36}{36}}
\put(592,1042){\ellipse{190}{190}}
\path(597,1587)(597,1137)
\path(552,957)(417,732)
\path(417,732)(327,462)(507,462)(417,732)
\path(777,732)(687,462)(867,462)(777,732)
\path(642,957)(777,732)
\put(477,1692){\arc{210}{1.5708}{3.1416}}
\put(477,1842){\arc{210}{3.1416}{4.7124}}
\put(717,1842){\arc{210}{4.7124}{6.2832}}
\put(717,1692){\arc{210}{0}{1.5708}}
\path(372,1692)(372,1842)
\path(477,1947)(717,1947)
\path(822,1842)(822,1692)
\path(717,1587)(477,1587)
\put(3402,1197){\arc{210}{1.5708}{3.1416}}
\put(3402,1617){\arc{210}{3.1416}{4.7124}}
\put(4182,1617){\arc{210}{4.7124}{6.2832}}
\put(4182,1197){\arc{210}{0}{1.5708}}
\path(3297,1197)(3297,1617)
\path(3402,1722)(4182,1722)
\path(4287,1617)(4287,1197)
\path(4182,1092)(3402,1092)
\put(2862,2097){\arc{210}{1.5708}{3.1416}}
\put(2862,2112){\arc{210}{3.1416}{4.7124}}
\put(3462,2112){\arc{210}{4.7124}{6.2832}}
\put(3462,2097){\arc{210}{0}{1.5708}}
\path(2757,2097)(2757,2112)
\path(2862,2217)(3462,2217)
\path(3567,2112)(3567,2097)
\path(3462,1992)(2862,1992)
\path(3027,1992)(2847,1722)
\path(2262,597)(2082,372)
\path(3342,1992)(3522,1722)
\put(3672,702){\arc{210}{1.5708}{3.1416}}
\put(3672,717){\arc{210}{3.1416}{4.7124}}
\put(4272,717){\arc{210}{4.7124}{6.2832}}
\put(4272,702){\arc{210}{0}{1.5708}}
\path(3567,702)(3567,717)
\path(3672,822)(4272,822)
\path(4377,717)(4377,702)
\path(4272,597)(3672,597)
\path(2532,597)(2712,372)
\put(2592,117){\arc{210}{1.5708}{3.1416}}
\put(2592,267){\arc{210}{3.1416}{4.7124}}
\put(3192,267){\arc{210}{4.7124}{6.2832}}
\put(3192,117){\arc{210}{0}{1.5708}}
\path(2487,117)(2487,267)
\path(2592,372)(3192,372)
\path(3297,267)(3297,117)
\path(3192,12)(2592,12)
\put(1647,117){\arc{210}{1.5708}{3.1416}}
\put(1647,267){\arc{210}{3.1416}{4.7124}}
\put(2247,267){\arc{210}{4.7124}{6.2832}}
\put(2247,117){\arc{210}{0}{1.5708}}
\path(1542,117)(1542,267)
\path(1647,372)(2247,372)
\path(2352,267)(2352,117)
\path(2247,12)(1647,12)
\path(4062,597)(4242,372)
\path(2577,1092)(2397,822)
\path(3837,1092)(4017,822)
\put(3897,117){\arc{210}{1.5708}{3.1416}}
\put(3897,267){\arc{210}{3.1416}{4.7124}}
\put(4497,267){\arc{210}{4.7124}{6.2832}}
\put(4497,117){\arc{210}{0}{1.5708}}
\path(3792,117)(3792,267)
\path(3897,372)(4497,372)
\path(4602,267)(4602,117)
\path(4497,12)(3897,12)
\put(2232,1197){\arc{210}{1.5708}{3.1416}}
\put(2232,1617){\arc{210}{3.1416}{4.7124}}
\put(3012,1617){\arc{210}{4.7124}{6.2832}}
\put(3012,1197){\arc{210}{0}{1.5708}}
\path(2127,1197)(2127,1617)
\path(2232,1722)(3012,1722)
\path(3117,1617)(3117,1197)
\path(3012,1092)(2232,1092)
\dashline{30.000}(1857,282)(2127,282)
\dashline{30.000}(4107,282)(4377,282)
\dashline{30.000}(2397,1182)(2667,1182)
\dashline{30.000}(2802,282)(3072,282)
\dashline{30.000}(3657,1632)(3927,1632)
\dashline{30.000}(2487,1632)(2757,1632)
\dashline{60.000}(237,1182)(147,732)(147,327)
	(1047,327)(1047,732)(957,1182)(237,1182)
\dashline{60.000}(192,2082)(12,1272)(12,192)
	(1182,192)(1182,1272)(1002,2082)(192,2082)
\put(2037,642){\makebox(0,0)[lb]{\smash{{\SetFigFontNFSS{12}{14.4}{\rmdefault}{\mddefault}{\updefault}$a$}}}}
\put(192,2172){\makebox(0,0)[lb]{\smash{{\SetFigFontNFSS{12}{14.4}{\rmdefault}{\mddefault}{\updefault}$S_{B}$}}}}
\put(372,1002){\makebox(0,0)[lb]{\smash{{\SetFigFontNFSS{12}{14.4}{\rmdefault}{\mddefault}{\updefault}$a$}}}}
\put(507,1722){\makebox(0,0)[lb]{\smash{{\SetFigFontNFSS{12}{14.4}{\rmdefault}{\mddefault}{\updefault}$B$}}}}
\put(2847,2037){\makebox(0,0)[lb]{\smash{{\SetFigFontNFSS{12}{14.4}{\rmdefault}{\mddefault}{\updefault}$u$}}}}
\put(3342,2037){\makebox(0,0)[lb]{\smash{{\SetFigFontNFSS{12}{14.4}{\rmdefault}{\mddefault}{\updefault}$v$}}}}
\put(3477,1542){\makebox(0,0)[lb]{\smash{{\SetFigFontNFSS{12}{14.4}{\rmdefault}{\mddefault}{\updefault}$u$}}}}
\put(4152,642){\makebox(0,0)[lb]{\smash{{\SetFigFontNFSS{12}{14.4}{\rmdefault}{\mddefault}{\updefault}$w$}}}}
\put(3657,642){\makebox(0,0)[lb]{\smash{{\SetFigFontNFSS{12}{14.4}{\rmdefault}{\mddefault}{\updefault}$u$}}}}
\put(3927,192){\makebox(0,0)[lb]{\smash{{\SetFigFontNFSS{12}{14.4}{\rmdefault}{\mddefault}{\updefault}$u$}}}}
\put(4422,192){\makebox(0,0)[lb]{\smash{{\SetFigFontNFSS{12}{14.4}{\rmdefault}{\mddefault}{\updefault}$w$}}}}
\put(3972,1542){\makebox(0,0)[lb]{\smash{{\SetFigFontNFSS{12}{14.4}{\rmdefault}{\mddefault}{\updefault}$v$}}}}
\put(2307,1542){\makebox(0,0)[lb]{\smash{{\SetFigFontNFSS{12}{14.4}{\rmdefault}{\mddefault}{\updefault}$u$}}}}
\put(2802,1542){\makebox(0,0)[lb]{\smash{{\SetFigFontNFSS{12}{14.4}{\rmdefault}{\mddefault}{\updefault}$v$}}}}
\put(2532,642){\makebox(0,0)[lb]{\smash{{\SetFigFontNFSS{12}{14.4}{\rmdefault}{\mddefault}{\updefault}$b$}}}}
\put(2172,192){\makebox(0,0)[lb]{\smash{{\SetFigFontNFSS{12}{14.4}{\rmdefault}{\mddefault}{\updefault}$b$}}}}
\put(3117,192){\makebox(0,0)[lb]{\smash{{\SetFigFontNFSS{12}{14.4}{\rmdefault}{\mddefault}{\updefault}$b$}}}}
\put(2712,1182){\makebox(0,0)[lb]{\smash{{\SetFigFontNFSS{12}{14.4}{\rmdefault}{\mddefault}{\updefault}$b$}}}}
\put(2262,1182){\makebox(0,0)[lb]{\smash{{\SetFigFontNFSS{12}{14.4}{\rmdefault}{\mddefault}{\updefault}$a$}}}}
\put(3432,1137){\makebox(0,0)[lb]{\smash{{\SetFigFontNFSS{12}{14.4}{\rmdefault}{\mddefault}{\updefault}$w$}}}}
\put(1902,1677){\makebox(0,0)[lb]{\smash{{\SetFigFontNFSS{12}{14.4}{\rmdefault}{\mddefault}{\updefault}$A$}}}}
\put(2622,2082){\makebox(0,0)[lb]{\smash{{\SetFigFontNFSS{12}{14.4}{\rmdefault}{\mddefault}{\updefault}$s$}}}}
\put(1677,192){\makebox(0,0)[lb]{\smash{{\SetFigFontNFSS{12}{14.4}{\rmdefault}{\mddefault}{\updefault}$a$}}}}
\put(2622,192){\makebox(0,0)[lb]{\smash{{\SetFigFontNFSS{12}{14.4}{\rmdefault}{\mddefault}{\updefault}$a$}}}}
\put(237,1272){\makebox(0,0)[lb]{\smash{{\SetFigFontNFSS{12}{14.4}{\rmdefault}{\mddefault}{\updefault}$S_a$}}}}
\put(1632,2172){\makebox(0,0)[lb]{\smash{{\SetFigFontNFSS{12}{14.4}{\rmdefault}{\mddefault}{\updefault}${\sf T}(B)$}}}}
\end{picture}
}

%% file: Figures/biConTriConTree2.eepic
\setlength{\unitlength}{0.00087489in}
\begingroup\makeatletter\ifx\SetFigFontNFSS\undefined%
\gdef\SetFigFontNFSS#1#2#3#4#5{%
  \reset@font\fontsize{#1}{#2pt}%
  \fontfamily{#3}\fontseries{#4}\fontshape{#5}%
  \selectfont}%
\fi\endgroup%
{\renewcommand{\dashlinestretch}{30}
\begin{picture}(5199,3054)(0,-10)
\path(2532,2127)(2442,1677)(2802,2127)
	(2712,1677)(2532,2082)
\path(4062,2127)(4287,1677)(4017,1677)
	(4332,2127)(4287,1677)
\dashline{30.000}(4062,2127)(4017,1677)
\path(2847,597)(2982,417)(3117,597)
\path(1902,597)(1902,462)(2037,417)
	(2172,462)(2172,597)
\path(4512,597)(4512,417)(4782,417)(4782,597)
\put(3522,2622){\blacken\ellipse{36}{36}}
\put(3522,2622){\ellipse{36}{36}}
\put(3252,2622){\blacken\ellipse{36}{36}}
\put(3252,2622){\ellipse{36}{36}}
\path(3252,2622)(3522,2622)
\put(2712,1677){\blacken\ellipse{36}{36}}
\put(2712,1677){\ellipse{36}{36}}
\put(2442,1677){\blacken\ellipse{36}{36}}
\put(2442,1677){\ellipse{36}{36}}
\put(2802,2127){\blacken\ellipse{36}{36}}
\put(2802,2127){\ellipse{36}{36}}
\put(2532,2127){\blacken\ellipse{36}{36}}
\put(2532,2127){\ellipse{36}{36}}
\put(2617,1892){\blacken\ellipse{36}{36}}
\put(2617,1892){\ellipse{36}{36}}
\put(4017,1677){\blacken\ellipse{36}{36}}
\put(4017,1677){\ellipse{36}{36}}
\put(4332,2127){\blacken\ellipse{36}{36}}
\put(4332,2127){\ellipse{36}{36}}
\put(4062,2127){\blacken\ellipse{36}{36}}
\put(4062,2127){\ellipse{36}{36}}
\put(4287,1677){\blacken\ellipse{36}{36}}
\put(4287,1677){\ellipse{36}{36}}
\put(4177,1902){\blacken\ellipse{36}{36}}
\put(4177,1902){\ellipse{36}{36}}
\put(2532,1047){\blacken\ellipse{36}{36}}
\put(2532,1047){\ellipse{36}{36}}
\put(2262,1047){\blacken\ellipse{36}{36}}
\put(2262,1047){\ellipse{36}{36}}
\path(2262,1047)(2532,1047)
\put(2097,1017){\arc{210}{1.5708}{3.1416}}
\put(2097,1032){\arc{210}{3.1416}{4.7124}}
\put(2697,1032){\arc{210}{4.7124}{6.2832}}
\put(2697,1017){\arc{210}{0}{1.5708}}
\path(1992,1017)(1992,1032)
\path(2097,1137)(2697,1137)
\path(2802,1032)(2802,1017)
\path(2697,912)(2097,912)
\put(2172,597){\blacken\ellipse{36}{36}}
\put(2172,597){\ellipse{36}{36}}
\put(1902,597){\blacken\ellipse{36}{36}}
\put(1902,597){\ellipse{36}{36}}
\put(3117,597){\blacken\ellipse{36}{36}}
\put(3117,597){\ellipse{36}{36}}
\put(2847,597){\blacken\ellipse{36}{36}}
\put(2847,597){\ellipse{36}{36}}
\put(1902,462){\blacken\ellipse{36}{36}}
\put(1902,462){\ellipse{36}{36}}
\put(2037,417){\blacken\ellipse{36}{36}}
\put(2037,417){\ellipse{36}{36}}
\put(2172,462){\blacken\ellipse{36}{36}}
\put(2172,462){\ellipse{36}{36}}
\put(2982,417){\blacken\ellipse{36}{36}}
\put(2982,417){\ellipse{36}{36}}
\put(4512,1047){\blacken\ellipse{36}{36}}
\put(4512,1047){\ellipse{36}{36}}
\put(4242,1047){\blacken\ellipse{36}{36}}
\put(4242,1047){\ellipse{36}{36}}
\path(4242,1047)(4512,1047)
\put(4782,597){\blacken\ellipse{36}{36}}
\put(4782,597){\ellipse{36}{36}}
\put(4512,597){\blacken\ellipse{36}{36}}
\put(4512,597){\ellipse{36}{36}}
\put(4512,417){\blacken\ellipse{36}{36}}
\put(4512,417){\ellipse{36}{36}}
\put(4782,417){\blacken\ellipse{36}{36}}
\put(4782,417){\ellipse{36}{36}}
\put(544,1000){\blacken\ellipse{36}{36}}
\put(544,1000){\ellipse{36}{36}}
\put(547,997){\ellipse{190}{190}}
\put(3087,2592){\arc{210}{1.5708}{3.1416}}
\put(3087,2607){\arc{210}{3.1416}{4.7124}}
\put(3687,2607){\arc{210}{4.7124}{6.2832}}
\put(3687,2592){\arc{210}{0}{1.5708}}
\path(2982,2592)(2982,2607)
\path(3087,2712)(3687,2712)
\path(3792,2607)(3792,2592)
\path(3687,2487)(3087,2487)
\dashline{30.000}(2442,1677)(2712,1677)
\dashline{30.000}(2532,2127)(2802,2127)
\put(3807,1692){\arc{210}{1.5708}{3.1416}}
\put(3807,2112){\arc{210}{3.1416}{4.7124}}
\put(4587,2112){\arc{210}{4.7124}{6.2832}}
\put(4587,1692){\arc{210}{0}{1.5708}}
\path(3702,1692)(3702,2112)
\path(3807,2217)(4587,2217)
\path(4692,2112)(4692,1692)
\path(4587,1587)(3807,1587)
\dashline{30.000}(4062,2127)(4332,2127)
\put(2277,1692){\arc{210}{1.5708}{3.1416}}
\put(2277,2112){\arc{210}{3.1416}{4.7124}}
\put(3057,2112){\arc{210}{4.7124}{6.2832}}
\put(3057,1692){\arc{210}{0}{1.5708}}
\path(2172,1692)(2172,2112)
\path(2277,2217)(3057,2217)
\path(3162,2112)(3162,1692)
\path(3057,1587)(2277,1587)
\path(3162,2487)(2982,2217)
\path(3612,2487)(3837,2217)
\dashline{60.000}(2037,2307)(3252,2307)(3567,1407)
	(3567,102)(12,102)(12,1407)(2037,2307)
\path(2307,912)(2127,687)
\path(2577,912)(2757,687)
\put(2637,432){\arc{210}{1.5708}{3.1416}}
\put(2637,582){\arc{210}{3.1416}{4.7124}}
\put(3237,582){\arc{210}{4.7124}{6.2832}}
\put(3237,432){\arc{210}{0}{1.5708}}
\path(2532,432)(2532,582)
\path(2637,687)(3237,687)
\path(3342,582)(3342,432)
\path(3237,327)(2637,327)
\put(1692,432){\arc{210}{1.5708}{3.1416}}
\put(1692,582){\arc{210}{3.1416}{4.7124}}
\put(2292,582){\arc{210}{4.7124}{6.2832}}
\put(2292,432){\arc{210}{0}{1.5708}}
\path(1587,432)(1587,582)
\path(1692,687)(2292,687)
\path(2397,582)(2397,432)
\path(2292,327)(1692,327)
\path(2577,1587)(2442,1137)
\dashline{30.000}(1902,597)(2172,597)
\dashline{30.000}(2847,597)(3117,597)
\put(4077,1017){\arc{210}{1.5708}{3.1416}}
\put(4077,1032){\arc{210}{3.1416}{4.7124}}
\put(4677,1032){\arc{210}{4.7124}{6.2832}}
\put(4677,1017){\arc{210}{0}{1.5708}}
\path(3972,1017)(3972,1032)
\path(4077,1137)(4677,1137)
\path(4782,1032)(4782,1017)
\path(4677,912)(4077,912)
\path(4467,912)(4647,687)
\path(4107,1587)(4422,1137)
\put(4302,432){\arc{210}{1.5708}{3.1416}}
\put(4302,582){\arc{210}{3.1416}{4.7124}}
\put(4902,582){\arc{210}{4.7124}{6.2832}}
\put(4902,432){\arc{210}{0}{1.5708}}
\path(4197,432)(4197,582)
\path(4302,687)(4902,687)
\path(5007,582)(5007,432)
\path(4902,327)(4302,327)
\dashline{30.000}(4512,597)(4782,597)
\path(552,1092)(2397,1587)
\path(507,912)(372,687)
\path(372,687)(282,417)(462,417)(372,687)
\path(732,687)(642,417)(822,417)(732,687)
\path(597,912)(732,687)
\dottedline{45}(1947,1227)(2937,1227)(3477,687)
	(3477,237)(1497,237)(1497,687)(1947,1227)
\dottedline{45}(327,1227)(192,687)(192,237)
	(912,237)(912,687)(777,1227)(327,1227)
\dottedline{105}(2127,2847)(4647,2847)(5187,1587)
	(5187,12)(1317,12)(1317,1587)(2127,2847)
\put(2082,957){\makebox(0,0)[lb]{\smash{{\SetFigFontNFSS{12}{14.4}{\rmdefault}{\mddefault}{\updefault}$a$}}}}
\put(3072,2532){\makebox(0,0)[lb]{\smash{{\SetFigFontNFSS{12}{14.4}{\rmdefault}{\mddefault}{\updefault}$u$}}}}
\put(3567,2532){\makebox(0,0)[lb]{\smash{{\SetFigFontNFSS{12}{14.4}{\rmdefault}{\mddefault}{\updefault}$v$}}}}
\put(2847,2577){\makebox(0,0)[lb]{\smash{{\SetFigFontNFSS{12}{14.4}{\rmdefault}{\mddefault}{\updefault}$s$}}}}
\put(2352,2037){\makebox(0,0)[lb]{\smash{{\SetFigFontNFSS{12}{14.4}{\rmdefault}{\mddefault}{\updefault}$u$}}}}
\put(2847,2037){\makebox(0,0)[lb]{\smash{{\SetFigFontNFSS{12}{14.4}{\rmdefault}{\mddefault}{\updefault}$v$}}}}
\put(2757,1677){\makebox(0,0)[lb]{\smash{{\SetFigFontNFSS{12}{14.4}{\rmdefault}{\mddefault}{\updefault}$b$}}}}
\put(2307,1677){\makebox(0,0)[lb]{\smash{{\SetFigFontNFSS{12}{14.4}{\rmdefault}{\mddefault}{\updefault}$a$}}}}
\put(3882,2037){\makebox(0,0)[lb]{\smash{{\SetFigFontNFSS{12}{14.4}{\rmdefault}{\mddefault}{\updefault}$u$}}}}
\put(4377,2037){\makebox(0,0)[lb]{\smash{{\SetFigFontNFSS{12}{14.4}{\rmdefault}{\mddefault}{\updefault}$v$}}}}
\put(3837,1632){\makebox(0,0)[lb]{\smash{{\SetFigFontNFSS{12}{14.4}{\rmdefault}{\mddefault}{\updefault}$w$}}}}
\put(2037,2352){\makebox(0,0)[lb]{\smash{{\SetFigFontNFSS{12}{14.4}{\rmdefault}{\mddefault}{\updefault}$S_A$}}}}
\put(1992,1992){\makebox(0,0)[lb]{\smash{{\SetFigFontNFSS{12}{14.4}{\rmdefault}{\mddefault}{\updefault}$A$}}}}
\put(1902,2892){\makebox(0,0)[lb]{\smash{{\SetFigFontNFSS{12}{14.4}{\rmdefault}{\mddefault}{\updefault}${\sf T}(B)$}}}}
\put(2577,957){\makebox(0,0)[lb]{\smash{{\SetFigFontNFSS{12}{14.4}{\rmdefault}{\mddefault}{\updefault}$b$}}}}
\put(2217,507){\makebox(0,0)[lb]{\smash{{\SetFigFontNFSS{12}{14.4}{\rmdefault}{\mddefault}{\updefault}$b$}}}}
\put(3162,507){\makebox(0,0)[lb]{\smash{{\SetFigFontNFSS{12}{14.4}{\rmdefault}{\mddefault}{\updefault}$b$}}}}
\put(1722,507){\makebox(0,0)[lb]{\smash{{\SetFigFontNFSS{12}{14.4}{\rmdefault}{\mddefault}{\updefault}$a$}}}}
\put(2667,507){\makebox(0,0)[lb]{\smash{{\SetFigFontNFSS{12}{14.4}{\rmdefault}{\mddefault}{\updefault}$a$}}}}
\put(4557,957){\makebox(0,0)[lb]{\smash{{\SetFigFontNFSS{12}{14.4}{\rmdefault}{\mddefault}{\updefault}$w$}}}}
\put(4062,957){\makebox(0,0)[lb]{\smash{{\SetFigFontNFSS{12}{14.4}{\rmdefault}{\mddefault}{\updefault}$u$}}}}
\put(4332,507){\makebox(0,0)[lb]{\smash{{\SetFigFontNFSS{12}{14.4}{\rmdefault}{\mddefault}{\updefault}$u$}}}}
\put(4827,507){\makebox(0,0)[lb]{\smash{{\SetFigFontNFSS{12}{14.4}{\rmdefault}{\mddefault}{\updefault}$w$}}}}
\put(327,957){\makebox(0,0)[lb]{\smash{{\SetFigFontNFSS{12}{14.4}{\rmdefault}{\mddefault}{\updefault}$a$}}}}
\put(192,1272){\makebox(0,0)[lb]{\smash{{\SetFigFontNFSS{12}{14.4}{\rmdefault}{\mddefault}{\updefault}$S_a$}}}}
\put(1992,1272){\makebox(0,0)[lb]{\smash{{\SetFigFontNFSS{12}{14.4}{\rmdefault}{\mddefault}{\updefault}$S_{(a,b)}$}}}}
\end{picture}
}

%% file: main.bbl
\begin{thebibliography}{ADK08}

\bibitem[AD04]{AD04}
Vikraman Arvind and Nikhil Devanur.
\newblock Symmetry breaking in trees and planar graphs by vertex coloring.
\newblock In {\em In Proc. The Nordic Combinatorial Conference, NORCOM}, 2004.

\bibitem[ADK08]{ADK08}
Vikraman Arvind, Bireswar Das, and Johannes K\"obler.
\newblock A logspace algorithm for partial 2-tree canonization.
\newblock In {\em CSR 2008: Computer Science Symposium in Russia}, pages
  40--51, 2008.

\bibitem[ADR05]{ADR05}
Eric Allender, Samir Datta, and Sambuddha Roy.
\newblock The directed planar reachability problem.
\newblock In {\em Proc. 25th annual Conference on Foundations of Software
  Technology and Theoretical Computer Science (FSTTCS)}, pages 238--249., 2005.

\bibitem[AK06]{AK06}
V.~Arvind and Piyush~P. Kurur.
\newblock Graph isomorphism is in spp.
\newblock {\em Information and Computation}, 204(5):835--852, 2006.

\bibitem[AM00]{AM00}
Eric Allender and Meena Mahajan.
\newblock The complexity of planarity testing.
\newblock In {\em STACS '00: Proceedings of the 17th Annual Symposium on
  Theoretical Aspects of Computer Science}, pages 87--98, 2000.

\bibitem[Art96]{Artin96}
M.~Artin.
\newblock Algebra.
\newblock {\em Prentice Hall, India, New Delhi}, 1996.

\bibitem[Bab95]{Babai95}
L\'{a}szl\'{o} Babai.
\newblock Automorphism groups, isomorphism, reconstruction.
\newblock {\em Handbook of combinatorics (vol. 2)}, pages 1447--1540, 1995.

\bibitem[BHZ87]{BHZ87}
R.~B. Boppana, J.~Hastad, and S.~Zachos.
\newblock Does co-{NP} have short interactive proofs?
\newblock {\em Inf. Process. Lett.}, 25(2):127--132, 1987.

\bibitem[BL83]{BL83}
L\'{a}szl\'{o} Babai and Eugene~M. Luks.
\newblock Canonical labeling of graphs.
\newblock In {\em STOC '83: Proceedings of the fifteenth annual ACM symposium
  on Theory of computing}, pages 171--183, 1983.

\bibitem[BTV07]{BTV07}
Chris Bourke, Raghunath Tewari, and N~V Vinodchandran.
\newblock Directed planar reachability is in unambiguous logspace.
\newblock In {\em to appear in Proceedings of {IEEE} Conference on
  Computational Complexity CCC}, pages~--, 2007.

\bibitem[Bus97]{Buss97}
Samuel~R. Buss.
\newblock Alogtime algorithms for tree isomorphism, comparison, and
  canonization.
\newblock In {\em KGC '97: Proceedings of the 5th Kurt G\"{o}del Colloquium on
  Computational Logic and Proof Theory}, pages 18--33, 1997.

\bibitem[Coo85]{Coo85}
Stephen~A. Cook.
\newblock A taxonomy of problems with fast parallel algorithms.
\newblock {\em Inf. Control}, 64(1-3):2--22, 1985.

\bibitem[DLN08]{DLN08}
Samir Datta, Nutan Limaye, and Prajakta Nimbhorkar.
\newblock $3$-connected planar graph isomorphism is in log-space.
\newblock In {\em FSTTCS (to appear)}, 2008.

\bibitem[HT73]{HT73}
John~E. Hopcroft and Robert~E. Tarjan.
\newblock Dividing a graph into triconnected components.
\newblock {\em SIAM Journal on Computing}, 2(3):135--158, 1973.

\bibitem[HT74]{HT74}
John~E. Hopcroft and Robert Tarjan.
\newblock Efficient planarity testing.
\newblock {\em J. ACM}, 21(4):549--568, 1974.

\bibitem[HW74]{HW74}
John~E. Hopcroft and J.~K. Wong.
\newblock Linear time algorithm for isomorphism of planar graphs (preliminary
  report).
\newblock In {\em STOC '74: Proceedings of the sixth annual ACM symposium on
  Theory of computing}, pages 172--184, 1974.

\bibitem[JT98]{MJT98}
Pierre McKenzie~Birgit Jenner and Jacobo Tor\'{a}n.
\newblock A note on the hardness of tree isomorphism.
\newblock In {\em COCO '98: Proceedings of the Thirteenth Annual IEEE
  Conference on Computational Complexity}. IEEE Computer Society, 1998.

\bibitem[KHC04]{KHC04}
Jacek~P. Kukluk, Lawrence~B. Holder, and Diane~J. Cook.
\newblock Algorithm and experiments in testing planar graphs for isomorphism.
\newblock {\em J. Graph Algorithms Appl.}, 8(2):313--356, 2004.

\bibitem[Lin92]{Lin92}
Steven Lindell.
\newblock A logspace algorithm for tree canonization (extended abstract).
\newblock In {\em STOC '92: Proceedings of the twenty-fourth annual ACM
  symposium on Theory of computing}, pages 400--404, 1992.

\bibitem[Mac37]{M37}
Saunders Maclane.
\newblock A structural characterization of planar combinatorial graphs.
\newblock {\em Duke Mathematical Journal}, 3:460--472, 1937.

\bibitem[MR91]{MR91}
Gary~L. Miller and John~H. Reif.
\newblock Parallel tree contraction part 2: further applications.
\newblock {\em SIAM J. Comput.}, 20(6):1128--1147, 1991.

\bibitem[RA97]{RA97}
Klaus Reinhardt and Eric Allender.
\newblock Making nondeterminism unambiguous.
\newblock In {\em {IEEE} Symposium on Foundations of Computer Science}, pages
  244--253, 1997.

\bibitem[Rei05]{Rei05}
Omer Reingold.
\newblock Undirected st-connectivity in log-space.
\newblock In {\em STOC '05: Proceedings of the thirty-seventh annual ACM
  symposium on Theory of computing}, pages 376--385, 2005.

\bibitem[RR90]{RR94}
Vijaya Ramachandran and John Reif.
\newblock Planarity testing in parallel.
\newblock Technical report, 1990.

\bibitem[Sch88]{S88}
Uwe Sch\"{o}ning.
\newblock Graph isomorphism is in the low hierarchy.
\newblock {\em J. Comput. Syst. Sci.}, 37(3):312--323, 1988.

\bibitem[Tor04]{Tor04}
Jacobo Tor\'{a}n.
\newblock On the hardness of graph isomorphism.
\newblock {\em SIAM J. Comput.}, 33(5):1093--1108, 2004.

\bibitem[TW08]{TW08}
Thomas Thierauf and Fabian Wagner.
\newblock The isomorphism problem for planar 3-connected graphs is in
  unambiguous logspace.
\newblock In {\em STACS}, pages 633--644, 2008.

\bibitem[Ver07]{Ver07}
Oleg Verbitsky.
\newblock Planar graphs: Logical complexity and parallel isomorphism tests.
\newblock In {\em STACS}, pages 682--693, 2007.

\bibitem[Wag07]{Wag07}
Fabian Wagner.
\newblock Hardness results for tournament isomorphism and automorphism.
\newblock In {\em MFCS}, pages 572--583, 2007.

\bibitem[Wei66]{Wei66}
H.~Weinberg.
\newblock A simple and efficient algorithm for determining isomorphism of
  planar triply connected graphs.
\newblock {\em Circuit Theory}, 13:142148, 1966.

\bibitem[Whi33]{Whi33}
H.~Whitney.
\newblock A set of topological invariants for graphs.
\newblock {\em American Journal of Mathematics}, 55:235--321, 1933.

\end{thebibliography}
